\documentclass[draft,nofootinbib,aps,pra,twocolumn,showpacs,groupaddress,preprintnumbers,floatfix]{revtex4-2}
\usepackage{dynlearn}
\usepackage{braket}
\usepackage{bbm}
\usepackage{comment}
\usepackage[toc,page]{appendix}
\usepackage{booktabs}
\usepackage{multirow}

\newcommand{\AX}{\mathcal{X}}
\newcommand{\AY}{\mathcal{Y}}
\newcommand{\AQ}{\mathcal{Q}}
\newcommand{\AS}{\mathcal{S}}
\newcommand{\AT}{\mathcal{T}}
\newcommand{\AM}{M}

\newcommand{\qer}{s}
\newcommand{\rhoL}{\rho_{0:\ell}}
\newcommand{\ketpsi}[1][]{\ket{\psi_{#1}}}
\newcommand{\brapsi}[1][]{\bra{\psi_{#1}}}
\newcommand{\ketPsi}[1][]{\ket{\Psi_{#1}}}

\newcommand{\QBiInfinity}{\ketPsi[-\infty:\infty]}
\newcommand{\cqChannel}{\mathcal{E}}
\newcommand{\qcChannel}{\mathcal{M}}
\newcommand{\ccChannel}{\mathcal{C}}
\newcommand{\HSpace}{\mathcal{H}}
\newcommand{\Meas}{M}
\newcommand{\Rq}{\mathbf{R}_q}
\newcommand{\Gq}{\mathbf{G}_q}
\newcommand{\MOrderq}{R_q}

\begin{document}

\def\ourTitle{Intrinsic and Measured Information\\
in\\
Separable Quantum Processes
}

\def\ourAbstract{Stationary quantum information sources emit sequences of correlated
qudits---that is, structured quantum stochastic processes. If an observer
performs identical measurements on a qudit sequence, the outcomes are a
realization of a classical stochastic process. We introduce
quantum-information-theoretic properties for separable qudit sequences that
serve as bounds on the classical information properties of subsequent measured
processes. For sources driven by hidden Markov dynamics we describe how an
observer can temporarily or permanently synchronize to the source's internal
state using specific positive operator-valued measures or adaptive measurement
protocols. We introduce a method for approximating an information source with
an independent and identically-distributed, Markov, or larger memory model
through tomographic reconstruction. We identify broad classes of separable
processes based on their quantum information properties and the complexity of
measurements required to synchronize to and accurately reconstruct them.
}

\author{David Gier}
\email{dgier@ucdavis.edu}
\affiliation{Complexity Sciences Center and Physics and Astronomy Department,
University of California at Davis, One Shields Avenue, Davis, CA 95616}

\author{James P. Crutchfield}
\email{chaos@ucdavis.edu}
\affiliation{Complexity Sciences Center and Physics and Astronomy Department,
University of California at Davis, One Shields Avenue, Davis, CA 95616}

\date{\today}
\bibliographystyle{unsrt}

\title{\ourTitle}

\begin{abstract}
\ourAbstract
\end{abstract}

\title{\ourTitle}
\date{\today}
\maketitle

\begin{spacing}{0.9}
\tableofcontents
\end{spacing}

\setstretch{1.1}

\section{Introduction}
\label{sec:introduction}

Determining a quantum system's state requires grappling with multiple sources
of uncertainty, including several that do not arise in classical physics.
Irreducible limits on measurement, in particular, have been a hallmark of
quantum physics since Heisenberg introduced the position-momentum uncertainty
principle in 1927 \cite{Heis27}. Similar incompatible measurements exist for
generic pure quantum states \cite{Pere91}.

For a $2$-level quantum system, or \emph{qubit}, it is impossible to
simultaneously measure the value of a spin in the $x$-, $y$-, and
$z$-directions. (Stated mathematically, the Pauli matrices $\sigma_x$,
$\sigma_y$, and $\sigma_z$ do not commute.) Additionally, a single measurement
in each basis is insufficient. One must measure many copies in each basis to
specify the distribution of outcomes. As a result, determining an unknown qubit
state through quantum state tomography requires measuring a large ensemble of
identical copies with a set of mutually unbiased bases \cite{Woot89} or a
single informationally-complete \emph{positive operator-valued measure} (POVM)
\cite{Rene04}.

These sources of uncertainty are familiar in quantum physics. Contrast them
with when an observer receives a sequence of correlated qubits. Measuring them
one by one, what will they see? And, what then can they infer about the
resources necessary to generate these qubit strings? The following answers
these questions by teasing apart the sources of apparent randomness and
correlation in measured quantum processes.

\subsection{Quantum and Classical Randomness} 

Also in 1927, von Neumann formulated quantum mechanics in terms of statistical
ensembles and quantified the entropy of these \emph{mixed quantum states}. In
doing so, he extended Gibbs' work on statistical ensembles and classical
thermodynamic entropies to the quantum domain \cite{VonN18}. A mixed quantum
state $\rho$ has an entropy $S(\rho) = -tr(\rho \log\rho)$, now known as the
\emph{von Neumann entropy}. ($tr(\cdot)$ is the trace operator.) $S(\rho) = 0$
if and only if $\rho$ is a pure (nonmixed) quantum state. On the one hand, the
von Neumann entropy is key to understanding quantum systems, particularly those
with \emph{entangled} subsystems that exhibit nonclassical correlations. On the
other, the uncertainty $S(\rho)$ quantifies a generic feature of statistical
ensembles. It does not correspond to any particular quantum mechanical effect.

These two forms of uncertainty---due to ensembles and to quantum
indeterminacy---are combined within the framework of quantum information theory,
which generalizes classical information theory to quantum observables
\cite{Wild17}. One notable example is noiseless coding. Shannon quantified the
information content produced by a noiseless classical independent and
identically-distributed (i.i.d.) information source---one that emits a state
drawn from the same distribution at each timestep \cite{Shan48a}. Schumacher's
quantum noiseless coding theorem generalized this to quantum information
sources. This gave a new physical interpretation of the von Neumann entropy:
For an i.i.d. quantum source emitting state $\rho$, $S(\rho)$ is the number of
qubits required for a reliable compression scheme \cite{Schu95}.

\subsection{Sources with Memory}

Non-i.i.d. stationary information sources inject additional forms of
uncertainty. For example, a source may have an internal memory that induces
correlations between sequential qubits and therefore between measurement
outcomes. Such correlations may be purely classical or uniquely quantal in
nature. As we will show, an experimenter who assumes (incorrectly) that such a
source is i.i.d. and then applies existing tomographic methods will not detect
these correlations and so will overestimate the source's randomness and
underestimate its compressibility.

Classical memoryful sources are described within the framework of computational
mechanics, in which stationary dynamical systems serve as information sources
with their own internal states and dynamic \cite{Crut12a}. Sequential
finite-precision measurements of a dynamical system form a discrete-time
stochastic process. The resulting process' statistics allow one to construct a
model of the source and calculate its asymptotic entropy rate, internal memory
requirements, and other physically-relevant properties \cite{Crut88a,Crut01a}.
Importantly, the uncertainty associated with sequential measurements of a
classical information source can be reduced, sometimes substantially, by an
observer capable of synchronizing to the process' internal states
\cite{Trav11a,Trav10b}.

Subjecting an open quantum system to sequential qudit probes presents a similar
but more general challenge, as the amount of information an observer can glean
from an individual qudit through measurement is limited. Recent results
established that applying particular measuring instruments to qubits induces
complex behavior in measurement sequences \cite{Vene19}. Here, we extend these
results by studying properties of the quantum states themselves in addition to
particular sequences of measurement outcomes.

\begin{figure*}
\includegraphics[width=\textwidth]{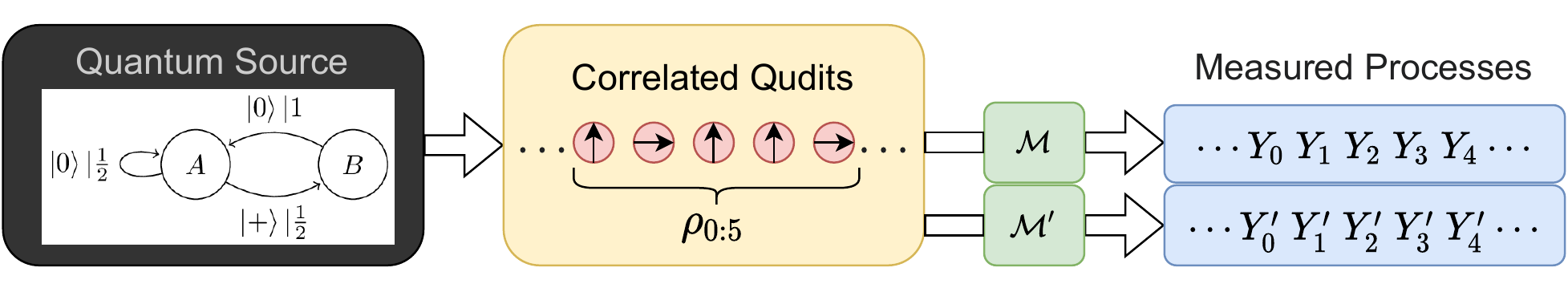}
\caption{A stationary quantum information source emits qudits that are
	correlated due to the source's internal memory. An experimenter measures
	these qudits in different ways ($\qcChannel$ or $\qcChannel'$) resulting in
	a family of classical stochastic processes.
	}
\label{fig:SPQS_Diagram}
\end{figure*}

The following introduces novel quantum-information-theoretic properties for
sequences of separable---i.e., nonentangled---qudits. We build on previous
results that focused on entropy rates, compression limits, and optimal coding
strategies for stationary quantum information sources
\cite{Hole98,Datt02,Petz01}, as well as on results for specific
experimentally-motivated deviations from the i.i.d. assumption \cite{Naga16}.
The approach is distinct from but complements recent efforts on quantum
stochastic processes in which an observer measures a quantum system directly.
This is complicated due to the latter's interaction with an inaccessible
environment that induces memory effects in sequential measurement outcomes
\cite{Poll18a,Poll18b,Tara19a,Tara21,Tara19c}.

Section \ref{sec:processes} introduces classical processes, separable qudit
processes, and methods of transforming from one to the other via
classical-quantum channels and measurement channels. Then Section
\ref{sec:quantum_properties}, in concert with App.
\ref{app:classical_properties}, defines the entropies associated with quantum
and classical processes, respectively. Adapting Ref. \cite{Crut01a}'s entropy
hierarchy, we employ discrete-time derivatives and integrals to obtain a family
of distinct quantitative measures of quantum process randomness and correlation.
We prove that, for projective or informationally-complete measurements, the
sequences of measurement outcomes form classical processes whose information
properties are bounded by those of the quantum process being measured.

Section \ref{sec:examples} then surveys examples of increasingly-structured
separable qubit and qutrit processes. Section \ref{sec:synchronization}
discusses how an observer can synchronize to a memoryful source---i.e.,
determine its internal state---through sequential measurement. Section
\ref{sec:tomography} uses the resulting catalog of possible process behaviors to
answer practical questions for an observer of a quantum process attempting to
perform tomography. Finally, Section \ref{sec:conclusion} draws out lessons and
proposes future directions and applications, most notably extending the results
to the experimentally-realizable generation of arbitrary entangled qudit states
\cite{Scho05a,Scho07a} and using correlations as a resource to perform
thermodynamic quantum information processing \cite{Boyd16c, Chap15}.

\section{Stochastic Processes}
\label{sec:processes}

We consider the output of an information source to be a discrete-time,
stationary stochastic process. If the source output is a classical random
variable---$\MeasSymbol_t$ for each timestep $t$---we can directly apply the
methods of computational mechanics \cite{Crut01a}. Our goal is to extend these
methods to describe separable sequences of qudits, each represented by a pure
state in $d$-dimensional Hilbert space: $\ketpsi[t] \in \HSpace^d$ at each
timestep $t$. Given such a qudit sequence, one can perform repeated, identical
measurements such that the outcomes form a classical stochastic process. Since
one can choose to measure qudit states in many different bases, the properties
of the classical measured process are determined by both the state of the
correlated qudits and the measurement choice. Thus, the relationship between a
quantum process and classical measured processes is one-to-many. Figure
\ref{fig:SPQS_Diagram} illustrates this setup.

\subsection{Classical Processes}

A \emph{classical stochastic process} is defined by a probability measure
$\Pr(\BiInfinity)$ over a chain of random variables:
\begin{align*}
\BiInfinity \equiv ... \MeasSymbol_{-2} \MeasSymbol_{-1} \Present
  \MeasSymbol_{1} \MeasSymbol_{2} \ldots
  ~,
\end{align*}
with each $\MeasSymbol_t$ taking on values drawn from a finite alphabet $\AX$. A
block of $\ell$ consecutive random variables is denoted $\LPresent{\ell} =
\MeasSymbol_{0} \MeasSymbol_{1} \cdots \MeasSymbol_{\ell-1} $. The indexing is
left-inclusive and right-exclusive. A particular bi-infinite process
\emph{realization} is denoted $\biinfinity \equiv ...\meassymbol_{-2}
\meassymbol_{-1} \present \meassymbol_{1} \meassymbol_{2}...$ with events
$\meassymbol_{t}$ taking values in a discrete set $\AX$. Realizations of a
block of length $\ell$ are known as \emph{words} and denoted $\lpresent{\ell}$.
The set of all words of length $\ell$ is $\AX^\ell$.

We consider processes that are \emph{stationary}, meaning that word
probabilities $\Pr(\LPresent{\ell})$ are time-independent:
\begin{align*}
\Pr(\LPresent{\ell}) = \Pr(\MeasSymbol_{t:t+\ell})
  ~, 
\end{align*}
for all $t \in \mathbb{Z}$ and $\ell \in \mathbb{Z}^+$. A stationary
process's statistics are fully described by the set of length-$\ell$ word
distributions $\Pr(\LPresent{\ell})$. A block of length $\ell$ has at most
$ \vert \AX \vert ^\ell$ possible realizations (words).

One important subclass of processes are \emph{independently and identically
distributed} (i.i.d.) processes. The joint block probabilities of an i.i.d.
process take the form:
\begin{align}
\Pr(\LPresent{\ell}) \equiv \Pr(\Present) \Pr(\MeasSymbol_{1}) \cdots
\Pr(\MeasSymbol_{\ell-1})
~,
\label{eq:iid}
\end{align}
for all $\ell \in \mathbb{Z}^+$. This factoring of the block probabilities
results in no statistical correlations between any random variables. Due to
stationarity $\Pr(\MeasSymbol_{t}) = \Pr(\Present)$ for all $t$ for an i.i.d.
process.

Another commonly-studied subclass consists of the \emph{Markov processes} for
which the distribution for each $\MeasSymbol_t$ depends only on the immediately
preceding random variable $\MeasSymbol_{t-1}$. For Markov processes the joint
probabilities for finite-length blocks factor as:
\begin{align}
\Pr(\LPresent{\ell}) \equiv \Pr(\Present) \Pr(\MeasSymbol_{1} \vert \Present) \cdots
\Pr(\MeasSymbol_{\ell-1} \vert \MeasSymbol_{\ell-2})
~,
\label{eq:markov}
\end{align}
where $\Pr(X \vert Y)$ is the probability distribution of random variable $X$
conditioned on random variable $Y$.
 
Finally, there is the markedly-larger subclass of \emph{hidden Markov
processes} that have an internal Markov dynamic that is not directly
observable. Though the joint probabilities do not factor as in Eq.
(\ref{eq:markov}), the internal Markov dynamic restricts the process statistics
as we describe next.

\subsection{Presentations}

A \emph{presentation} of a process is a model consisting of a set of internal
states and a transition dynamic between those states that together reproduce
the process' statistics exactly. A given process may have many presentations. We
focus on those depicted with state transition diagrams (directed graphs) that
generate stationary, discrete-time stochastic processes in a natural way.

A \emph{Markov chain} is a process presentation defined by the pair $(\AX,T)$:
\begin{itemize}
      \setlength{\topsep}{0pt}
      \setlength{\itemsep}{0pt}
      \setlength{\parsep}{0pt}
	\item A finite alphabet $\AX$ of $m$ symbols $\meassymbol \in \AX$, and
	\item A $m \times m$ transition matrix $T$. That is, if the source emits	symbol
	$\meassymbol_i$, with probability $T_{ij} =
	\Pr(\meassymbol_j \vert \meassymbol_i)$ it emits symbol $\meassymbol_j$ next.
\end{itemize} 
The stationary distribution for a Markov chain is denoted $\pi$, and it is a
distribution over causal states in $\AS$ that satisfies $\pi = \pi T$. For a
Markov chain the set of internal states is exactly the set of emitted symbols
since the probability distribution for the next symbol is completely determined
by the previous symbol. We represent each state as a node in a graph and each
transition as a directed edge between nodes labeled by the associated
probability.

Markov chains are sufficient to represent Markov processes, but we can describe
the more general class of hidden Markov processes by allowing for internal
states not directly observable. These processes are generated by
\emph{hidden Markov chains} (HMCs), defined by the triple, $(\AS,\AX, \AT)$:
\begin{itemize}
      \setlength{\topsep}{0pt}
      \setlength{\itemsep}{0pt}
      \setlength{\parsep}{0pt}
	\item A finite set $\AS = \{\sigma_1, \ldots, \sigma_n\}$ of internal states,
	\item A finite alphabet $\AX$ of $m$ symbols $\meassymbol \in \AX$, and
	\item A set $\AT = \{T^x : x \in \AX\}$ of $m$ $n \times n$ symbol-labeled
	transition matrices. That is, if the source is in state $\sigma_i$, with
	probability $T^x_{ij} = \Pr(x,\sigma_j \vert \sigma_i)$ it emits symbol $x$
	while transitioning to state $\sigma_j$.
\end{itemize}
We represent each possible transition between states as an edge between their
nodes labeled with the emitted symbol and the transition probability. An HMC's
stationary distribution $\pi$ uniquely satisfies $\pi = \pi \sum_x T^x$.

Any HMC that exactly reproduces a process' statistical features is a
\emph{generative} HMC. This is an important distinction, since only some of
those also belong to the more restrictive class of \emph{predictive} HMCs. An
HMC is predictive if its state at time $t+1$ is completely determined by the
state at time $t$ and the emitted symbol. This property is known as
\emph{unifilarity}.

At this point we must emphasize the difference between a process and a
particular presentation of that process. This distinction is critical when
designating processes and models to be `classical' or `quantum'. A discrete-time
classical stochastic processes is classical because it consists of a chain of
\emph{classical} random variables. Markov chains and HMCs are classical models
because their internal states are dynamics are both \emph{classical}. One may
instead construct a presentation of a classical process with a set of
\emph{quantum} states that the model transitions between via some \emph{quantum}
dynamic. An observer can recover the classical process' statistics by taking
sequential measurements on either the system or on ancilla qudits that interact
with the system at each timestep. The simulation of classical stochastic
processes with quantum resources is the objective of \emph{quantum computational
mechanics}. There, a class of quantum models (q-simulators) shows advantage in
terms of memory requirements over provably-minimal classical predictive models
(\eMs) \cite{Gu12a, Maho15a, Suen22, Bind18, Loom19}. Likewise different
presentations of a quantum processes may have an underlying dynamic that is
either classical or quantum. We turn now to quantum processes and their presentations.

\subsection{Quantum Processes}

Discrete-time classical stochastic processes consist of one classical random
variable for each timestep. Likewise discrete-time quantum stochastic processes
consist of one quantum state $\ketpsi[t] \in \HSpace^d$ at each timestep. We
first describe an i.i.d. quantum information source and then generalize to
sources with memory.

\subsubsection{Memoryless}

A discrete-time \emph{quantum information source} emits a $d$-level quantum
system or \emph{qudit} at each timestep. The statistical mixture of the infinite
qudit sequences emitted by a source is a \emph{quantum process}. As in the
classical setting, different classes of quantum processes are distinguished by
their temporal correlations. Now, however, for quantum sources we must use
quantum information theory to account for both classical and quantal
correlations.

First, consider the output of an i.i.d. (memoryless) quantum information
source. Let $\HSpace$ be a $d$-dimensional Hilbert space with pure states
$\ketpsi[\meassymbol] \in \HSpace$. A $d$-level i.i.d. quantum information source consists of
a set $\AQ$ of pure qudit states and a probability distribution over those
states such that $\Pr(\ketpsi[\meassymbol]) > 0$ for all $\ketpsi[\meassymbol] \in \AQ$. We refer to $\AQ$
as a quantum alphabet and consider only quantum alphabets with a finite number
of pure states.

At each discrete timestep $t$, the source emits state $\ketpsi[\meassymbol]$ with probability
$\Pr(\ketpsi[\meassymbol])$. The resulting ensemble is described by the $d \times d$ density
matrix:
\begin{align}
\rho_{iid} = \sum_{\ketpsi[\meassymbol] \in \AQ}
	\Pr(\ketpsi[\meassymbol]) \ketpsi[\meassymbol] \brapsi[\meassymbol]
~.
\label{eq:iid_density_matrix}
\end{align}
This particular pure-state decomposition of $\rho_{iid}$ is not unique.
Moreover, an observer cannot determine $\AQ$ through observations---unless
$\AQ$ consists of only one state---since many pure-state ensembles correspond
to the same density matrix.

If an i.i.d. source emits $\rho_{iid}$ at each timestep then the quantum
process generated by the source is simply the infinite tensor product state:
\begin{align}
\ketPsi[iid] = \cdots \otimes \rho_{iid} \otimes \rho_{iid}
  \otimes \rho_{iid} \otimes \cdots 
  ~.
\label{eq:iid_quantum_process}
\end{align}

\subsubsection{Memoryful}

We cannot describe non-i.i.d. sources using a single probability distribution
over $\AQ$ but must introduce a probability distribution over \emph{sequences}
of states drawn from $\AQ$. We do this by associating each element of $\AQ$ with
an element in the symbol alphabet $\AX$ of an underlying classical stochastic
process $\BiInfinity$. Infinite qudit sequences then inherit probabilities from
$\BiInfinity$. This construction results in qudit sequences that are
\emph{separable}---i.e., not entangled.

We express the relationship between symbols and pure quantum states via a
memoryless classical-quantum channel $\cqChannel: \AX \to \HSpace$, taking $x
\to \ketpsi[\meassymbol]$. This is also known as a \emph{preparation channel} (or
encoder); see App. \ref{appendix:cq_channels} for more.

Preparation channels are dual to measurement channels, described later, that
map quantum states to classical probability distributions and, via sampling, to
particular symbols.

For the classical process $\BiInfinity$ whose realizations consist of symbols
$x \in \AX$, the associated quantum alphabet $\AQ$ is constructed by passing
each element of $\AX$ through $\cqChannel$ such that $\AQ = \{\cqChannel(x)\}$.
Thus, $\AQ$ is completely determined by $\AX$ and $\cqChannel$. For example, in
the i.i.d. case Eq. (\ref{eq:iid_density_matrix}) can also be written
$\rho_{iid} = \cqChannel(X)$, and each possible pure-state decomposition of
$\rho_{iid}$ can now be interpreted as a different combination of classical
random variable $X$ and preparation channel $\cqChannel$.

In a slight abuse of notation we write $\QBiInfinity = \cqChannel(\BiInfinity)$
to indicate that quantum process $\QBiInfinity$ is formed by passing each random
variable of $\BiInfinity$ through the classical-quantum channel $\cqChannel$.

Note that an infinite qudit sequence (separable or entangled) can be viewed as a
one-dimensional lattice of qudits indexed by $t \in \mathbb{Z}$. These
possibly-entangled states can be described in full generality using an
operator-algebraic approach. We ground our formal definition of quantum
processes in this mathematical setting. (Reference \cite{Petz01} provides a more
detailed treatment of observable algebras for entangled qudit sequences over
$\mathbb{Z}$.)

Let $\mathcal{B}_t$ be the $d$-dimensional matrix algebra describing all
possible observables on lattice site $t$. (For $d=2$, the space of observables
is spanned by the identity and the $2 \times 2$ complex Pauli (Hermitian,
unitary) matrices.) The state of the qudit at site $t$ can be described by the
density matrix $\rho_t$ acting on $\HSpace_t$ of dimension $d$. For a block of
$\ell$ consecutive qudits, all observables can be described by the joint algebra
over $\ell$ sites of the lattice: $\mathcal{B}_{0:\ell} =
\bigotimes_{t=0}^{\ell-1} \mathcal{B}_t$, and the state of this block is $\rhoL$
acting on $\HSpace_{0:\ell} = \bigotimes_{t=0}^{\ell-1} \HSpace_t$, a Hilbert
space of dimension $d^\ell$. Combining all local algebras allows one to define
an algebra $\mathcal{B}$ over the infinite lattice. A quantum process is a
particular state over the infinite lattice and can be written as $\QBiInfinity$.

As a necessary first step and to more readily adapt information-theoretic tools
from classical processes, we return to the more restricted case: separable
sequences of qudits drawn from a finite alphabet $\AQ$ of pure qudit states.
Given a classical word $w = x_0 x_1 \ldots x_{\ell-1} \in \AX^\ell$ and a
preparation channel $\cqChannel: \AX \to \AQ$, a qudit sequence takes the form:
\begin{align}
\ketpsi[w] &= \cqChannel(w) \nonumber \\
 &= \ketpsi[\meassymbol_0] \otimes \ketpsi[\meassymbol_1] \otimes \cdots
	\otimes \ketpsi[\meassymbol_{\ell-1}]
  ~,
\label{eq:separable_word}
\end{align}
where $\ell$ is the length of the sequence and $\ketpsi[w] \in
\HSpace_{0:\ell}$. Note that $\dim(\HSpace_{0:\ell}) = d^\ell$, $\AQ =
\{\cqChannel(x)\}$, and the number of possible qudit sequences of length $\ell$
is $ \vert \AQ \vert ^\ell$ or---assuming all $\ketpsi[\meassymbol]$ are distinguishable---$ \vert \AX \vert ^\ell$.

A \emph{separable quantum process} is then defined by $\QBiInfinity =
\cqChannel(\BiInfinity)$. Different preparations---i.e., different combinations
of $\BiInfinity$ and $\cqChannel$---may produce the same quantum process.

The set of length-$\ell$-block density matrices for a quantum process is given
by:
\begin{align}
\rhoL & = \cqChannel(\MeasSymbol_{0:\ell}) \nonumber \\
& = \sum_{w \in \AX^\ell}
  \Pr(\ketpsi[w]) \ketpsi[w] \brapsi[w]
  ~,
\label{eq:finite_rho}
\end{align}
where $\ketpsi[w]$ are the separable vectors given in Eq.
(\ref{eq:separable_word}). Conveniently, their probabilities are determined by
those of the underlying classical stochastic process $\Pr(\BiInfinity)$:
$\Pr(\ketpsi[w]) = \Pr(w)$. Each $\rhoL$ is a finite subsystem of the pure
quantum state $\QBiInfinity$ over the infinite lattice. We use
left-inclusive/right-exclusive indexing for density matrices as well.

For a given $\rhoL$, one can also obtain a purification in a finite-dimensional
Hilbert space \cite{QCQI}. It is important to note that, since $\rhoL$ does not
have a unique pure-state decomposition, one cannot generally reconstruct the
probabilities $\Pr(\ketpsi[w])$ from it. Rather, $\rhoL$ contains only
information accessible to an observer. And, if $\AQ$ contains nonorthogonal
qudit states ($\braket{\psi_x \vert \psi_{x'}} \neq 0$ for some
$\ketpsi[x]$,$\ketpsi[x'] \in \AQ$) then an observer cannot unambiguously
distinguish them.

In addition to separability, we also focus on \emph{stationary} quantum
processes, meaning:
\begin{align}
\rhoL = \rho_{t:\ell+t}
  ~,
\label{eq:stationarity}
\end{align}
for all $\ell \in \mathbb{Z}^+$ and $t \in \mathbb{Z}$. If $\BiInfinity$ is
stationary, then $\QBiInfinity = \cqChannel(\BiInfinity)$ will be stationary by
construction.

For an i.i.d. quantum process the joint probabilities of $\BiInfinity$ factor as
in Eq. (\ref{eq:iid}), giving the quantum process the form of Eq.
(\ref{eq:iid_quantum_process}). The length-$\ell$-block density matrix is
represented by a product state:
\begin{align}
\rhoL = \bigotimes_{t=0}^{\ell-1} \rho_{iid}
~,
\label{eq:iid_quantum_block}
\end{align}
with $\rho_{iid}$ taking the form in Eq. (\ref{eq:iid_density_matrix}).

For an underlying classical process $\BiInfinity$ that is Markov, there are
additional subtleties. Joint probabilities of $\BiInfinity$ factor as in Eq.
(\ref{eq:markov}), so the joint probabilities of $\QBiInfinity =
\cqChannel(\BiInfinity)$ also factor so that:
\begin{align}
\Pr(\ketpsi[w]) & = \Pr(\ketpsi[\meassymbol_0])
  \Pr(\ketpsi[\meassymbol_1] \vert \ketpsi[\meassymbol_0]) \nonumber \\
  & \qquad \cdots \Pr(\ketpsi[\meassymbol_{\ell-1}]  \vert  \ketpsi[\meassymbol_{\ell-2}]) 
  ~.
\label{eq:quantum_markov}
\end{align}
However, an observer cannot reliably distinguish between different states
$\ketpsi[\meassymbol]$ when measuring a quantum process, and the underlying Markov dynamic
is hidden from observation. Thus, the general setting for memoryful quantum
processes is that of hidden Markov processes. These are best introduced using
concrete models that directly represent a process' structure.

\subsection{Presentations of Quantum Processes}

A \emph{presentation} for a quantum process is a model with internal states and
a transition dynamic between them that emits pure quantum states, rather than
classical symbols. As for presentations of classical processes, we depict them
with state transition diagrams. When $\BiInfinity$ is a Markov or hidden Markov
process, $\QBiInfinity = \cqChannel(\BiInfinity)$ can be represented with an
extension of Ref. \cite{Vene19}'s classically-controlled qubit sources (cCQS),
as follows.

A \emph{hidden Markov chain quantum source} (HMCQS) is a triple
$(\AS,\AQ, \AT)$ consisting of:
\begin{itemize}
      \setlength{\topsep}{0pt}
      \setlength{\itemsep}{0pt}
      \setlength{\parsep}{0pt}
\item A finite set $\AS = \{\sigma_1, \ldots, \sigma_n\}$ of
	internal states,
\item A finite alphabet $\AQ = \{\ketpsi[0], \ldots, \ketpsi[m-1]\}$
	of pure qudit states, with each $\ketpsi[\meassymbol] \in \HSpace^d$, and
\item A set $\AT = \{T^x : \ketpsi[\meassymbol] \in \AQ\}$ of $m$ $n \times n$
	transition matrices. That is, if the source is in state $\sigma_i$, with
	probability $T^x_{ij} = \Pr(\ketpsi[\meassymbol],\sigma_j \vert \sigma_i)$ it emits
	qudit $\ketpsi[\meassymbol]$ while transitioning to internal state $\sigma_j$.
\end{itemize}
As with HMCs, the stationary distribution for an HMCQS satisfies $\pi = \pi
\sum_{x} T^x$.

Any HMCQS that exactly reproduces a quantum process is a \emph{generative}
HMCQS, or generator of the process. Though quantum models cannot be predictive
in the same sense as classical models, we can define an analog to classical
unifilarity. A HMCQS is \emph{quantum unifilar} if, for every state $\sigma \in
\AS$, there exists a possible measurement such that the internal state at time
$t+1$ is completely determined if the state at time $t$ is $\sigma$. We discuss
several implications of quantum unifilarity later.

We call a HMCQS a \emph{classical} controller of a quantum process since there
is nothing quantal about its internal states or transition dynamic. This is in
contrast to related classes of quantum models that evolve a finite quantum
system according to a quantum operation (defined via a set of Kraus operators)
at each timestep. These include \emph{Quantum Markov Chains} (QMCs)
\cite{Gudd08} and \emph{Hidden Quantum Markov Models} (HQMMs) \cite{Monr10,
Wies08, Srin18}. While HMCQS emit separable quantum states, QMC and HQMM
generate sequences of measurement outcomes (each corresponding to a particular
Kraus operator) that form classical stochastic processes.

Anticipating future effort, we consider it worthwhile to draw out several
observations on entanglement between successive qudits at this point.
Entanglement means that finite-length qudit sequences are not separable and so
are not described by Eq. (\ref{eq:separable_word}). Moreover, their sequence
probabilities cannot be straightforwardly defined with reference to an
underlying classical stochastic process.

That said, there are systematic ways of defining stationary $\QBiInfinity$ such
that the set $\rhoL$ of marginals describe all measurements over blocks of
$\ell$ qudits. For example, if the source's internal structure consists of a
$D$-dimensional quantum system interacting unitarily with one qudit per
timestep, it generates a matrix product state (MPS) with a maximum bond
dimension of $D$ \cite{Pere06}. If the source operates stochastically (rather
than unitarily), then many different MPSs can be emitted with varying
probabilities. The collection is then described by \emph{matrix-product density
operators} (MPDO) \cite{Vers04}. We refer to these as \emph{entangled qudit
processes}. Their dynamical and informational analyses are left for elsewhere.
The present goal is to layout the basics for those efforts.

\subsection{Measured Processes}

An agent observing a quantum process has many ways to measure it. Let
$\Meas$ represent a measurement applied to the qudit in state $\rho$. In
general, $\Meas$ is a \emph{positive operator-valued measure} (POVM)
described by a set of positive semi-definite Hermitian operators $\{E_y\}$ on
the Hilbert space $\HSpace$ of dimension $d$. Each $E_y$ corresponds to a
possible measurement outcome $y$, and POVM elements must sum to the identity:
\begin{align*}
\sum_{y} E_{y} = \mathbb{I}
  ~.
\end{align*}

\emph{Projection-valued measures} (PVMs) are an important subclass of POVMs with
an additional constraint: operators $E_{y}$ must be orthogonal projectors. PVMs
have at most $d$ elements. A PVM consisting only of rank-one projectors on
$\HSpace^d$ is a \emph{von Neumann measurement} and has exactly $d$ elements
\cite{VonN18}.

A set of measurements applied to a block of $\ell$ qudits are described by some
block POVM $\qcChannel_{0:\ell}$ with elements $\{E_{y_{0:\ell}}\}$ on the Hilbert
space $\HSpace_{0:\ell}$ of dimension $d^\ell$. $\qcChannel_{0:\ell}$ may include
measurements in the joint basis of multiple qudits\textemdash measurements
essential for fully characterizing entangled processes.

For separable processes we focus on ``local'' measurements---operators on a
single qudit. The measurement operator for a block of $\ell$ qudits then takes a
tensor product structure:
\begin{align}
\qcChannel_{0:\ell} = \bigotimes_{t=0}^{\ell-1} \Meas_t
  ~,
\label{eq:local_measurement}
\end{align}
where each $\Meas_t$ is a POVM on $\HSpace_t$. 

If we apply the same local POVM $\Meas$ to each qudit, then
\begin{align}
\qcChannel_{0:\ell} = \bigotimes_{t=0}^{\ell-1} \Meas
  ~.
\label{eq:repeated_measurement}
\end{align}
We refer to this as a \emph{repeated POVM measurement}.

An observer can also have multiple POVMs at their disposal and apply different
measurements at different time steps according to some \emph{measurement
protocol}. We describe measurement protocols in more detail shortly.

For simplicity, the following ignores $\rhoL$'s post-measurement state and
considers only the measurement outcomes $y_{0:\ell}$. Thus, we take
$\qcChannel_{0:\ell}$ to be a stochastic map $\rhoL \to Y_{0:\ell}$, the
random variables representing measurement outcomes.

When applying a measurement of the form of Eq. (\ref{eq:repeated_measurement})
to a finite block of $\ell$ qudits, the outcomes factor into a block of $\ell$
classical random variables:
\begin{align*}
Y_{0:\ell} & = Y_{0} Y_{1} \cdots Y_{\ell-1} \\
   & = \qcChannel_{0:\ell}(\rhoL)
  ~,
\end{align*}
where the possible values of each $Y_t$ are the POVM measurement outcomes $y \in
\AY$. There are $ \vert \AY \vert ^\ell$ possible realizations of $Y_{0:\ell}$. We write a
realization (word) of length-$\ell$ as $y_{0:\ell}$.

The probability of any particular measurement outcome for a block of
$\ell$ qudits in state $\rhoL$ is:
\begin{align*}
\Pr(y_{0:\ell}) = tr(E_{y_{0:\ell}}\rhoL)
~.
\end{align*}
For $\rhoL$ with the separable form of Eq. (\ref{eq:finite_rho}) and identical
POVM measurements on each qudit as in Eq. (\ref{eq:repeated_measurement}), we
can decompose $E_{y_{0:\ell}}$ into $\ell$ local operators $E_{y_t}$:
\begin{align}
\Pr(y_{0:\ell}) & = tr\left(E_{y_{0:\ell}}
  \sum_{w \in \AX^\ell }
  \!\! \Pr(w) \ketpsi[w]\brapsi[w] \right)
  \nonumber \\
  & = tr\left( \sum_{w \in \AX^\ell }
  \!\! \Pr(w) \prod_{t=0}^{\ell-1}
  E_{y_t}\ketpsi[\meassymbol_t]\brapsi[\meassymbol_t]\right)
  .
\label{eq:measured_probs}
\end{align}

For a separable qudit process, a sequence $y_{0:\ell}$ of local measurement
outcomes can also be interpreted as the result of sending random variables
$X_{0:\ell}$ from $\BiInfinity$ over the same memoryless noisy channel
$\ccChannel: \AX \to \AY$. $\ccChannel$
decomposes into the deterministic preparation $\cqChannel$ and our 
stochastic measurement $\qcChannel$:
\begin{align*}
\ccChannel = \qcChannel \circ \cqChannel
  ~.
\end{align*}
(Appendix \ref{appendix:cq_channels} presents a more thorough description of the
classical-quantum channels $\cqChannel$ and $\qcChannel$.)

This construction makes it clear that measurement outcomes correspond to
classical random variables $Y_t$ that take values $y \in \AY$ and form a
classical process $\OutputBi$ with probabilities defined by Eq.
(\ref{eq:measured_probs}). To express the relationship between a quantum
process and a measured classical process we write:
\begin{align*}
\OutputBi = \qcChannel(\QBiInfinity)
  ~,
\end{align*}
where $\qcChannel$ is a repeated, local POVM. If the qudit process is
separable, we can also write:
\begin{align*}
\OutputBi = \ccChannel(\BiInfinity)
  ~.
\end{align*}

\subsection{Adaptive Measurement Protocols}

An observer does not need to repeat the same measurement on every qudit but may
apply different POVMs at different time steps according to some algorithm. If
the agent uses past measurement outcomes to inform their choice of POVM we say
they are using an \emph{adaptive measurement protocol}.

The following limits discussion to measurement protocols that have a
deterministic finite automata (DFA) as their underlying controller. Similar
constructions combining quantum measurement and DFAs have appeared in the
context of quantum grammars \cite{Moor97a,Qiu12,Zhen12}. 

A \emph{deterministic quantum measurement protocol}
(DQMP) is defined by the quintuple $(\AS, S_0,\AM,\AY, \delta)$:
\begin{itemize}
      \setlength{\topsep}{0pt}
      \setlength{\itemsep}{0pt}
      \setlength{\parsep}{0pt}
	\item A finite set $\AS = \{\sigma_1, \ldots, \sigma_n\}$ of internal states,
	\item A unique start state $S_0 \in \AS$,
	\item A set of POVMs $\AM = \{\Meas_s\}_{\sigma_s \in \AS}$, one for each internal state,
	\item An alphabet $\AY$ of $m$ symbols corresponding to different measurement
	outcomes, and
	\item A deterministic transition map $\delta: \AS \times \AY \to \AS$.
\end{itemize}

If $\AS$ consists of only $S_0$, then the DQMP is a repeated POVM measurement
for POVM $\Meas_{S_0}$. When $\AS$ has more than one internal state, the POVMs
corresponding to different states may have the same or a different number of
elements. Likewise the symbol sets corresponding to their measurement outcomes
may be disjoint or symbols may be repeated.

We can place the following bounds on the size of the set $\AY$: $m \leq \sum_s
\vert \{E_{s,y}\} \vert $, where $\{E_{s,y}\}$ is the set of operators
corresponding to POVM $\Meas_s$ and $m \geq \max_s \vert \{E_{s,y}\} \vert $,
the size of the POVM with the most elements.

For DQMP $\qcChannel$ and qudit process $\QBiInfinity$, obtaining a measured
process $\OutputBi = \qcChannel(\QBiInfinity)$ is generically more difficult
than for the case of repeated POVM measurements. When an observer begins using
protocol $\qcChannel$ at $t=0$ they experience two distinct operating
regimes: first the \emph{transient} dynamic, then the \emph{recurrent} dynamic.
We briefly outline this process and return to the subject when we describe
synchronization---a task deeply related to the transient dynamic---in Section
\ref{sec:synchronization}.

$\qcChannel$ begins in state $S_0$ at $t=0$. For a given (stationary, ergodic)
input $\QBiInfinity$, as $t \to \infty$ the DQMP approaches a stationary
distribution over a subset of its internal states $\pi = \{\Pr(\sigma_i) > 0,
\text{~for~all~} \sigma_i \in \AS_r\}$, where $\AS_r \subseteq \AS$ is the set
of recurrent states. This distribution (and even which states are in $\AS_r$)
depends on $\QBiInfinity$. The recurrent dynamic is determined by this
stationary distribution and the transition probabilities between states in
$\AS_r$. Any state not in $\AS_r$ is in the transient state set $\AS_t$.

The transient dynamic describes how $\qcChannel$ goes from $S_0$ at $t=0$ to its
recurrent dynamic over $\AS_r$, which may occur at a finite time $t = t_{sync}$
or only asymptotically as $t \to \infty$. In general two dynamics produce two
distinct measured processes $\OutputBi_r$, which is stationary and ergodic by
construction, and $Y_{0:t_{sync}}$ which is not. The final measured process has
two components---i.e. $\OutputBi = \{\OutputBi_r, Y_{0:t_{sync}}\}$.

\subsection{Discussion}

These nested layers of complication suggest working through a concrete example
and restating the overall goals.

Imagine an observer measures a single qubit from a quantum source that emitted
state $\rho_t$, using a projective measurement in the computational basis
$\Meas_{01} = \{\ket{0}\bra{0},\ket{1}\bra{1}\}$. The possible measurement
outcomes $y_0 = 0$ and $y_0 = 1$ occur with probabilities:
\begin{align*}
\Pr(y_0 = 0) & = tr(\ket{0}\bra{0}\rho_t) \\
\Pr(y_0 = 1) & = tr(\ket{1}\bra{1}\rho_t)
  ~,
\end{align*}
respectively. These two values determine the distribution for the random
variable $Y_0$ and, by applying the same projective measurement to $\rhoL$, we
completely determine the statistics $\Pr(Y_{0:\ell})$ of the measured block.
Continuing this procedure for $\ell \to \infty$ defines the measured process
$\OutputBi$.

Naturally, the observer can also choose to apply measurements in another
basis; e.g., $\Meas_{\pm} = \{\ket{+}\bra{+},\ket{-}\bra{-}\}$ where $\ket{\pm}
= \frac{1}{\sqrt{2}}(\ket{0} \pm \ket{1})$. This typically results in a
measured process $\OutputBi'$ with radically-different statistical features.

Finally, an observer could use an adaptive measurement protocol $\qcChannel$.
They start in state $s = S_0$ and measure with $\Meas_{01}$. If $y_0=0$ they
stay in $S_0$ and continue using $\Meas_{01}$. If $y_0 = 1$, they 
transition to a new internal state and use $\Meas_{\pm}$ on the next qubit.
Regardless of the outcome of $\Meas_{\pm}$ they return to $S_0$ and measure
the next qubit with $\Meas_{01}$. The measured process $\OutputBi''$ will be
distinct from both $\OutputBi$ and $\OutputBi'$ and may consist of both a
transient and recurrent component.

With this setting laid out, we can now more precisely state the questions the
following development answers:
\begin{enumerate}
      \setlength{\topsep}{0pt}
      \setlength{\itemsep}{0pt}
      \setlength{\parsep}{0pt}
\item Given the density matrices $\rhoL$ describing sequences of
	$\ell$ separable qudits, what are the general properties of sequences
	$Y_{0:\ell}$ of measurement outcomes? This is Section
	\ref{sec:quantum_properties}'s focus. There, $\rhoL$'s quantum
	information properties bound the classical information properties of
	measurement sequences $Y_{0:\ell}$ for certain classes of measurements.
\item Given a hidden Markov chain quantum source, when is an observer with
	knowledge of the source able to determine the internal state (synchronize)?
	Can the observer remain synchronized at later times? Section
	\ref{sec:synchronization} addresses this.
\item If an observer encounters an unknown qudit source, how accurately can the
	observer estimate the informational properties of the emitted process through
	tomography with limited-resources? How can they build approximate models of
	the source if they reconstruct $\rhoL$ for some finite $\ell$? This is
	Section \ref{sec:tomography}'s subject.
\end{enumerate}
Additionally, Section \ref{sec:examples} illustrates these general results and
the required analysis methods using specific examples of qudit processes.

\section{Information in Quantum Processes}
\label{sec:quantum_properties}

We wish to develop an information-theoretic analysis of quantum processes for
which the observed sequences depend on the observer's choice of measurement.
(Much of this parallels the classical information measures reviewed in App.
\ref{app:classical_properties}.) This requires a more general approach using
density matrices $\rhoL$ that contain all the information necessary to describe
the outcome of any measurement performed on $\ell$-qudit blocks. We use quantum
information theory to study properties of the set of $\rhoL$'s and then relate
them to classical properties of measurement sequences described in App.
\ref{app:classical_properties}. We begin by briefly reviewing several basic
quantities in quantum information theory. References \cite{Wild17} and
\cite{QCQI} give a more complete picture of the subject.

\subsection{von Neumann Entropy}

In quantum information theory the von Neumann entropy plays a role similar to
that of the Shannon entropy in classical information theory. Given a quantum
mixed quantum state $\rho$, the \emph{von Neumann entropy} is:
\begin{align}
S(\rho) & = -tr(\rho \log_2\rho) \nonumber \\
  & = -\sum_{i} \lambda_i \log_2 \lambda_i
  ~,
\label{eq:vne}
\end{align}
where $\lambda_i$'s are the eigenvalues of the density matrix $\rho$. $S(\rho) =
0$ if and only if $\rho$ is a pure state. We use $\log_2(\cdot)$, therefore the
units of the von Neumann entropy will be \emph{bits}.

From Eq. (\ref{eq:vne}), the von Neumann entropy is the Shannon entropy of the
eigenvalue distribution of density matrix $\rho$. Therefore:
\begin{align}
S(\rho) = \min_\Meas \H{\Meas(\rho)}
  ~,
\label{eq:shan_von}
\end{align}
where $\H{\cdot}$ is the Shannon entropy and the minimum is taken over the set
of all rank-one POVMs. The minimum will always be a PVM with projectors that
compose $\rho$'s eigenbasis \cite{Wild17}. We use brackets to indicate that
$M(\rho)$ is a classical probability distribution over measurement outcomes.

To monitor correlations between two quantum systems we use the quantum
relative entropy:
\begin{align}
S(\rho \Vert \sigma) \equiv tr(\rho \log_2 \rho) - tr(\rho \log_2 \sigma)
~,
\label{eq:quantum_relative_entropy}
\end{align}
where $\rho$ and $\sigma$ are the density operators of the two systems.
The quantum relative entropy is nonnegative:
\begin{align}
S(\rho \Vert \sigma) \geq 0
~,
\label{eq:quantum_relative_entropy_nonnegative}
\end{align}
with equality if and only if $\rho = \sigma$, a result known as \emph{Klein's
Inequality} \cite{QCQI}.

The \emph{joint quantum entropy} for a state $\rho^{AB}$ of a bipartite system
$AB$ is:
\begin{align*}
S(A,B) & = S(\rho^{AB}) \\
  & = -tr(\rho^{AB} \log_2 \rho^{AB})
~,
\end{align*}

We can further define a \emph{conditional quantum entropy} of system $A$
conditioned on system $B$ as:
\begin{align}
S(A\vert B) = S(A,B) - S(B)
\label{eq:cond_vn_ent}
~,
\end{align}
where $S(B) = S(\rho^B)$. Note that $S(A \vert B) \neq S(B \vert A)$.

In contrast to the classical case, the conditional quantum entropy may be
negative---a phenomenon leveraged in super-dense coding protocols \cite{Wild17}.
Equivalently, the conditional quantum entropy can be written using the quantum
relative entropy as:
\begin{align}
S(A\vert B) &= -S(\rho^{AB} \Vert \mathbb{I}^A \otimes \rho^B) \nonumber \\
& = \log_2(d_A)
-S \left(\rho^{AB} \big\Vert \frac{\mathbb{I}^A}{d_A} \otimes \rho^B \right)
\label{eq:conditional_quantum_entropy_relative}
~,
\end{align}
where $\mathbb{I}^A$ is the identity operator on Hilbert space $\HSpace^A$
with dimension $d_A$. 

The \emph{quantum mutual information} between quantum subsystems $A$ and $B$ is
given by:
\begin{align}
S(A{:}B) & = S(A) - S(A \vert B) \nonumber \\
    & = S(A) + S(B) - S(A,B)
~.
\label{eq:quantum_mutual_information}
\end{align}
The quantum mutual information is symmetric and nonnegative. If the joint
system $AB$ is in a pure state, then $S(A,B)$ will be zero, and $S(A) = S(B)$.
It can also be expressed as a quantum relative entropy as:
\begin{align}
S(A{:}B) & = S\left(\rho^{AB} \Vert \rho^A \otimes \rho^B\right)
~.
\label{eq:quantum_mutual_information_relative}
\end{align}

A few additional well-known properties of the von Neumann entropy facilitate
later results. First, each $\rhoL$ for separable qudit sequences is a finite
mixture of states formed from length-$\ell$ words of an underlying classical
process, so the following will be useful:

\begin{Lem}
Consider a random variable $X$ that takes values $x \in \{0,1,\ldots,n\}$ with
corresponding probabilities $\{p_0, p_1,\ldots p_n\}$. Given a set of density
matrices $\{\rho_0, \rho_1, \ldots, \rho_n\}$, the following inequality holds
\cite{QCQI}:
\begin{align*}
    S\left(\sum_{x=0}^n p_x \rho_x \right) \leq \H{X} + \sum_{x=0}^n p_x S(\rho_x)
~,
\end{align*}
with equality if and only if the all $\rho_x$ have support on orthogonal
subspaces.
\label{lem:mixture_entropy}
\end{Lem}

Second, since we use quantum channels to both prepare and measure a qudit process, we make use of the fact that the quantum relative entropy is
monotonic \cite{Wild17}:
\begin{align}
S(\rho \Vert \sigma) \geq S(\cqChannel(\rho) \Vert \cqChannel(\sigma))
~,
\label{eq:monotonicity_qre}
\end{align}
where $\cqChannel$ is any quantum channel. This inequality becomes an equality
if and only if there exists a recovery map $\mathcal{R}$ such that
$\mathcal{R}(\cqChannel(\rho)) = \rho$ and $\mathcal{R}(\cqChannel(\sigma)) =
\sigma$ \cite{Jung18}.

\subsection{Quantum Block Entropy}

Since stationary qudit processes are correlated across time, we explore how the
von Neumann entropy for qudit blocks scales with block size. The following gives
bounds on the possible measurement sequences one can observe from a quantum
information source. As the von Neumann entropy generalizes Shannon entropy, the
results here (and many of the proofs) are natural generalizations of those in
App. \ref{app:classical_properties}. We also note that exactly determining
$\rhoL$ becomes practically infeasible for large $\ell$. And so, Section
\ref{sec:tomography} addresses how to approximate properties and models for
qudit processes when restricted to measurements of finite length blocks.

For a qudit process we define the quantum block entropy as the von Neumann
entropy of a block of $\ell$ consecutive qudits:
\begin{align*}
S(\ell) \equiv -tr(\rhoL \log_2 \rhoL)
  ~.
\end{align*}
If $\rhoL$ is a pure state, $S(\ell) = 0$.
By the same logic as the classical case, $S(0) \equiv 0$.

Many properties of the classical block entropy hold for $S(\ell)$.
\begin{Prop}
For a stationary qudit process $S(\ell)$ is a nondecreasing function of $\ell$.
\end{Prop}

\begin{ProProp}
As a consequence of the strong subadditivity of the von Neumann entropy
\cite{QCQI}:
\begin{align}
S(\rho^A) + S(\rho^C) \leq S(\rho^{AB}) + S(\rho^{BC})
  ~.
\label{subadd2}
\end{align}

Let $\rho^{ABC} = \rho_{0:2\ell+1}$, where $\rho^A = \rhoL$, $\rho^B =
\rho_\ell$, and $\rho^C = \rho_{\ell+1:2\ell+1}$.

Incorporating qudit process stationarity, we rewrite Eq. (\ref{subadd2}):
\begin{align}
S(\rhoL) + S(\rho_{\ell+1:2\ell+1}) &\leq S(\rho_{0:\ell+1}) + S(\rho_{\ell:2\ell+1}) \nonumber \\
S(\ell) + S(\ell) &\leq S(\ell+1) + S(\ell+1) \label{stat_nd} \\
S(\ell) &\leq S(\ell+1) \nonumber
~,
\end{align}
where Eq. (\ref{stat_nd}) follows from stationarity.

Thus, $S(\ell) \leq S(\ell+1)$, for all $\ell \geq 0$, and $S(\ell)$ is a
nondecreasing function of $\ell$.
\end{ProProp}

\begin{Prop}
For a stationary qudit process $S(\ell)$ is concave.
\end{Prop}

\begin{ProProp}
The von Neumann entropy is strongly subadditive  \cite{QCQI}, meaning that:
\begin{align}
S(\rho^{ABC}) + S(\rho^B) \leq S(\rho^{AB}) + S(\rho^{BC})
  ~.
\label{subadd}
\end{align}

For $\ell \geq 3$, let $\rho^{ABC} = \rhoL$ where $\rho^A = \rho_0$, $\rho^B = \rho_{1:\ell-1}$, and $\rho^C = \rho_{\ell-1}$

We can rewrite Eq. (\ref{subadd}) by incorporating the stationarity of qudit processes:
\begin{align}
S(\rhoL) + S(\rho_{1:\ell-1}) &\leq S(\rho_{0:\ell-1}) + S(\rho_{1:\ell}) \nonumber \\
S(\ell) + S(\ell-2) &\leq S(\ell-1) + S(\ell-1) \label{conc} \\
S(\ell) - 2 S(\ell-1) + S(\ell-2) &\leq 0 \nonumber
~,
\end{align}
where Eq. (\ref{conc}) follows from stationarity.

Thus, $S(\ell)$ is concave.
\end{ProProp}

For a separable qudit process formed by passing a classical process through a
classical-quantum channel, their block entropies are related in the following
way:

\begin{Prop}
Let $\QBiInfinity = \cqChannel(\BiInfinity)$.
The block entropies of $\QBiInfinity$ and $\BiInfinity$ obey:
\begin{align*}
S(\rhoL) \leq \H{X_{0:\ell}}
~,
\end{align*}
for all $\ell$, with equality if and only if $\AQ$ consists of $\vert \AX \vert$
orthogonal pure states in $\HSpace$ of dimension $d \geq {\vert \AX \vert}$.
\label{prop:block_entropy_separable}
\end{Prop}

\begin{ProProp}
Recall from Eq. (\ref{eq:finite_rho}) that for separable qudit processes:
\begin{align*}
 \rhoL = \sum_{w \in \AX^\ell} \Pr(w) \ketpsi[w] \bra{\psi_w}
~,
\end{align*}
with each $\ketpsi[w]$ taking the separable form of Eq.
(\ref{eq:separable_word}) and $w = x_{0:\ell}$.

We first note that, for all symbols in $\AX$ to be associated with orthogonal
qudit states, the minimum dimension of the Hilbert space is $d_{min} = \vert \AX
\vert$.

With $\rhoL$ written as a mixture of separable qudit words, we apply Lemma
\ref{lem:mixture_entropy} to obtain:
\begin{align*}
    S(\rhoL) & \leq \H{X_{0:\ell}} + \sum_{w \in \AX^\ell}
	\Pr(w) S(\ketpsi[w]) \\
    & \leq \H{X_{0:\ell}}
~,
\end{align*}
where $w = x_{0:\ell}$. The second term evaluates to zero in the final line
since each $\ketpsi[w]$ is a pure state. That is, $S(\ketpsi[w]) = 0$ for all $w
\in \AX^\ell$.

Equality occurs if and only if the states $\ketpsi[w]$ have support on
orthogonal subspaces, which requires that $d \geq \vert \AX \vert$ and all
elements of $\AQ$ are orthogonal.
\end{ProProp}

We cannot use $S(\rhoL)$ to bound the block entropy of a measured process
$\OutputBi$ for \emph{general} POVM measurements. For the case where the
measurement $\qcChannel_{0:\ell}$ consists only of rank-one POVMs (including all
PVMs), however, the following holds:
\begin{align}
S(\rhoL) \leq \H{\qcChannel_{0:\ell}(\rhoL)}
~,
\label{eq:measured_block}
\end{align}
with equality if and only if the measurement is performed in the minimum-entropy
(eigen)basis of $\rhoL$. This follows directly from Eq. (\ref{eq:shan_von}).
\begin{Prop}
Let $\OutputBi = \qcChannel(\QBiInfinity)$, where $\qcChannel$ is a repeated
rank-one POVM measurement.
The block entropies of $\QBiInfinity$ and $\OutputBi$ then obey:
\begin{align*}
S(\rhoL) \leq \H{Y_{0:\ell}}
  ~,
\end{align*}
for all $\ell$, with equality if and only if $\QBiInfinity$ is a separable
process with an orthogonal alphabet $\AQ$, and $\qcChannel$ uses a POVM whose
operators include one projector for each element in $\AQ$.
\label{prop:block_entropy_measured}
\end{Prop}

\begin{ProProp}
The bound follows directly from Eq. (\ref{eq:measured_block}) because repeated
rank-one POVM measurements are a subclass of the more general measurement
sequence $\qcChannel_{0:\ell}$.

The condition for equality also follows from Eq. (\ref{eq:measured_block}) but
requires more justification. First, we consider measuring the single-qubit
marginal $\rho_0$ with POVM $\Meas$. For equality each element of the eigenbasis
of $\rho_0$ ($\{\ket{e_i}\}$) must have a corresponding operator in $\Meas$
that is a projector on that eigenspace ($E_i = \ket{e_i}\bra{e_i}$).

We can write $\rho_0 = \sum_i p_i \ket{e_i}\bra{e_i}$. Since we apply the same
POVM to each qudit, all blocks of length-$\ell$ must have eigenstates of the
form $\bigotimes_{t = 0}^{\ell-1} \ket{e_t}$---i.e., they must take the
separable form of Eq. (\ref{eq:separable_word})---making $\QBiInfinity$ a
separable process with a quantum alphabet $\AQ$ of orthogonal states. The
$\Meas$ consisting of projectors onto the states in $\AQ$ is then the
minimum-entropy measurement over blocks $\rhoL$. Note that there may be other
elements of the POVM that are not projectors if the probability of those
measurement outcomes is $0$ when applied to $\rhoL$. (If the process does not
make use of that part of Hilbert space, it does not matter how it is measured.)
\end{ProProp}

To summarize, in the case of separable qudit processes, $S(\ell)$ is
upper-bounded by the underlying classical process' block entropy
$\H{X_{0:\ell}}$. For repeated measurement with rank-one POVMs, $S(\ell)$ serves
as a lower bound on the block entropy of all classical measured processes
$\H{Y_{0:\ell}}$. There is no direct relationship between $\H{X_{0:\ell}}$ and
$\H{Y_{0:\ell}}$. Rather, it depends on the specifics of $\cqChannel$ and
$\qcChannel$.

\subsection{von Neumann Entropy Rate}

The von Neumann entropy rate of a qudit process is:
\begin{align}
\qer = \lim_{\ell \to \infty} \frac{S(\ell)}{\ell}
~.
\label{eq:quantum_entropy_rate}
\end{align}
The units of $\qer$ are \emph{bits~per~timestep}. This quantity is equivalent to
the \emph{mean entropy}, first introduced in the context of quantum statistical
mechanics \cite{Lanf68}. The limit exists for all stationary processes
\cite{Ohya93}. Operationally, $\qer$ is the optimal coding rate for a stationary
quantum source \cite{Petz01}.

\begin{Prop}
For a stationary qudit process we can equivalently write the von Neumann
entropy rate as:
\begin{align}
\qer = \lim_{\ell \to \infty} S(\rho_0 \vert \rho_{-\ell:0})
  ~.
\label{eq:quantum_entropy_rate_conditional}
\end{align}
\end{Prop}

\begin{ProProp}
This proof closely follows the proof for classical entropy rate in Ref.
\cite{Cove91a}. We begin by showing that $\lim_{\ell \to \infty}
S(\rho_0 \vert \rho_{-\ell:0})$ exists and then that it is equivalent to
the limit in Eq. (\ref{eq:quantum_entropy_rate}). The limit exists if
$S(\rho_0 \vert \rho_{-\ell:0})$ is a decreasing, nonnegative function of $\ell$:
\begin{align}
S(\rho_0 \vert \rho_{-\ell:0}) & = S(\rho_{-\ell:1}) - S(\rho_{-\ell:0}) \nonumber \\
                        & = S(\ell+1) - S(\ell) \label{stationary_s} \\
                        & \geq 0 \label{stationary_s_2}
~,
\end{align}
where Eq. (\ref{stationary_s}) follows from stationarity and Eq.
(\ref{stationary_s_2}) follows from the nondecreasing nature of $S(\ell)$.
This, combined with the fact that $S(\ell)$ is concave, means $\lim_{\ell \to
\infty} S(\rho_0 \vert \rho_{-\ell:0})$ exists.

Now, we establish that $\lim_{\ell \to \infty} S(\rho_0 \vert \rho_{-\ell:0}) =
\lim_{\ell \to \infty} S(\ell) / \ell$. Through repeated application of Eq.
(\ref{eq:cond_vn_ent}) to a block of length-$\ell$ we obtain the following chain
rule for the von Neumann entropy:
\begin{align}
S(\rhoL) = \sum_{j=0}^{\ell-1} S(\rho_j  \vert  \rho_{j-1} ... \rho_{0})
~.
\label{eq:vn_chain_rule}
\end{align}

We can modify indices (due to stationarity) and divide both sides by $\ell$ to
obtain:
\begin{align}
\frac{S(\rhoL)}{\ell} = \frac{1}{\ell} \sum_{i=1}^{\ell} S(\rho_i  \vert  \rho_{i-1} ... \rho_{1})
~.
\label{eq:qer_conditional_sum}
\end{align}

The final steps require the following result, known as the Ces\'aro mean
\cite{Cove91a}:
\begin{Lem}
If $a_n \to a$ and $b_n = \frac{1}{n}\sum_{i=1}^{n}a_i$, then $b_n \to a$.
\label{lem:cesaro_mean}
\end{Lem}

Taking the limit of both sides of Eq. (\ref{eq:qer_conditional_sum}) and applying
Lemma \ref{lem:cesaro_mean}, we find:
\begin{align}
\lim_{\ell \to \infty} \frac{S(\rhoL)}{\ell} = \lim_{\ell \to \infty} S(\rho_\ell  \vert  \rho_{\ell-1} ... \rho_{1})
~.
\label{eq:qer_conditional_result}
\end{align}

Using stationarity, $\lim_{\ell \to \infty} S(\rho_\ell  \vert  \rho_{\ell-1}
\ldots \rho_{1}) = \lim_{\ell \to \infty} S(\rho_0  \vert  \rho_{-\ell:0})$.
When combined with Eq. (\ref{eq:qer_conditional_result}) this proves that our
two definitions of $\qer$ are equivalent.
\end{ProProp}

\begin{figure}
\centering
\includegraphics[width=\columnwidth]{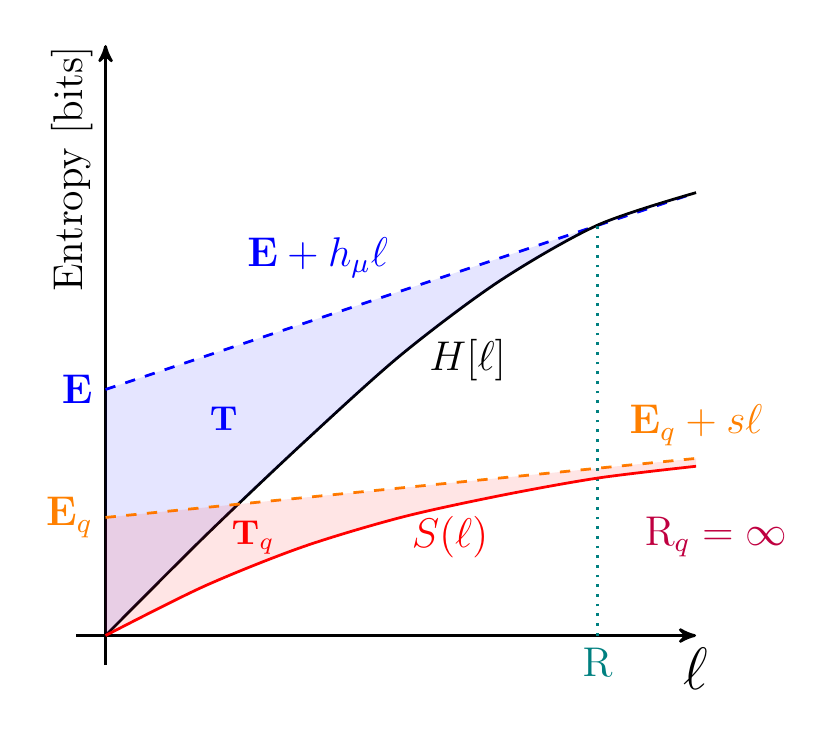}
\caption{Convergence of the block entropies to their linear asymptotes.
$\H{\ell}$ is the block entropy for a finitary classical process with Markov
Order $\MOrder$ (see App. \ref{app:classical_properties}), and $S(\ell)$ is the
quantum block entropy for a finitary quantum process with infinite Quantum
Markov order $\MOrderq$. For the classical process $\EE$ is the excess entropy,
and $\hmu$ is its Shannon entropy rate. Similarly, for a quantum process $\EE_q$
is the quantum excess entropy, and $\qer$ is its von Neumann entropy rate. The
area of the blue shaded region is the classical transient information $\TI$ and
the area of the red shaded region is the quantum transient information $\TI_q$.
	}
\label{fig:process_properties}
\end{figure}

To motivate a number of the following results it is important to appreciate
that simply because a process has a von Neumann entropy rate given by Eq.
(\ref{eq:quantum_entropy_rate}) does not imply that an observer is able to
perform a measurement on any qudit such that the uncertainty in that individual
measurement is $\qer$. Rather, obtaining $\qer$ corresponds to the measurement
basis over the entire chain of qudits for which the distribution of outcomes
has the minimal Shannon entropy. This basis is highly nonlocal or otherwise
experimentally infeasible for many nontrivial examples. As in the classical
case, $\qer$ appears graphically as the slope of $S(\ell)$ as $\ell \to
\infty$, as shown in Fig. \ref{fig:process_properties}.

For separable qudit processes, we can also relate $\qer$ to the classical
entropy rate of the underlying process $\BiInfinity$, as follows:

\begin{Prop}
Let $\QBiInfinity = \cqChannel(\BiInfinity)$.
The von Neumann entropy rate $\qer$ of $\QBiInfinity$ then obeys
the bound:
\begin{align*}
    \qer \leq \hmu^X
    ~,
\end{align*}
where $\hmu^X$ is the Shannon entropy rate of $\BiInfinity$. Equality occurs if
and only if $\AQ$ consists of $\vert \AX \vert$ orthogonal pure states in
$\HSpace^d$ of dimension $d \geq \vert \AX \vert$.
\label{prop:qer_separable}
\end{Prop}

\begin{ProProp}
Divide both sides of Prop. \ref{prop:block_entropy_separable} by $\ell$ and take
the limit of both sides as $\ell \to \infty$ to obtain:
\begin{align*}
\lim_{\ell \to \infty} \frac{S(\rhoL)}{\ell} \leq \lim_{\ell \to \infty}
\frac{\H{\MeasSymbol_{0:\ell}}}{\ell}
~.
\end{align*}
From Eq. (\ref{eq:quantum_entropy_rate}), the left side is $\qer$, and from Eq.
(\ref{eq:hmu}) the right side $\hmu^X$. The condition for equality is inherited
from Prop. \ref{prop:block_entropy_separable}, concluding the proof.
\end{ProProp}

Restricting once again to repeated measurements with rank-one POVMs, we can
prove the following bound for measured processes:
\begin{Prop}
Let $\OutputBi = \qcChannel(\QBiInfinity)$ and let $\qcChannel$ be a repeated
rank-one POVM. The measured entropy rate $\hmu^Y$ then obeys:
\begin{align*}
\qer \leq \hmu^Y 
~,
\end{align*}
where $\qer$ is the von Neumann entropy rate of $\QBiInfinity$, with equality if
and only if $\QBiInfinity$ is a separable process with an orthogonal alphabet
$\AQ$, and $\qcChannel$ uses a POVM whose operators include a projector for each
element in $\AQ$.
\label{prop:qer_measured}
\end{Prop}

\begin{ProProp}
Divide both sides of Prop. \ref{prop:block_entropy_measured} by $\ell$ and take
the limit of both sides as $\ell \to \infty$ to obtain:
\begin{align*}
\lim_{\ell \to \infty} \frac{S(\rhoL)}{\ell} \leq \lim_{\ell \to \infty}
\frac{\H{Y_{0:\ell}}}{\ell}
~.
\end{align*}
The left side is $\qer$ and the right side $\hmu^Y$. The conditions for equality
are inherited from Prop. \ref{prop:block_entropy_measured}, concluding the
proof.
\end{ProProp}

\subsection{Quantum Redundancy}

Unlike a classical process, the maximum entropy rate for a qudit process depends
on the size of the Hilbert space rather than on the size of the alphabet $\AQ$.
For Hilbert space of dimension $d$, the largest possible value of $\qer$ is
$\log_2(d)$, corresponding to an i.i.d. sequence of qudits, each in a
maximally-mixed state $\rho_{iid} = \mathbb{I}/d$.

A qudit process can be compressed down to its von Neumann entropy rate $\qer$.
The amount that it can be compressed is the \emph{quantum redundancy}:
\begin{align}
\Rq \equiv \log_2(d) - \qer
~.
\label{eq:quantum_redundancy}
\end{align}
Statistical biases in individual qudits and temporal correlations between them
offer opportunities for compression. $\Rq$ includes the effects of
both.

For separable qudit processes we can bound the quantum redundancy using
properties of the underlying classical process:

\begin{Prop}
Let $\BiInfinity$ be a classical process with redundancy $\mathbf{R}^X$,
symbol alphabet $\AX$, and entropy rate $\hmu^X$, and let $\QBiInfinity =
\cqChannel(\BiInfinity)$ be a qudit process with redundancy $\Rq$, Hilbert space
of dimension $d$, and entropy rate $\qer$. 

For $d \geq  \vert \AX \vert $:
\begin{align*}
    \Rq \geq \mathbf{R}^X
  ~,
\end{align*}
with equality if and only if $d = \vert \AX \vert$ and $\AQ$ consists of $\vert
\AX \vert$ orthogonal pure states.

For $d <  \vert \AX \vert $,
\begin{align*}
    \Rq < \mathbf{R}^X + (\hmu^X - \qer)
  ~,
\end{align*}
where the term $(\hmu^X - \qer)$ is always positive from Prop.
(\ref{prop:qer_separable}). \label{prop:redundancy_classical_quantum}
\end{Prop}

\begin{ProProp}
First, consider $d = \vert \AX \vert$:
\begin{align*}
    \Rq & = \log_2( \vert \AX \vert ) - \qer \\
            & = \mathbf{R}^X + \hmu^X - \qer \\ 
            & \geq \mathbf{R}^X
~,
\end{align*}
The final line comes from Prop. \ref{prop:qer_separable}, as does
the condition for equality.

For $d > \vert \AX \vert$:
\begin{align*}
    \Rq & = \log_2(d) - \qer \\
            & > \log_2( \vert \AX \vert ) - \qer \\ 
            & > \mathbf{R}^X
  ~.
\end{align*}
There is no opportunity for equality. In this case, $\AQ$ will not span
$\HSpace$, naturally leading to more redundancy.

Finally, for $d < \vert \AX \vert$:
\begin{align*}
    \Rq & < \log_2(d) - \qer \\
            & < \log_2( \vert \AX \vert ) - \qer \\
            & < \mathbf{R}^X + (\hmu^X - \qer)
  ~,
\end{align*}
concluding the proof.
\end{ProProp}

We can also compare the classical redundancy of a measured process (obtained
through repeated use of a rank-one POVM) to the quantum redundancy of the qudit
process being measured.
\begin{Prop}
Let $\OutputBi$ be a measured process such that $\OutputBi = \qcChannel
(\QBiInfinity)$, and $\qcChannel$ a repeated rank-one POVM. Let $\OutputBi$
have redundancy $\mathbf{R}^Y$, and let $\QBiInfinity$ have quantum redundancy
$\Rq$.
Then:
\begin{align}
\Rq \leq \mathbf{R}^Y
  ~,
\end{align}
with equality if and only if $d = \vert \AY \vert$, $\QBiInfinity$ is a
separable process with an orthogonal alphabet $\AQ$, and $\qcChannel$ uses a
POVM whose operators include a projector for each element in $\AQ$.
\label{prop:redundancy_measured}
\end{Prop}

\begin{ProProp}
A rank-one POVM on $\HSpace$ must have at least $d$ elements, therefore $\vert
\AY \vert \geq d$:
\begin{align*}
    \Rq & = \log_2(d) - \qer \\
            & \leq \log_2(\vert \AY \vert) - \qer \\ 
            & \leq \mathbf{R}^Y + \hmu^Y - \qer \\
            & \leq \mathbf{R}^Y
~.
\end{align*}
Going from the first line to the second provides a condition for equality: $d =
\vert \AY \vert $. Proposition \ref{prop:qer_measured} is used in the final
line and provides the other conditions for equality.
\end{ProProp}

\subsection{Quantum Entropy Gain}

We can take discrete-time derivatives of $S(\ell)$, as was done for $\H{\ell}$
in \cite{Crut01a}. This process is summarized in App.
\ref{app:classical_properties}. We call the first derivative of $S(\ell)$ the
\emph{quantum entropy gain}:
\begin{align}
\Delta S(\ell) \equiv S(\ell) - S(\ell - 1)
  ~,
\label{eq:quantum_entropy_gain}
\end{align}
for $\ell > 0$. The units for the quantum entropy gain are
\emph{bits~per~timestep}, and we set the boundary condition $\Delta S(0) =
\log_2(d)$. Since $S(\ell)$ is monotone-increasing, $\Delta S(\ell) \geq 0$.

The quantum entropy gain is the amount of additional uncertainty introduced by
including the $\ell\textsuperscript{th}$ qudit in a block, where that
uncertainty is quantified by the von Neumann entropy.

By combining Eq. (\ref{eq:quantum_entropy_gain}) and Eq. (\ref{eq:cond_vn_ent}),
we can write $\Delta S(\ell)$ as:
\begin{align*}
\Delta S(\ell) = S(\rho_0 \vert \rho_{-\ell:0})
  ~.
\end{align*} 

This allows relating the quantum entropy gain and the von Neumann entropy rate
as follows:
\begin{align}
\qer = \lim_{\ell \to \infty} \Delta S(\ell)
  ~.
\label{eq:quantum_entropy_gain_limit}
\end{align}

Thus, paralleling the classical case, the quantum entropy gain serves as a
finite-$\ell$ approximation of the von Neumann entropy rate:
\begin{align}
 \qer(\ell) & \equiv \Delta S(\ell) 
  ~.
\label{eq:qer_est}
\end{align}
$ \qer(\ell)$ serves as the best estimate for the entropy rate of a qudit
process that can be made by an observer who only has access to measurement
statistics for length-$\ell$ blocks of qudits.

\begin{figure}
\centering
\includegraphics[width=\columnwidth]{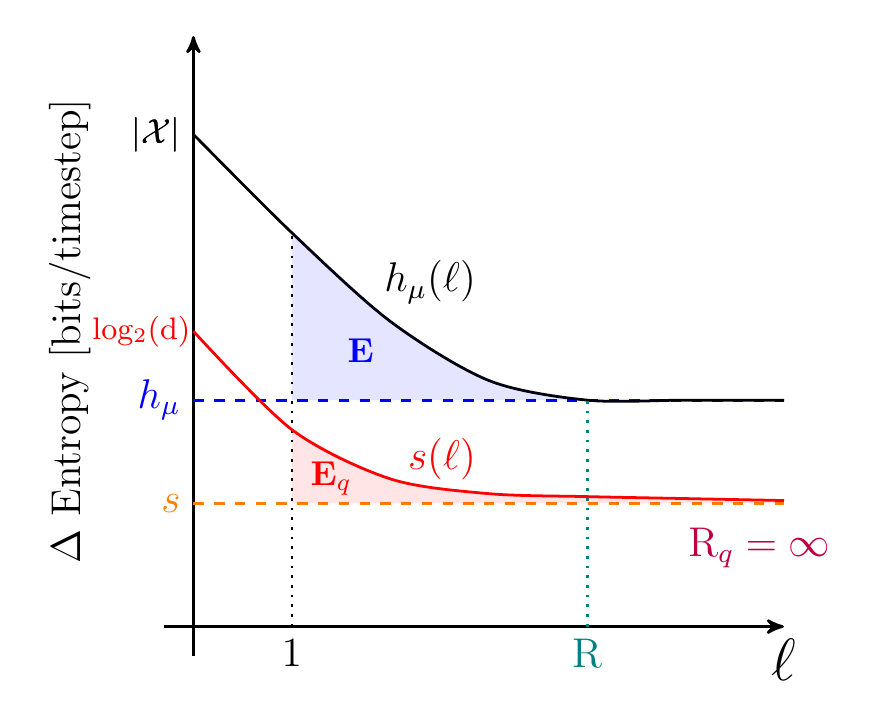}
\caption{Convergence of $\Delta H(\ell)$ (for a finitary classical process with Markov
order $\MOrder$) and $\Delta S(\ell)$ (for a finitary quantum process with infinite quantum Markov
order $\MOrderq$) to the processes' entropy rates, $\hmu$ and $\qer$. The
shaded areas are the classical (blue) and quantum (red) excess entropies.
	}
\label{fig:entropy_convergence_properties}
\end{figure}

The way in which the entropy rate estimate converges and its relationship to
other information properties of a qudit process are summarized in Fig.
\ref{fig:entropy_convergence_properties}.

\subsection{Quantum Predictability Gain}

\begin{figure}
\centering
\includegraphics[width=\columnwidth]{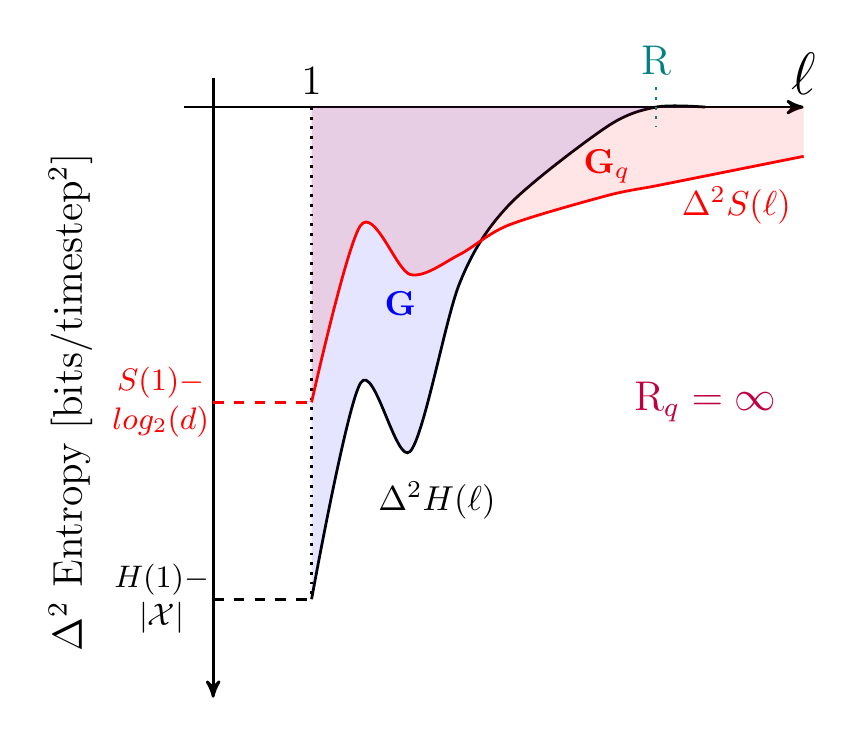}
\caption{Convergence of the predictability ($\Delta^2 H(\ell)$ and $\Delta^2
S(\ell)$) to $0$ for a finitary classical process with Markov order $\MOrder$
and a finitary quantum process with infinite Quantum Markov order $\MOrderq$.
Note that the predictability is not monotonic (unlike the block entropy and its
first derivative). The overlapping shaded areas represent the magnitude of the
classical (blue) and quantum (red) predictabilities, $\textbf{G}$ and $\Gq$,
which are negative by convention.
	}
\label{fig:predictability_convergence_properties}
\end{figure}

We call the second derivative of $S(\ell)$ the \emph{quantum predictability
gain}, given by:
\begin{align*}
\Delta^2 S(\ell) & \equiv \Delta  \qer(\ell) \\
    & =  \qer(\ell) - s(\ell-1)
~,
\end{align*} 
for $\ell > 0$. The units of $\Delta^2 S(\ell)$ are
\emph{bits~per~timestep$^2$}. Since $S(\ell)$ is concave, then $\Delta^2 S(\ell)
\leq 0$. $ \vert \Delta^2 S(\ell) \vert $ quantifies how much an observer's estimate of the
von Neumann entropy rate $\qer$ improves if they are enlarge their observations
from blocks of $\ell-1$ to blocks of $\ell$ qudits. The generic convergence
behavior of $\Delta^2 S(\ell)$ is shown in Fig.
\ref{fig:predictability_convergence_properties}.

For all higher-order discrete derivatives of $S(\ell)$ (as with the classical
block entropy):
\begin{align*}
\lim_{\ell \to \infty} \Delta^n S(\ell) = 0, n \geq 2
  ~.
\end{align*}
This follows directly from the existence of the limit in Eq.
(\ref{eq:quantum_entropy_rate}) for stationary quantum states.

\subsection{Total Quantum Predictability}

Up to this point, introducing new information-theoretic characteristics of
separable quantum processes proceeded by taking discrete-time derivatives of the
von Neumann block entropy. We can likewise integrate the functions $\Delta^n
S(\ell)$, as is done for the classical case with Eq.
(\ref{eq:discrete_integral}). While this starts off straightforwardly, a number
of interesting new informational quantities emerge.

These properties of qudit processes take the general form:
\begin{align*}
\mathcal{I}^q_n \equiv \sum_{\ell=\ell_0}^\infty [\Delta^n S (\ell) - \lim_{\ell \to \infty} \Delta^n S(\ell)]
~,
\end{align*}
where $\ell_0$ is the first value of $\ell$ for which $\Delta^n S(\ell)$ is defined.

$\mathcal{I}^q_2$, the first of these, monitors the convergence of the quantum
predictability gain $\Delta^2 S(\ell)$ to its limit of $0$ for $\ell \to
\infty$. We use $\ell_0 = 1$ to get the \emph{total quantum predictability}
$\Gq$:
\begin{align}
\Gq & \equiv \mathcal{I}^q_2 \nonumber \\
  & = \sum_{\ell=1}^\infty \Delta^2 S (\ell)
    ~.
\label{eq:total_quantum_predictability}
\end{align}
The units of $\Gq$ are \emph{bits~per~timestep}. Note that
$\Gq \leq 0$ because $\Delta^2 S(\ell) \leq 0$ for all $\ell$.

$\Gq$ can be interpreted by relating it to a previously-established
property of qudit processes: the quantum redundancy.
\begin{Prop}
For a stationary qudit process:
\begin{align*}
\Gq = -\Rq
    ~.
\end{align*}
\label{prop:predictability_redundancy}
\end{Prop}

\begin{ProProp}
Applying Eq. (\ref{eq:discrete_integral}) to Eq. (\ref{eq:total_quantum_predictability}) we find that:
\begin{align*}
\Gq & = \lim_{\ell \to \infty} \Delta S(\ell) - \Delta S(0)  \\
            & = \qer- \log_2(d) \\
            & = -\Rq
  ~.
\end{align*}
Here, the second line follows from Eq. (\ref{eq:quantum_entropy_gain_limit})
and the third from Eq. (\ref{eq:quantum_redundancy}).
\end{ProProp}

Thus, $\vert \Gq \vert$ is the total amount of predictable information
per timestep for a qudit process.

Rather immediately, one sees that the amount of information in an individual
qudit decomposes into:
\begin{align*}
    \log_2(d) = \vert \Gq \vert + s
    ~.
\end{align*}
$\vert \Gq \vert$ is the amount of quantum information within a qudit
that is predictable. Whereas, $\qer$ is the amount of information that is
irreducibly random.

The relation $\vert \Gq \vert =
\Rq$ can be combined with Props. \ref{prop:redundancy_classical_quantum} and
\ref{prop:redundancy_measured} to prove two corollaries.

\begin{Cor}
Let $\BiInfinity$ be a classical process with total predictability
$\mathbf{G}^X$ and symbol alphabet $\AX$, and entropy rate $\hmu^X$, and let
$\QBiInfinity = \cqChannel(\BiInfinity)$ be a qudit process with total quantum
predictability $\Gq$, Hilbert space of dimension $d$ and entropy rate $\qer$.

For $d \geq  \vert \AX \vert $:
\begin{align*}
   \vert \Gq \vert \geq \vert \mathbf{G}^X \vert
  ~,
\end{align*}
with equality if and only if $d =  \vert \AX \vert $ and $\AQ$ consists of $\vert \AX \vert$
orthogonal pure states.

For $d <  \vert \AX \vert $,
\begin{align*}
    \vert \Gq \vert < \vert \mathbf{G}^X \vert + (\hmu^X - \qer)
  ~,
\end{align*}
where the term $(\hmu^X - \qer)$ is always positive from Prop.
(\ref{prop:qer_separable}). \label{prop:predictability_classical_quantum}

\end{Cor}

\begin{ProCor}
This follows immediately from combining Props.
\ref{prop:redundancy_classical_quantum} and
\ref{prop:predictability_redundancy}.
\end{ProCor}

\begin{Cor}

Let $\OutputBi$ be a measured process such that $\OutputBi = \qcChannel
(\QBiInfinity)$, and $\qcChannel$ is a repeated rank-one POVM. Let $\OutputBi$
have redundancy $\mathbf{G}^Y$, and let $\QBiInfinity$ have quantum redundancy
$\Gq$.
Then:
\begin{align*}
\vert \Gq \vert \leq \vert \mathbf{G}^Y \vert
  ~,
\end{align*}
with equality if and only if $d =  \vert \AY \vert $, $\QBiInfinity$ is a separable process
with an orthogonal alphabet $\AQ$, and $\qcChannel$ uses a POVM whose operators
include a projector for each element in $\AQ$.
\label{cor:predictability_measured}
\end{Cor}

\begin{ProCor}
This follows immediately from combining Props. $\ref{prop:redundancy_measured}$
and $\ref{prop:predictability_redundancy}$.
\end{ProCor}

Graphically, the total quantum predictability is the area between the
predictability gain curve and its linear asymptote of $0$, as seen in Fig.
\ref{fig:predictability_convergence_properties}. The von Neumann entropy rate
and total predictability lend insight into compression limits for a stationary
sources. They do not say, however, whether that compression is achievable due to
bias within individual qudit states or correlations between qudits. For that we
must continue our way back up the entropy hierarchy.

\subsection{Quantum Excess Entropy}

The convergence of $ \qer(\ell)$ to the true von Neumann entropy rate $\qer$ is
quantified with the \emph{quantum excess entropy}:
\begin{align}
\EE_q \equiv \mathcal{I}^q_1 & =  \sum_{\ell=1}^\infty \left[ \Delta S(\ell) -
\lim_{\ell \to \infty} \Delta S(\ell) \right] \nonumber \\
        & = \sum_{\ell=1}^\infty \left[  \qer(\ell) - \qer \right]
~.
\label{eq:quantum_excess_entropy}
\end{align}
The units for $\EE_q$ are \emph{bits}.
Paralleling the classical case, we refer to any qudit process with finite
$\EE_q$ as \emph{finitary} and those with infinite $\EE_q$ as
\emph{infinitary}.

We can further express $\EE_q$ in terms of the asymptotic behavior of $S(\ell)$.
\begin{Prop}
The quantum excess entropy can be written as:
\begin{align}
\EE_q = \lim_{\ell \to \infty} \left[ S(\ell) - \qer \ell \right]
  ~.
\label{eq:block_excess_entropy}
\end{align}
\label{prop:block_excess_entropy}
\end{Prop}

\begin{ProProp}
We evaluate the discrete integral in Eq. (\ref{eq:quantum_excess_entropy}) with
Eq. (\ref{eq:discrete_integral}) using partial sums:
\begin{align*}
\EE_q & = \lim_{\ell \to \infty} \sum_{m=1}^\ell \left[ \Delta s(m) - \qer \right] \\
	& = \lim_{\ell \to \infty} \left[ S(\ell) - S(0) - \qer \ell \right] \\
	& = \lim_{\ell \to \infty} \left[ S(\ell) - \qer \ell \right]
~,
\end{align*}
since $S(0) = 0$ by definition.

\end{ProProp}

For finitary quantum processes, $\EE_q$ is the area between the entropy gain
curve and its asymptote $\qer$ as seen in Fig.
\ref{fig:entropy_convergence_properties}. It also appears in Fig.
\ref{fig:process_properties} as the vertical offset of the linear asymptote to
the $S(\ell)$ curve.

This leads to a natural scaling of the quantum block entropy as:
\begin{align*}
    S(\ell) \sim \EE_q + \qer \ell
	~,
\end{align*}
as $\ell \to \infty$.

A clearer interpretation of $\EE_q$ as a quantum mutual information is provided
by the following proposition:
\begin{Prop}
The quantum excess entropy can be written as:
\begin{align*}
\EE_q = \lim_{\ell \to \infty} S(\rho_{-\ell:0}{:}\rhoL)
  ~,
\end{align*} 
where $\rho_{-\ell:0}$ and $\rhoL$ are two blocks of $\ell$ consecutive
qudits with a shared boundary.
\label{prop:Eq_mutual_info}
\end{Prop}

\begin{ProProp}
The quantum mutual information, from Eq.
(\ref{eq:quantum_mutual_information}),
between two neighboring blocks of $\ell$ qudits can be expressed as:
\begin{align*}
    S(\rhoL{:}\rho_{-\ell:0}) & = S(\rhoL) - S(\rhoL \vert \rho_{-\ell:0}) \\
    & = S(\ell) - \sum_{t=0}^{\ell-1} S(\rho_t  \vert  \rho_{-\ell:t-1})
  ~,
\end{align*}
where the final line is obtained through repeated application of Eq.
(\ref{eq:cond_vn_ent}). 

Taking $\ell \to \infty$:
\begin{align*}
\lim_{\ell \to \infty} S(\rhoL{:}\rho_{-\ell:0})
    & = \lim_{\ell \to \infty} \left[ S(\ell) - \sum_{t=0}^{\ell-1} S(\rho_t  \vert  \rho_{-\ell:t-1}) \right] \\
    & = \lim_{\ell \to \infty} \left[ S(\ell) - \qer \ell \right]
~,
\end{align*} where the final line follows from Eq. (\ref{eq:qer_conditional_sum})
and stationarity.

This final expression is equivalent to the form of $\EE_q$ derived in Prop.
\ref{prop:block_excess_entropy}, concluding the proof.
\end{ProProp}

$\EE_q$ is therefore a measure of all the correlations between two halves of
the infinite sequence of qudits. $\EE_q = 0$, if and only if a source is
i.i.d. (with $S(\ell) = s\ell$).

We can relate $\EE_q$ for a separable qudit process to $\EE^X$ of the
underlying classical process:
\begin{Prop}
Let $\BiInfinity$ be a classical process with alphabet $\AX$ and excess
entropy $\EE^X$, and let $\QBiInfinity = \cqChannel(\BiInfinity)$ be a qudit
process with  alphabet $\AQ$ and quantum excess entropy $\EE_q$.

Then:
\begin{align*}
\EE_q \leq \EE^X 
~,
\end{align*}
with equality if and only if $\AQ$ consists of $\vert \AX \vert$ orthogonal states.

\label{prop:Eq_separable}
\end{Prop}

\begin{ProProp}
Consider $\MeasSymbol_{-\ell:\ell}$, a block of length $2\ell$ of the classical
process $\BiInfinity$. We can write realizations of $\MeasSymbol_{-\ell:\ell}$
into a classical register to form the state:
\begin{align}
\rho^C_{-\ell:\ell} = \sum_{x_{-\ell:\ell}} \Pr(x_{-\ell:\ell})\ket{x_{-\ell:\ell}}\bra{x_{-\ell:\ell}}
~,
\end{align}
where all $\ket{x_{-\ell:\ell}}$ are orthogonal. 
Then we pass each symbol through the preparation channel $\cqChannel$ to obtain
blocks of our qudit process $\rho_{-\ell:\ell} =
\cqChannel^{2\ell}(\rho^C_{-\ell:\ell})$, where $\cqChannel^{2\ell} =
\bigotimes_{i=0}^{2\ell} \cqChannel$.

We can express the quantum mutual information as a quantum relative entropy,
Eq. (\ref{eq:quantum_mutual_information_relative}), giving the following
relation:
\begin{align}
\I{\MeasSymbol_{-\ell:0}{:}\MeasSymbol_{0:\ell}}
	& = S(\rho^C_{-\ell:0}{:}\rho^C_{0:\ell})
	\nonumber
	\\
	& \geq S(\cqChannel^\ell(\rho^C_{-\ell:0}){:}\cqChannel^\ell(\rho^C_{0:\ell})) \label{eq:mutual_info_channel} \\
	& \geq S(\rho_{-\ell:0}{:}\rhoL)
	\nonumber
  ~,
\end{align}
where Eq. (\ref{eq:mutual_info_channel}) comes from the monotonicity of the
quantum relative entropy in Eq. (\ref{eq:monotonicity_qre}). The condition for
equality comes from Eq. (\ref{eq:monotonicity_qre}) as well. The set of states
for which the recovery map must exist is $\AQ$, and this is only possible if all
$\ketpsi[\meassymbol]$ are distinguishable---i.e., orthogonal.

Using Eq. (\ref{eq:EE_mutual_info}), to write the excess entropy of 
$\BiInfinity$ as a limit we see that:
\begin{align*}
\lim_{\ell \to \infty} \I{\MeasSymbol_{-\ell:0}{:}\MeasSymbol_{0:\ell}}
	& \geq \lim_{\ell \to \infty}S(\rho_{-\ell:0}{:}\rhoL) \\
	\EE^X & \geq \EE_q
  ~.
\end{align*}
\end{ProProp}

A similar bound appears when we apply a repeated POVM measurement to the quantum
process to obtain a classical process.
\begin{Prop}
Let $\OutputBi$ be a measured process such that $\OutputBi = \qcChannel
(\QBiInfinity)$, and $\qcChannel$ is a repeated rank-one POVM. Let $\OutputBi$
have excess entropy $\EE^Y$, and let $\QBiInfinity$ have quantum excess entropy
$\EE_q$. Then:
\begin{align*}
\EE_q \geq \EE^Y 
~,
\end{align*}
with equality if and only if $\QBiInfinity$ is a separable process with an
orthogonal alphabet $\AQ$ and $\qcChannel$ uses a POVM whose operators include a
projector for each element in $\AQ$.
\label{prop:measured_Eq}
\end{Prop}

\begin{ProProp}
Consider $\rho_{-\ell:\ell}$---a block of length $2\ell$ of the quantum process
$\QBiInfinity$---and let $\qcChannel$ be a repeated measurement of rank-one POVM
$\Meas$ with elements $\{E_y\}$ so that $\Meas(\rho_t) = \sum_y \Pr(y) \ket{y}
\bra{y}$, $\Pr(y) = tr(E_y \rho_t)$, and all $\ket{y}$ are orthogonal. The
repeated measurement applied over a block of length $\ell$ is then
$\qcChannel_{0:\ell} = \bigotimes_{i=0}^{\ell-1} \Meas$.

We express the quantum mutual information as a
quantum relative entropy (using Eq.
(\ref{eq:quantum_mutual_information_relative})), and apply Eq.
(\ref{eq:monotonicity_qre}) to obtain:
\begin{align}
\nonumber
S(\rho_{-\ell:0}{:}\rhoL)
	& \geq S \left( \qcChannel_{0:\ell}(\rho_{-\ell:0}){:}\qcChannel_{0:\ell}(\rhoL) \right) \\
\begin{split}
& \geq S \Bigg( \sum_{y_{-\ell:0}} \Pr(y_{-\ell:0}) \bigotimes_{t = -\ell}^0 \ket{y_t}\bra{y_t} {:} \Bigg. \\
& \qquad \qquad \Bigg. \sum_{y_{0:\ell}} \Pr(y_{0:\ell}) \bigotimes_{t=0}^{\ell-1} \ket{y_t}\bra{y_t} \Bigg)
\label{eq:EEq_trace}
\end{split}
\\
& \geq \I{Y_{-\ell:0}:Y_{0:\ell}}
  \nonumber
~.
\end{align}

The condition for equality comes from Eq. (\ref{eq:monotonicity_qre}) as well.
The set of states for which the recovery map must exist is $\{\ket{y}\}$, and
this is only possible if all $\ketpsi[\meassymbol] \in \AQ$ are orthogonal, and $\Meas$
contains a projector for each $\ketpsi[\meassymbol]$. By the same argument as Prop.
\ref{prop:block_entropy_measured}, $\QBiInfinity$ must be a separable process,
and $\AQ$ must consist of orthogonal states.

Taking the limit $\ell \to \infty$ we see that:
\begin{align*}
\lim_{\ell \to \infty} S(\rho_{-\ell:0}{:}\rhoL) & \geq \lim_{\ell \to \infty}\I{Y_{-\ell:0}:Y_{0:\ell}} \\
\EE_q & \geq \EE^Y
.~
\end{align*}

\end{ProProp}

Combining the above proofs we obtain the following corollary relating the excess
entropies of the underlying classical process $\BiInfinity$ and the measured
process $\OutputBi$:
\begin{Cor}
Let $\BiInfinity$ be a classical process with excess entropy $\EE^X$ and alphabet $\AX$, let
$\QBiInfinity = \cqChannel(\BiInfinity)$ be a separable qudit process with alphabet $\AQ$, and let
$\OutputBi$ be a measured process with excess entropy $\EE^Y$ such that
$\OutputBi = \qcChannel(\QBiInfinity)$ where $\qcChannel$ is a repeated rank-one
POVM.

Then:
\begin{align*}
   \EE^X \geq \EE^Y 
  ~,
\end{align*}
with equality if and only if $\AQ$ consists of $\vert \AX \vert$ orthogonal
states and $\qcChannel$ uses a POVM whose operators include a projector for each
element in $\AQ$.
\label{cor:classical_excess_entropies}
\end{Cor}

\begin{ProCor}
This follows immediately from combining Props.
\ref{prop:Eq_separable} and
\ref{prop:measured_Eq}.
\end{ProCor}

An exact value of $\EE_q$ typically requires characterizing infinite-length
sequences of qudits. However, we can write a finite-$\ell$ estimate of $\EE_q$
using Eq. (\ref{eq:block_excess_entropy}):
\begin{align}
\EE_q(\ell) = S(\ell) - \ell  \qer(\ell)
\label{eq:Eq_est}
  ~, 
\end{align}
that generally underestimates $\EE_q$'s true value.

\subsection{Quantum Transient Information}

We now turn to look at how the quantum block entropy curve converges to its
linear asymptote $\EE_q + \qer \ell$. We define the quantum transient
information as:
\begin{align}
\TI_q \equiv -\mathcal{I}^q_0 = \sum_{\ell = 1}^{\infty} [\EE_q + \qer \ell - S(\ell)]
~.
\label{eq:quantum_transient_info}
\end{align}
The units of $\TI_q$ are \emph{bits$\times$time steps}.

$\TI_q$ is represented graphically as the area between the $S(\ell)$ curve
and its linear asymptote for $\ell \to \infty$, as seen in Fig.
\ref{fig:process_properties}. We will see that $\TI_q$ distinguishes
between periodic qudit processes that cannot be distinguished with previous
information quantities such as $\EE_q$ and $\qer$.

\begin{Prop}
The transient quantum information $\TI_q$ can be written:
\begin{align*}
\TI_q = \sum_{\ell=1}^\infty \ell \left[  \qer(\ell) - \qer \right]
~.
\end{align*}
\end{Prop}

\begin{ProProp}
The proof reduces to the straightforward proof for transient information of a
classical stochastic process in Ref. \cite{Crut01a}. It depends only upon Eq.
(\ref{eq:quantum_transient_info}) and Eq. (\ref{eq:discrete_integral}),
that have the same form in the quantum case as the classical case.
\end{ProProp}

This expression allows us to estimate $\TI_q$ for a given quantum process as:
\begin{align}
\TI_q(\ell) = \sum_{m=1}^{\ell-1} m \left[  \qer(m) - \qer(\ell) \right]
~,
\label{eq:Tq_est}
\end{align}
that generally underestimates the true value of $\TI_q$.

$\TI_q$ is related to the minimal amount of information necessary for an
observer to synchronize to a HMCQS. We say an observer is \emph{synchronized}
when they are able to determine a source's internal state. If $S(\ell)$
converges to its linear asymptote at finite $\ell$, then there exists an
optimal POVM on $\rhoL$ (in the eigenbasis of $\rhoL$) that exactly determines
the HMCQS's internal state. Note that this is not guaranteed to be a
repeated POVM or even consist of local POVMs. The information within that
measurement that is useful for synchronization is quantified by $\TI_q$. In
contrast, if $S(\ell)$ does not converge for finite-$\ell$, then no such POVM
over any finite block of qudits exists, and an observer can (at best) only
converge asymptotically to the source's internal state. We will see in Section
\ref{sec:synchronization} that even this is not possible for most sources when
we restrict ourselves to local measurements.

\subsection{Quantum Markov Order}

The quantum Markov order corresponds to the number of previous qudits on which
the next qudit is conditionally dependent. A process has quantum Markov order
$\MOrderq$ if $\MOrderq$ is the smallest value for which the following
property holds:
\begin{align*}
S(\rho_{0}  \vert  \rho_{-\MOrderq:0}) = S(\rho_{0}  \vert  \rho_{-\infty:0}).
\end{align*}

A graphical interpretation is that when the block size reaches $\MOrderq$,
$S(\ell)$ levels off and has a constant slope---which is $\qer$, as seen in Fig.
\ref{fig:process_properties}. As a consequence $\MOrderq$ is the value of $\ell$
for which $\Delta S(\ell)$ and $\Delta^2 S(\ell)$ converge to their asymptotic
values, as seen in Figs. \ref{fig:entropy_convergence_properties} and
\ref{fig:predictability_convergence_properties} respectively.

Note that a classical process is referred to as `Markov' if $\MOrder = 1$. If a
separable qudit process has an underlying classical process that is Markov, then
it obeys the property in Eq. (\ref{eq:quantum_markov}) but does not generally
have $\MOrderq = 1$.

Consider a separable quantum process $\QBiInfinity = \cqChannel(\BiInfinity)$
where $\BiInfinity$ has Markov order $\MOrder^X$ and $\QBiInfinity$ has quantum
Markov order $\MOrderq$. $\MOrderq$ can be equal to, less than, or greater than
$\MOrder^X$. We will give a simple example of each case:
\begin{itemize}
\item $\MOrderq = \MOrder^X$ trivially if $\AQ$ consists of orthogonal states,
in which case they have identical block entropy curves, via Prop.
\ref{prop:block_entropy_separable}.

\item $\MOrderq < \MOrder^X$ if $\MOrder^X > 0$ and all symbols in $\AX$ are
mapped to the same pure state $\ketpsi[\meassymbol]$. In this case $\MOrderq = 0$, as the
resulting process is i.i.d.

\item $\MOrderq > \MOrder^X$ when $\MOrder^X > 0$, $\vert \AX \vert = \vert \AQ
\vert$ and $\AQ$ consists of nonorthogonal states. Frequently, $\MOrderq = 
\infty$ since arbitrary sequences of nonorthogonal states cannot reliably be
distinguished with a finite POVM.
\end{itemize}

Similar rules apply when comparing $\MOrderq$ to the Markov order $\MOrder^Y$ of
measured process $\OutputBi = \qcChannel(\QBiInfinity)$:
\begin{itemize}
\item $\MOrderq = \MOrder^Y$ if $\QBiInfinity$ is a separable process with an
orthogonal alphabet $\AQ$, and $\qcChannel$ uses a POVM whose operators include
a projector for each element in $\AQ$, via Prop.
\ref{prop:block_entropy_measured}.

\item $\MOrderq < \MOrder_Y$ if $\AQ$ consists of orthogonal states, $\MOrderq =
1$ and $\qcChannel$ is a repeated rank-one POVM that does not include
projectors onto the states in $\AQ$.

\item $\MOrderq > \MOrder_Y$ if $\MOrderq > 0$ and $\qcChannel$ is a repeated
POVM measurement with the one-element POVM, $\mathbb{I}$. Note that this is not
a rank-one POVM.
\end{itemize}

Now, with a toolbox of quantum information properties in hand, the following
section calculates (or estimates) their values for paradigmatic examples of
qudit processes.

\section{Qudit Processes}
\label{sec:examples}

We now present a variety of separable qudit processes, organized roughly by
increasing structural complexity and demonstrate the extent to which their
behavior can be quantified with Section \ref{sec:quantum_properties}'s
informational measures. Their properties are summarized in Table
\ref{table:example_processes} near the end when we have their informational
measures in hand.

\subsection{I.I.D. Processes}

Recall that an i.i.d. (independent and identically-distributed) qudit process
has no classical or quantum correlation between any of the qudits, and its
length-$\ell$ density matrices are in a product state such that:
\begin{align*}
\rhoL = \bigotimes_{i=0}^{\ell-1} \rho_{iid}
~,
\end{align*}
where $\rho_{iid} \in \HSpace$ is the density matrix for a single timestep.

The quantum block entropy takes the form $S(\ell) = \ell S(\rho_{iid})$, thus
the von Neumann entropy rate is $\qer = S(\rho_{iid})$. This implies that a
repeated projective measurement exists for which the measured process
$\OutputBi$ has a classical entropy rate of $\hmu = S(\rho_{iid})$. The
measurement that realizes this bound consists of orthogonal projectors $P_y$,
each of which is constructed from an eigenvector of $\rho_{iid}$. In the special
case where $\rho_{iid}$ is the maximally-mixed state, $\qer = \log_2(d)$ and any
set of orthogonal projectors on a block $\rhoL$ gives a uniform distribution
over all measurement outcomes $y_{0:\ell}$. Since there are no correlations
between qudits $\EE_q = 0$, $\TI_q = 0$, and $\MOrderq = 0$ trivially. The
output of a single-qudit source with uncorrelated noise can be considered an
i.i.d. qudit process.

\subsection{Quantum Presentations of Classical Processes}

Any classical process with alphabet $\AX$ can be represented by a qudit process
with orthogonal alphabet $\AQ$ where each symbol $x \in \AX$ corresponds to a
pure state $\ketpsi[\meassymbol] \in \AQ$. In this case the encoding
$\cqChannel$ is trivial and one can recover the underlying process $\BiInfinity$
via repeated measurement with orthogonal projectors $\{P_x =
\ketpsi[\meassymbol]\brapsi[\meassymbol]\}$. This requires that $d \geq \vert
\AX \vert$.

The information measures for the underlying process, the qudit process, and
measurement outcomes obey:
\begin{align*}
\H{X_{0:\ell}} & = S(\rhoL) \\
	& \leq \H{Y_{0:\ell}}
~,
\end{align*}
with equality for repeated measurement with a rank-one POVM whose elements
include the projector set $\{P_x\}$.

From this relation we can see that many quantum information quantities such as
$\qer$, $\EE_q$, $\TI_q$, and $\MOrderq$ are equal to the classical
properties of $\BiInfinity$. Exceptions include quantities that depend on the
relationship between $d$ and $\vert \AX \vert$, such as the quantum redundancy.

Since the quantum-classical channel $\cqChannel$ is trivial, the output process
$\OutputBi$ is the result of passing $\BiInfinity$ through a noisy classical
channel, where the level of noise depends on the particular measurement scheme
$\qcChannel$.

For a repeated rank-one POVM $\H{X_{0:\ell}} \leq \H{Y_{0:\ell}}$ for all $\ell$
from Props. (\ref{prop:block_entropy_separable}) and
(\ref{prop:block_entropy_measured}) . Similar inequalities explicitly relate
other classical process properties such as the entropy rates ($\hmu^X \leq
\hmu^Y$), the predictabilities ($\vert \textbf{G}^X \vert \geq \vert
\textbf{G}^Y \vert$), and the excess entropies (see Cor.
\ref{cor:classical_excess_entropies}). 

\subsection{Periodic Processes}

A classical stochastic process $\BiInfinity$ is periodic with period $p$ if it
consists of repetitions of a template sequence---a length-$p$ block of symbols.
A periodic separable qudit process $\QBiInfinity = \cqChannel(\BiInfinity)$ is
one for which the underlying classical process $\BiInfinity$ is periodic.

For classical periodic processes the block entropy curve reaches a maximal value
at its Markov order $\MOrder = p$ and thereafter remains constant with
increasing $\ell$ ($\hmu = 0$). That maximal value is the excess entropy,
which is entirely determined by the period according to the formula $\EE =
\log_2(p)$. Finally, an observer attempting to synchronize to different
length-$p$ templates may encounter more or less uncertainty in the process
depending on the template itself, a feature captured by the transient
information $\TI$ \cite{Crut01a}.

\begin{figure}
\centering
\includegraphics[width=\columnwidth]{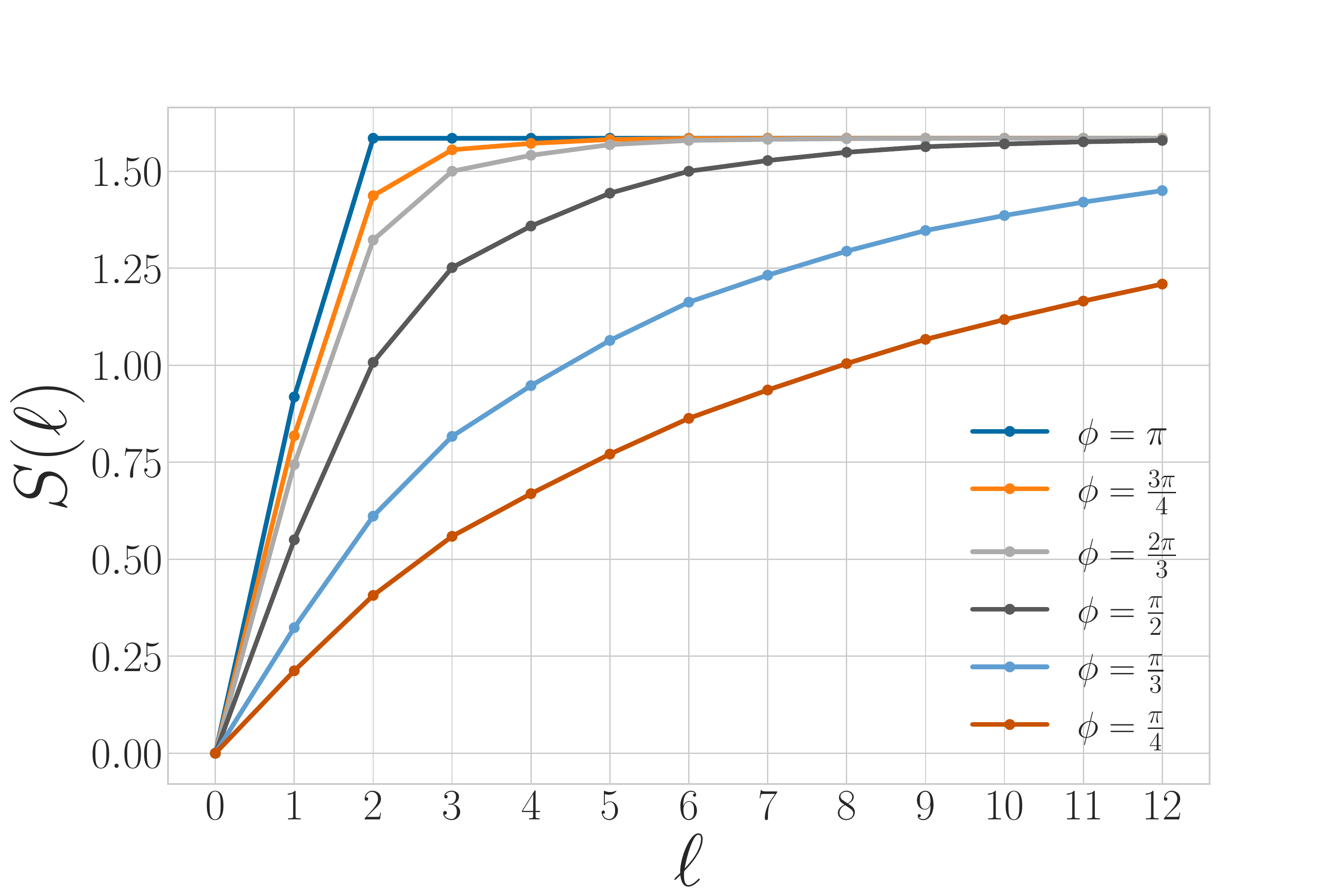}
\caption{Quantum block entropies $S(\ell)$ versus length $\ell$ for the periodic
process emitting the state $\ketpsi[00\phi]$ with different values of $\phi$.
Each curve approaches a maximum value of $\EE_q = \log_2{3}$. Larger values of
$\phi$ correspond to more distinguishable alphabet states and lower values of
$\TI_q$.
	}
\label{fig:period3_block_entropies}
\end{figure}

Periodic qudit processes share many of these properties (as we show via Section
\ref{sec:quantum_properties}'s results), but also exhibit richer behavior since
they can consist of nonorthogonal qudit states. Fig.
\ref{fig:period3_block_entropies} shows the quantum block entropies for the
period-$3$ process consisting of the repeated quantum word $\ketpsi[00\phi] =
\ket{0} \otimes \ket{0} \otimes \ket{\psi(\phi)}$ where $\ket{\psi(\phi)} =
\cos{\frac{\phi}{2}}\ket{0} + \sin{\frac{\phi}{2}}\ket{1}$. For $\phi$ =
$\pi$ we recover the block entropy for the classical period-$3$ word `$001$'. As
$\phi$ decreases $\ket{\psi(\phi)}$ becomes less distinguishable from
$\ket{0}$.

From Props. \ref{prop:qer_separable} and \ref{prop:Eq_separable} it follows that
periodic qudit processes with period $p$ have $\qer = 0$ and $\EE_q \leq
\log_2(p)$. (And, $\EE_q = \log_2(p)$ unless two classical symbols in $\AX$ are
sent to the same pure state in $\AQ$ in a way that reduces the effective period
of the qudit process to less than $p$.) However, the quantum block entropy curve
does not necessarily reach its maximal value of $\EE_q$ for $\ell = p$ if $\AQ$
contains nonorthogonal states. In this case, $\MOrderq \to \infty$ since an
observer cannot unambiguously distinguish where one length-$p$ block begins and
another one ends with any finite measurement.

Though all period-$p$ qudit processes have the same von Neumann entropy rate and
quantum excess entropy, they may be distinguished by their values for the
quantum transient information in two different ways.

First, different quantum alphabets $\AQ$ give different values of $\TI_q$. Fig.
\ref{fig:period3_block_entropies} demonstrates that, for a 2-state qubit
alphabet, as the states become more less distinguishable, the quantum transient
information increases. For $\phi = \pi$ (orthogonal alphabet), $\TI_q \approx
2.33$ bits$\times$time steps, whereas for $\phi = \frac{\pi}{2}$, $\TI_q \approx
4.22$ bits$\times$time steps. These values of $\TI_q$ (and others in this
section) are numerically approximated using Eq. (\ref{eq:Tq_est}) with $\ell =
12$.

Second, $\TI_q$ can distinguish between different length-$p$ words. Reference
\cite{Crut01a} shows that $\TI$ can distinguish between different period-5
classical words (`$00001$', `$00011$', and `$00101$'), and $\TI_q$ generalizes
this behavior. Whereas all period-3 words are equivalent to `$001$' under global
bit swap and translations, the same is not true for period-$5$ words. Section
\ref{sec:synchronization} discusses synchronizing to period-5 qudit sources in
more detail and relates that task to the value of $\TI_q$.

\subsection{Quantum Golden Mean Processes}

The classical Golden Mean process consists of all binary strings with no
consecutive `$1$'s. It is a Markov process ($\MOrder = 1$) since the joint
probabilities $\Pr(\MeasSymbol_{0:\ell})$ for blocks factor as in Eq.
(\ref{eq:markov}) with $\Pr(0 \vert 0) = \frac{1}{2}$, $\Pr(1 \vert 0) =
\frac{1}{2}$, $\Pr(0 \vert 1) = 1$, and $\Pr(1 \vert 1) = 0$. For the classical
Golden Mean $\hmu = \frac{2}{3}$ bits per symbol and $\EE \approx 0.2516$ bits.

\begin{figure}
\centering
\includegraphics[width=\columnwidth]{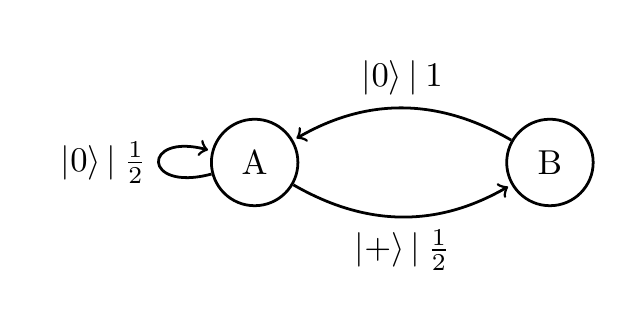}
\caption{$\ket{0}$-$\ket{+}$ Quantum Golden Mean Process generator.}
\label{fig:0+_qgm}
\end{figure}

Replacing the classical symbol alphabet with the quantum alphabet $\AQ =
\{\ket{0},\ket{+}\}$, where $\ket{+} = \frac{1}{\sqrt{2}} \left( \ket{0} +
\ket{1} \right)$, gives the $\ket{0}$-$\ket{+}$ Quantum Golden Mean Process,
introduced in Ref. \cite{Vene19}. Figure \ref{fig:0+_qgm} shows its generator.

\begin{figure}
\centering
\includegraphics[width=\columnwidth]{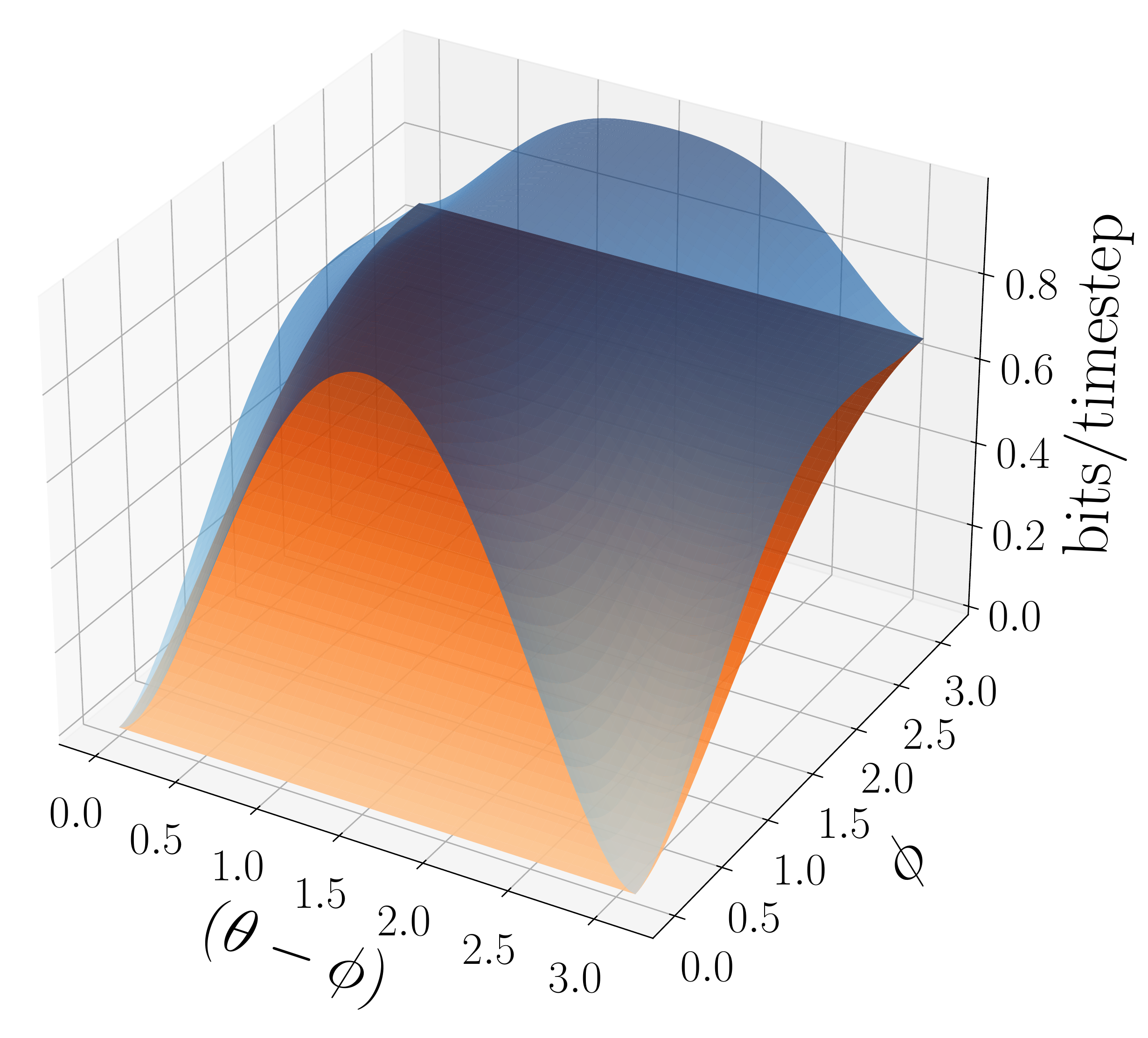}
\caption{von Neumann entropy rate $\qer$ (lower, orange surface) and measured
	entropy rate $\hmu^Y$ (higher, blue surface) for the $\ket{0}$-$\ket{\phi}$
	Quantum Golden Mean process measured with the repeated PVM $\Meas_\theta$.
	$\qer$ increases as $\phi$ does and the alphabet becomes more
	distinguishable. For $\phi = \pi$ and $\theta = 0,\pi$ we recover the classical
	Golden Mean. For $\left( \theta - \phi \right) = \frac{\pi}{2}$, $\Meas_\theta$
	applied to $\ket{\phi}$ is a maximum entropy PVM (distribution of measurement
	outcomes is 50-50). Maxima of $\hmu^Y$ lie in this region.
}
\label{fig:qgm_0theta_entropy_rate}
\end{figure}

We can further generalize this process to the $\ket{0}$-$\ket{\psi(\phi)}$
Quantum Golden Mean process with quantum alphabet $\{ \ket{0}, \ket{\psi(\phi)}
= \cos{\frac{\phi}{2}}\ket{0} + \sin{\frac{\phi}{2}}\ket{1}\}$. This process'
quantum entropy rate is shown in Fig. \ref{fig:qgm_0theta_entropy_rate} for different
values of $\phi$. For $\phi = \pi$, $\ket{\psi(\phi)} = \ket{1}$ and we recover the classical Golden Mean process. As
$\phi$ decreases to $0$, the states in $\AQ$ become less distinguishable and
$\qer$ decreases, as expected from Prop. \ref{prop:qer_separable}.

Also in Fig. \ref{fig:qgm_0theta_entropy_rate} we see the entropy rate
$\hmu^Y$ of the measured processes obtained by applying a repeated PVM to the
$\ket{0}$-$\ket{\psi(\phi)}$ Quantum Golden Mean process. This PVM consists of
projectors parametrized by the angle $\theta$, where $\Meas_\theta =
\{\ket{\psi(\theta)}\bra{\psi(\theta)},\ket{\psi(\theta+\pi)}\bra{\psi(\theta+\pi)}\}$.
As mandated by Prop. \ref{prop:qer_measured}, $\hmu^Y \geq \qer$ for all $\phi$
and $\theta$.

\begin{figure}
\centering
\includegraphics[width=\columnwidth]{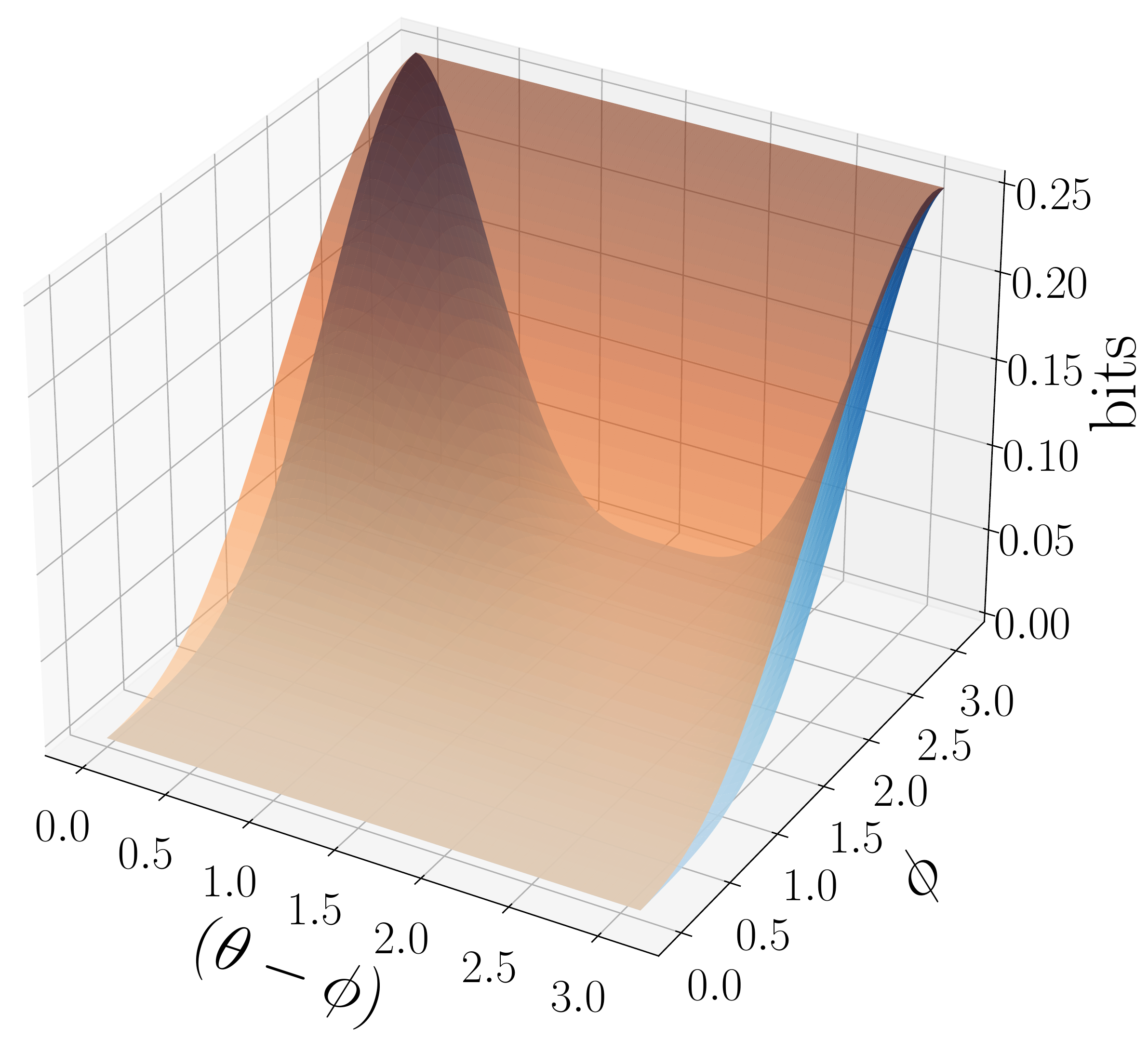}
\caption{Quantum excess entropy $\EE_q$ (higher, orange surface) and measured
	excess entropy $\EE^Y$ (lower, blue surface) for the $\ket{0}$-$\ket{\phi}$
	Quantum Golden Mean process measured with repeated PVM $\Meas_\theta$. $\EE_q$
	increases with $\phi$ since $\ket{0}$ and $\ket{\phi}$ become more
	distinguishable. $\EE_Y$ is maximized for $\left( \theta - \phi \right) = 0,
	\pi$, since $\Meas_\theta$ can best determine if $\ket{\phi}$ was emitted.
}
\label{fig:qgm_0theta_excess_entropy}
\end{figure}

Figure \ref{fig:qgm_0theta_excess_entropy} demonstrates how another of Section
\ref{sec:quantum_properties}'s quantum information properties, the quantum
excess entropy $\EE_q$, relates to the excess entropy $\EE^Y$ of the classical
measured process. For all $\phi$ and $\theta$ the bound from Prop.
\ref{prop:measured_Eq} ($\EE_q \geq \EE^Y$) holds, and $\EE^Y$ has maxima at
$\phi = \pi$, and $\theta = 0, \pi$, when $\AQ = \{\ket{0},\ket{1}\}$ and
$\Meas = \Meas_{01}$. These quantities were estimated using $\ell = 10$.

Underlying the $\ket{0}$-$\ket{\psi(\phi)}$ Quantum Golden Mean is the classical
Golden Mean process, which is Markov. Thus, it obeys the quantum Markov property
of Eq. (\ref{eq:quantum_markov}) despite the fact that it has infinite quantum
Markov order for most values of $\phi$. This has implications for an observer's
ability to synchronize to a Quantum Golden Mean source, which Section
\ref{sec:synchronization} explores.

\subsection{$3$-Symbol Quantum Golden Mean}

\begin{figure}
\centering
\includegraphics[width=\columnwidth]{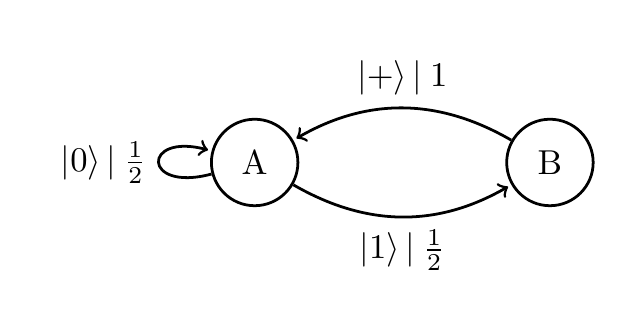}
\caption{$3$-Symbol Quantum Golden Mean Process Generator.
	}
\label{fig:3s_qgm}
\end{figure}

Figure \ref{fig:3s_qgm} shows the generator of the 3-Symbol Quantum Golden Mean
Process with alphabet $\AQ = \{\ket{0}, \ket{1}, \ket{+}\}$. Though its
generator shares the same internal states and transition probabilities as the
$\ket{0}$-$\ket{+}$ Quantum Golden Mean, the 3-Symbol Quantum Golden Mean does
not have a one-to-one correspondence between the quantum alphabet $\AQ$ and the
generator states $\AS$ ($\ket{0} \to A$ and $\ket{+} \to B$). Instead $\vert \AQ
\vert = 3$, and these three states cannot all be orthogonal with $d=2$.

However, unlike the $\ket{0}$-$\ket{+}$ Quantum Golden Mean, we can calculate
the quantum entropy rate directly from the generator because (1) it has the
property of \emph{quantum unifilarity} and (2) it is possible to synchronize to
it. Both are discussed at length in Section \ref{sec:synchronization}. For now,
a HMCQS is \emph{quantum unifilar} if and only if, for every $\sigma \in \AS$
there exists some POVM $\Meas_\sigma$ such that an observer knowing $\sigma$ and
the outcome of $\Meas_\sigma$ can uniquely determine the internal state to which
the HMCQS transitioned. The generator in Fig. \ref{fig:3s_qgm} meets this
criterion with $\Meas_A = \Meas_{01}$. $\Meas_B$ can be any POVM.

For a classical, unifilar HMC, $\hmu$ can be calculated as:
\begin{align}
\hmu = \sum_{\sigma_i, \sigma_j \in \AS} \pi_{i} \sum_{x \in \AX} T^{x}_{ij} \log_2 T^{x}_{ij}
~,
\end{align}
where $\pi$ is the stationary state distribution. This result dates back to the
foundations of information theory \cite{Shan48a}.

Similarly, for a quantum unifilar HMCQS to which one can synchronize, we can
write:
\begin{align}
\qer = \sum_{\sigma_i, \sigma_j \in \AS} \pi_{i} \sum_{x \in \AX} T^{x}_{ij} \log_2 T^{x}_{ij}
~.
\end{align}

Let's walk through this logic for the 3-Symbol Quantum Golden Mean. In state $A$
($\pi_A = \frac{2}{3}$), the generator emits a qubit either in state $\ket{0}$
or $\ket{1}$, each with probability $\frac{1}{2}$. The density matrix describing
this qubit is a maximally-mixed state, and any measurement performed on it
involves $1$ bit of irreducible randomness. If it is in state $B$ ($\pi_B =
\frac{1}{3}$), it emits a qubit in state $\ket{+}$ deterministically. Thus,
averaging over the states, the von Neumann entropy rate is $\qer = \frac{2}{3}$
bits/timestep.

\begin{figure}
\centering
\includegraphics[width=\columnwidth]{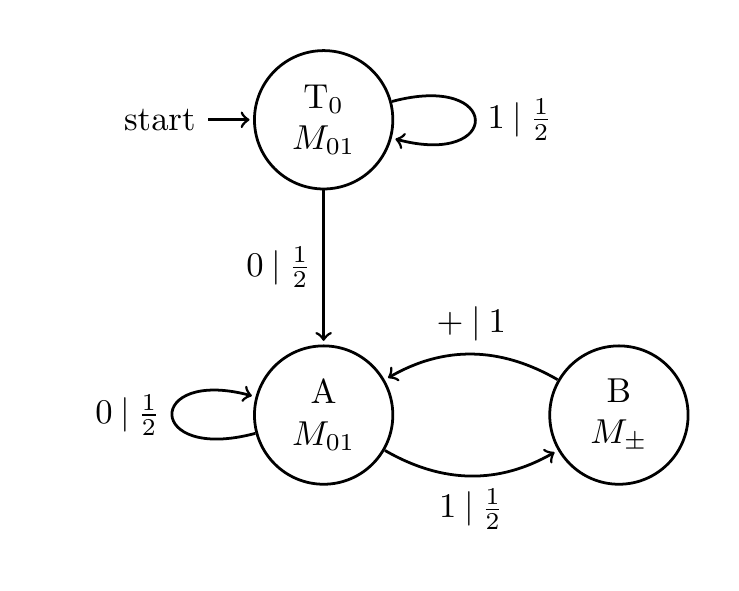}
\caption{Adaptive Measurement Protocol (in the form of a DQMP) for the
	$3$-Symbol Quantum Golden Mean process. To synchronize, an observer starts
	in $T_0$ (a transient state) and measures with $\Meas_{01}$. The
	probability of observing exactly $n$ `$1$'s is $\frac{1}{2^n}$. Upon
	observing a `$0$', the observer synchronizes. States $A$ and $B$ correspond
	exactly to internal states $A$ and $B$ of the source in Fig.
	\ref{fig:3s_qgm}. The `$-$' transition is not displayed because it has
	probability $0$. The source is quantum unifilar, thus one stays
	synchronized for future times.
}
\label{fig:3s_qgm_synch}
\end{figure}

Despite this, when restricted to measuring with a repeated rank-one POVM, the
entropy rate of the observed classical process cannot reach the lower bound of
$\frac{2}{3}$ bits~per~symbol because the minimum entropy basis when in state
$A$ is $\Meas_{01}$ while the minimum entropy basis when in state $B$ is
$\Meas_{\pm}$. However, an experimenter using the adaptive measurement protocol
in Fig. \ref{fig:3s_qgm_synch} would observe symbol sequences with an entropy
rate of $\frac{2}{3}$ bits~per~symbol once they have synchronized.

They start by measuring in the $\Meas_{01} = \{\ket{0},\ket{1}\}$ basis and use
the outcome of the initial measurement to select a new basis. If the outcome is
`$0$', they are able to synchronize to the process generator which is
necessarily in state $A$. They continue in state $A$ measuring with POVM
$\Meas_{01}$ until they observe a `$1$' at which point they transition to $B$
and measure with $\Meas_{\pm}$, a zero-entropy measurement. They observe a `$+$'
and return to state $A$.

In this way, when using an adaptive measurement protocol the recurrent part
$\OutputBi_r$ of the measurement sequence may have a lower entropy rate than
$\hmu^Y$ for any repeated rank-one POVM measurement, even if the associated DQMP
uses only rank-one PVMs, as in this example. These ideas are formalized and
expanded upon in the next section.

\subsection{Unifilar and Nonunifilar Qubit Sources}

\begin{figure}
\centering
\includegraphics[width=\columnwidth]{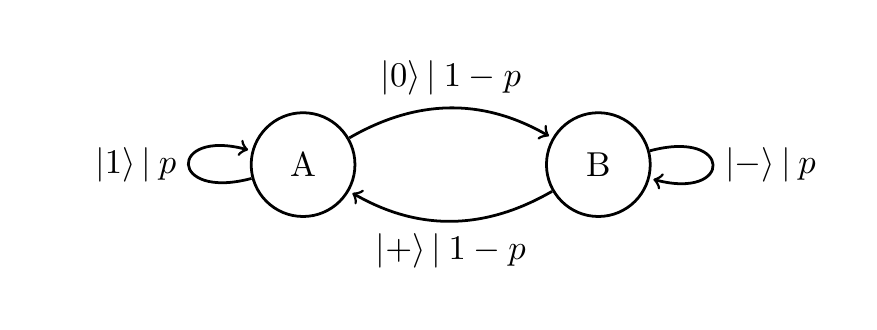}
\caption{Unifilar Qubit Source. Each internal state emits one of two orthogonal
	states and then transitions---e.g., $A$ emits either $\ket{0}$ or $\ket{1}$
	that can be distinguished by measurement $\Meas_{01}$---giving this source
	the property of quantum unifilarity. $p$ is a parameter that takes values
	from $0$ to $1$. Other processes correspond to particular p-values: for
	example, a nonorthogonal period-$2$ process ($p = 0$), the maximally-mixed
	i.i.d. process ($p = \frac{1}{2}$) and a deterministic sequence of either
	$\ket{1}$ or $\ket{-}$ ($p = 1$).
	}
\label{fig:unifilar_qubit_source}
\end{figure}

\begin{figure}
\centering
\includegraphics[width=\columnwidth]{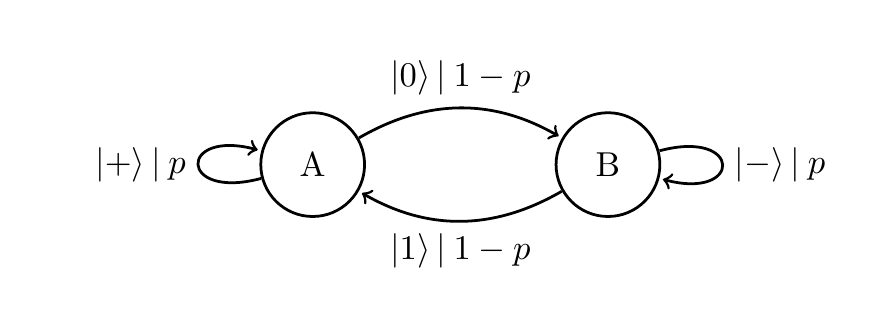}
\caption{Nonunifilar Qubit Source. Each internal state emits one of two
nonorthogonal states and then transitions. An observer will not be able to
determine which state the source transitioned to with any POVM. $p$ takes values
from $0$ to $1$. Other processes correspond to particular p-values: for example
an orthogonal period-$2$ process ($p = 0$), the maximally-mixed i.i.d. process
($p = \frac{1}{2}$) and a deterministic sequence of either $\ket{1}$ or
$\ket{-}$ ($p = 1$).}
\label{fig:nonunifilar_qubit_source}
\end{figure}

The last two classes of qubit processes to discuss are generated by the
unifilar and nonunifilar qubit sources shown in Figs.
\ref{fig:unifilar_qubit_source} and \ref{fig:nonunifilar_qubit_source},
respectively. Both of these generators consist of two internal states ($\AS =
\{A,B\}$) and emit four possible pure-qubit states ($\AQ = \{ \ket{0}, \ket{1},
\ket{+}, \ket{-}\}$) that form two orthogonal pairs. However, for each internal
state of the unifilar qubit source, the two possible emitted states are
orthogonal and one can unambiguously determine the next internal state. In other
words, it has the property of quantum unifilarity. The same is not true of
nonunifilar qubit source. The next section illustrates how this difference
strongly affects an observer's ability to synchronize.

By varying the parameter $p$, one can interpolate between one of the several
simpler processes already analyzed. Starting with the unifilar qubit source in
Fig. \ref{fig:unifilar_qubit_source}, for $p = 0$, the generator becomes a
period-$2$ source that emits the word $\ketpsi[0+]$. For $p=\frac{1}{2}$ we
obtain a two-state model that generates the i.i.d. maximally-mixed process. And,
as $p \to 1$ the generator emits longer strings of either only $\ket{1}$ or only
$\ket{-}$ qubits, depending on whether the source is in $A$ or $B$. At $p = 1$
the process becomes nonergodic (here demonstrated by the disconnection between
internal states), and the source will only emit either $\ket{1}$ or $\ket{-}$
deterministically.

Similarly, the nonunifilar qubit source in Fig.
\ref{fig:nonunifilar_qubit_source} simplifies for certain $p$'s. For $p = 0$ we
obtain a period-$2$ source, this time with sequence $\ketpsi[01]$, and for $p =
\frac{1}{2}$ it also generates the i.i.d. maximally-mixed process. As $p \to 1$
it emits long sequences of either all $\ket{+}$ or all $\ket{-}$ and becomes
nonergodic for $p = 1$.

\subsection{Unifilar Qutrit Source}

\begin{figure}
\centering
\includegraphics[width=\columnwidth]{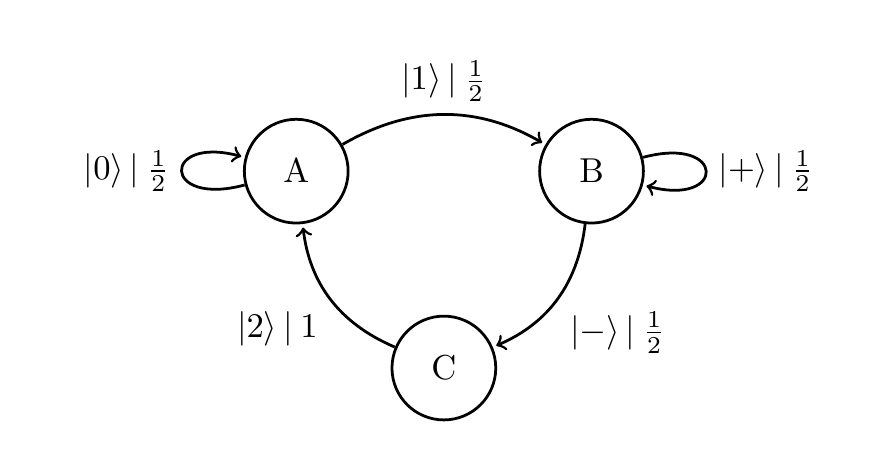}
\caption{Unifilar Qutrit Source. When in internal states $A$ and $B$ it emits a
qutrit in the subspace of Hilbert space spanned by $\ket{0}$ and $\ket{1}$. When
in $C$ it emits $\ket{2}$, which can always be distinguished from all other
states in $\AQ$. This demonstrates additional opportunities for synchronization
in higher-dimensional Hilbert spaces.
	}
\label{fig:unifilar_qutrit_source}
\end{figure}

Expanding beyond processes over qubits, we consider a single example of a
qutrit ($d = 3$) process whose generator is shown in Fig.
\ref{fig:unifilar_qutrit_source} and that employs a five-qubit alphabet $\AQ =
\{\ket{0}, \ket{1}, \ket{2}, \ket{+}, \ket{-}\}$. Using a higher-dimensional
Hilbert space makes more measurements available to an observer for
synchronization. As a consequence this example process exhibits behavior that is
impossible with qubit processes alone: a subspace of Hilbert space (occupied by
$\ket{2}$) is always distinguishable from all other states in $\AQ$ and can be
reserved for synchronization. The other subspace (including $\ket{0}$,
$\ket{1}$, $\ket{+}$ and $\ket{-}$) consists of states that cannot be
reliably distinguished. Note that this process is quantum unifilar, but one
cannot remain synchronized by measuring in a single basis. The next section
discusses multiple adaptive measurement protocols that can do so.

\begin{table*}[t]
\centering
\begin{tabular}{c|c|c|c|c|c|}
\cline{2-6}
& $\qer$ & $\vert \Gq \vert$ & $\EE_q$ & $\TI_q$ & $\MOrder_q$ \\ \cline{2-6}
& \multicolumn{2}{c|}{(\emph{bits/timestep})} & (\emph{bits})& (\emph{bits$\times$time steps}) & (\emph{time steps}) \\ \hline
\multicolumn{1}{|c|}{I.I.D. Qubit Process} & $S(\rho_{iid})$ & $1- S(\rho_{iid})$ & 0 & 0 & 0 \\ 
\multicolumn{1}{|c|}{Period-3 Process ($\phi = \pi$)} & 0 & 1 & $\log_2(3)$ & $2.33$ & $3$  \\ 
\multicolumn{1}{|c|}{Period-3 Process ($\phi = \pi /2$)} & 0 & 1  & $\log_2(3)$ & $4.22$ & $\infty$            \\ 
\multicolumn{1}{|c|}{Quantum Golden Mean ($\phi = \pi$)} & $2/3$ & $1/3$  & $ 0.2516$ & $1/3$ & $1$            \\ 
\multicolumn{1}{|c|}{Quantum Golden Mean ($\phi = \pi/2$)} & $0.4495$ & 0.5505  & $0.1092$ & $0.5687$ & $\infty$            \\ 
\multicolumn{1}{|c|}{3-Symbol Quantum Golden Mean} & 0.6667 & 0.3333  & $ 0.4652 $ & $0.8855$ & $\infty$            \\ 
\multicolumn{1}{|c|}{Unifilar Qubit Source ($p = 1/3$)} & 0.9184 & 0.0816  & $ 0.0808 $ & $0.1976$ & $\infty$            \\ 
\multicolumn{1}{|c|}{Nonunifilar Qubit Source ($p = 1/3$)} & 0.7306 & 0.2614  & $ 0.3217 $ & $0.4090$ & $\infty$            \\ 
\multicolumn{1}{|c|}{Unifilar Qutrit Source} & 0.8002 & 0.7848  & $ 1.290 $ & $2.156$ & $\infty$            \\ \hline
\end{tabular}
\caption{Information Properties for Example Quantum Processes/Sources. Decimal
	values were numerically estimated using Eqs. (\ref{eq:qer_est}),
	(\ref{eq:Eq_est}), and (\ref{eq:Tq_est}) with $\ell = 8$ for the Unifilar
	Qutrit Source, $\ell = 10$ for the Unifilar and Nonunifilar Qubit Sources,
	and $\ell = 12$ for all other processes. Other values were calculated
	analytically.
	}
\label{table:example_processes}
\end{table*}

\subsection{Discussion} 

Table \ref{table:example_processes} summarizes information properties for the
above examples with either analytic results or numerical estimates. Together
these examples illustrate a range of different features of separable qudit
processes. They demonstrate how the information properties defined and
characterized in Section \ref{sec:quantum_properties} are both indicative of
underlying structural features of quantum information sources and strongly
influenced by the distinguishability of states in a process' quantum alphabet.
Our analysis of the convergence of the quantum block entropy to its linear
asymptote thus gives meaningful and interpretable ways of quantifying
the randomness and correlation in a separable quantum process.

We continue by discussing two tasks that an observer might wish to perform when
faced with a quantum information source emitting a separable quantum process.
First, if they have prior knowledge of the internal structure of the source,
they may want to determine the internal state it occupies during a given
timestep. This task is \emph{synchronization}, and we discuss it in the next
section. Second, if they have no knowledge of the source, they may want to
measure the process it produces to infer its internal structure. This task is
\emph{system identification}, and we demonstrate how an observer can use a
tomographic protocol to perform it in Section \ref{sec:tomography}.

\section{Synchronizing to a Quantum Source}
\label{sec:synchronization}

How does an observer of a process with knowledge of its quantum generator
determine its internal state? When an observer is certain about the internal
state the observer is \emph{synchronized} to the quantum source. The following
explores synchronizing to quantum processes---both the manner in which
observations lead to inferring the source's state and quantitative measures of
partial and full synchronization.

Quantum measurement adds a subtlety to this task in comparison to the task of
synchronizing to a classical process given knowledge of its minimal unifilar
model---the \eM---as described in Refs. \cite{Trav11a, Trav12a}. 

\subsection{States of Knowledge}

Recall that a hidden Markov chain quantum source (HMCQS) consists of a set of
internal states ($\AS$), a pure-state alphabet ($\AQ$), and a set of labeled
transition matrices ($\AT$). We assume an observer has complete knowledge of the
HMCQS that generates a process but can only infer the internal state at time
$t=\ell$ by applying block measurement $\qcChannel_{0:\ell}$ and observing
outcome $y_{0:\ell}$. They have no access to the qudits that were emitted before
$t=0$.

An observer's best guess for the internal state of a source given different
sequences of observations can be represented as distributions over the source's
internal states, known as \emph{mixed states} (not to be confused with mixed
quantum states, represented by density matrices). Classically, after observing a
particular length-$\ell$ word $w = \meassymbol_{0:\ell}$, the observer is in the
mixed state $\eta(w) = \left(p_A, p_B, \ldots \right)$. After the next
observation $\MeasSymbol_{\ell}$, they will transition to one of a set of
new mixed states depending on the outcome $\eta(w,\meassymbol_\ell =
\text{0})$, $\eta(w,\meassymbol_\ell = \text{1})$, and so on. The word
corresponding to the new mixed state is a concatenation of $w$ and the new
observation $\meassymbol_\ell$. The set of mixed states for a classical process
and the dynamic between them define a process' \emph{mixed state presentation}
(MSP), which is unifilar by construction \cite{Crut08b}.

Using a classical process' MSP rather than a nonunifilar generator of the
process has many computational advantages. Two of interest are: it allows
one to calculate the entropy rate for processes without finite unifilar
presentations \cite{Jurg21} and it allows one to calculate the uncertainty an
observer experiences while attempting to synchronize to a process' generator
\cite{Crut13a}.

For quantum processes there is no unique MSP but rather a multiplicity of
possible MSPs, each corresponding to a different choice of measurement protocol.
For a given source and given measurement protocol $\qcChannel$ we can define a
set of mixed states with each corresponding to the possible measurement sequence
one can observe.

Consider the mixed states corresponding to length-$\ell$ sequences of
observations. We restrict $\qcChannel_{0:\ell}$ to consist of local POVMs,
allowing for adaptive measurement. Given that an observer has applied
measurement $\qcChannel_{0:\ell}$ and seen measurement outcomes $y_{0:\ell}$,
their best guess about the generator's internal state is represented by the
conditional distribution $\eta(y_{0:\ell} \vert \qcChannel_{0:\ell}) = \{\Pr(\sigma \vert
y_{0:\ell}, \qcChannel_{0:\ell}) \; \text{for~all} \; \sigma\}$.

For $t = 0$, an observer has no measurement outcomes with which to inform their
prediction about the source's internal state. However, they do know the
stationary state distribution $\pi$ of the model (since it can be calculated
directly from $\AT$). This serves as a `best guess' of the source's internal
state absent any measurements. If the initial state distribution
$\mathcal{S}_0$ is not $\pi$ and the observer is aware of this fact, the mixed
states for that process are $\eta(y_{0:\ell} \vert \qcChannel, \mathcal{S}_0)$.
We omit the conditioning on $\mathcal{S}_0$ if $\mathcal{S}_0 = \pi$.

For most repeated PVM measurements of qubit processes there are an
uncountably-infinite number of mixed-states \cite{Vene19}. This
\emph{measurement-induced complexity} appears even for $\vert \AS \vert = 2$.
The examples in this section step back from this complexity to focus on
measurements that are sufficiently informative to allow an observer to
synchronize with finite observation sequence $y_{0:\ell}$ and result in only
finite or (at most) a countably-infinite set of recurrent mixed states.

\subsection{Average State Uncertainty and Synchronization Information}

To compare synchronization behavior for different measurement protocols it is
useful to use entropic quantities rather than working with the set of mixed
states and their dynamic directly. In particular, we look at the entropy of the
possible mixed states for length-$\ell$ sequences of measurements.

An observer's uncertainty about the source's internal state after applying
measurement sequence $\qcChannel_{0:\ell}$ (according to some protocol
$\qcChannel$) and observing outcome $y_{0:\ell}$ is given by the Shannon entropy
of the corresponding mixed-state distribution:
\begin{align*}
\H{\eta(y_{0:\ell} \vert \qcChannel)} & = \H{\Pr(\sigma \vert y_{0:\ell}, \qcChannel)} \\
& = -\sum_{\sigma \in \mathbf{\mathcal{S}}} \Pr(\sigma \vert y_{0:\ell}, \qcChannel) \log_2
\Pr(\sigma \vert y_{0:\ell}, \qcChannel)
~.
\end{align*}

One can average over all measurement outcomes on a block of $\ell$ qudits to
find the \emph{average state uncertainty}:
\begin{align*}
\mathcal{H}(\ell \vert \qcChannel) & \equiv \H{\eta(Y_{0:\ell} \vert \qcChannel)} \\
                & \equiv -\sum_{y_{0:\ell}} \Pr(y_{0:\ell}\vert \qcChannel) \H{\eta(y_{0:\ell}\vert \qcChannel)}
~.
\end{align*}

For a given measurement protocol $\qcChannel$, this quantity generally converges
to a finite value in the $\ell \to \infty$ limit. This limit does not
necessarily exist if is not ergodic; for example, if there is some underlying
periodicity. If such a limit exists, the \emph{asymptotic state uncertainty} is:
\begin{align*}
C_\infty(\qcChannel) = \lim_{\ell \to \infty} \mathcal{H}(\ell \vert \qcChannel)
~.
\end{align*}
Both $\mathcal{H}(\ell|\qcChannel)$ and $C_\infty(\qcChannel)$ are measured in
bits.

If $\H{\eta(y_{0:\ell} \vert \qcChannel)} = 0$, then $y_{0:\ell}$ is a
\emph{synchronizing observation}, and an observer who sees outcome $y_{0:\ell}$
can precisely identify the source's internal state. If $\mathcal{H}(\ell \vert
\qcChannel) = 0$, then any measurement outcome seen when applying protocol
$\qcChannel$ to $\ell$ qudits allows the observer to synchronize to the
source. For classical and quantum processes, the Markov order is the first value
of $\ell$ for which $\mathcal{H}(\ell)$ vanishes (for some $\qcChannel_{0:\ell}$
in the quantum case). This generally does not occur for HMCQS with nonorthogonal
states in $\AQ$ except in the $\ell \to \infty$ limit since $\MOrderq$ is
generally infinite.

Being synchronized after measuring $\ell$ qudits does not guarantee the observer
remains synchronized after measuring qudit $\ell + 1$. In classical processes
for which this occurs, persistent synchronization requires that the HMC of the
process that the observer uses satisfies the additional condition of
\emph{unifilarity}. In the quantum setting, synchronization persists if and only
if the underlying HMCQS is \emph{quantum unifilar} and, when the observer is in
internal state $\sigma$, their measurement protocol ensures they apply the
measurement for which the internal state at time $t+1$ is completely determined.

This is equivalent to the statement `There exists some protocol $\qcChannel$ with
measurements $\qcChannel_{0:\ell}$ and $\Meas_\ell$ such that:
\begin{align}
 & \H{\Pr(\sigma \vert y_{0:\ell},\qcChannel_{0:\ell})} = 0 \nonumber \\
 & \qquad \implies \H{\Pr(\sigma \vert y_{0:\ell}y_{\ell},\qcChannel_{0:\ell}\otimes\Meas_\ell)} = 0
   ~,
\label{eq:quantum_unifilar}
\end{align}
for all $y_\ell \in \AY$.'

The measurement that maintains synchronization when the source is in one
internal state ($\sigma_i$) does not need to be the same as the measurement that
maintains synchronization when the source is in another internal state
($\sigma_j$). When the measurement required to maintain synchronization depends
upon the HMCQS's current state, then \emph{adaptive measurement protocols} are
capable of maintaining synchronization even when no fixed-basis measurement can,
as the following demonstrates with multiple examples.

Finally, the total amount of state uncertainty that an observer encounters
while synchronizing to a process using a given measurement protocol $\qcChannel$
is the \emph{synchronization information}, given by:
\begin{align}
\mathbf{S}(\qcChannel) = \sum_{\ell = 0}^{\infty} \mathcal{H}(\ell | \qcChannel)
~.
\end{align}
Note that, if the asymptotic state uncertainty $C_\infty(\qcChannel)$ is greater
than 0, $\mathbf{S}(\qcChannel)$ is infinite. When $C_\infty(\qcChannel) = 0$,
we can estimate $\mathbf{S}(\qcChannel)$ by terminating the sum at some finite
$\ell$.

For classical processes the synchronization information is closely tied to the
transient information $\TI$. We will similarly connect $\mathbf{S}(\qcChannel)$
to the quantum transient information $\TI_q$ and use it as a way to compare
synchronization via different measurement protocols.

The remainder of this section explores synchronization to various models
from Section $\ref{sec:examples}$. We pair each with a variety of measurements,
both repeated POVMs and adaptive protocols, to demonstrate the way
$\mathcal{H}(\ell | \qcChannel)$ is affected by this choice.

\subsection{Synchronizing to Quantum Presentations of Classical Processes}

A HMCQS with a qudit alphabet consisting entirely of orthogonal states
($\braket{\psi_x \vert \psi_{x'}} = 0$, for all $\ketpsi[\meassymbol]$ and
$\ket{\psi_{x'}} \in \AQ$), is a quantum presentation of the underlying
classical process $\BiInfinity$. An observer can perform a measurement using
orthogonal projectors $P_x = \ketpsi[\meassymbol]\brapsi[\meassymbol]$ for all
$\ketpsi[\meassymbol] \in \AQ$ that unambiguously discriminates between the pure
qudit states the source emits.

The task of synchronizing to such an HMCQS is equivalent to synchronizing to a
source emitting classical symbols. If it is quantum unifilar, then the
underlying classical HMC is unifilar and synchronization is exponentially
fast (on average) \cite{Trav11a, Trav12a}. Absent unifilarity, an observer may
not be able to synchronize even asymptotically---i.e., $C_\infty(\qcChannel) >
0$ for all $\qcChannel$---despite measuring with orthogonal projectors.

\subsection{Synchronizing to Periodic Processes}

All periodic quantum processes have vanishing von Neumann entropy rate---i.e.,
$\qer = 0$---and each state only has one incoming and one outgoing transition.
Due to these simplifications, to synchronize an observer simply determines the
source's \emph{phase}; i.e., which of the $p$ internal states the source
occupies at time $t$. Once this phase is determined for any $t$, it is also
determined for all other $t$. The initial phase uncertainty is equivalent to
the initial state uncertainty $\H{\pi} = \log_2 (p)$ since $\pi$ is uniform.

Any measurement protocol $\qcChannel$ that distinguishes between states in
$\AQ$ gives information about the phase and allows an observer to synchronize
to the source asymptotically. If two states in $\AQ$ cannot be distinguished by
$\qcChannel$, then synchronization may not be possible, even with an infinite
number of measurements.

For classical periodic processes, an observer synchronizes at the Markov order
$\ell = p$, and the synchronization information and the transient information
are equal: $\mathbf{S} = \TI$ \cite{Crut01a}. In contrast, quantum periodic
processes generically have infinite Markov order, if there are nonorthogonal
states in $\AQ$, and only asymptotically synchronize. Thus,
$\mathbf{S}(\qcChannel) \geq \TI_q$. The condition for equality is that
$\qcChannel$ is the optimal measurement over the process. For $\MOrder_q =
\infty$ this means that $\qcChannel$ must be a global measurement over the
bi-infinite chain of qudits.

\begin{figure}
\centering
\includegraphics[width=\columnwidth]{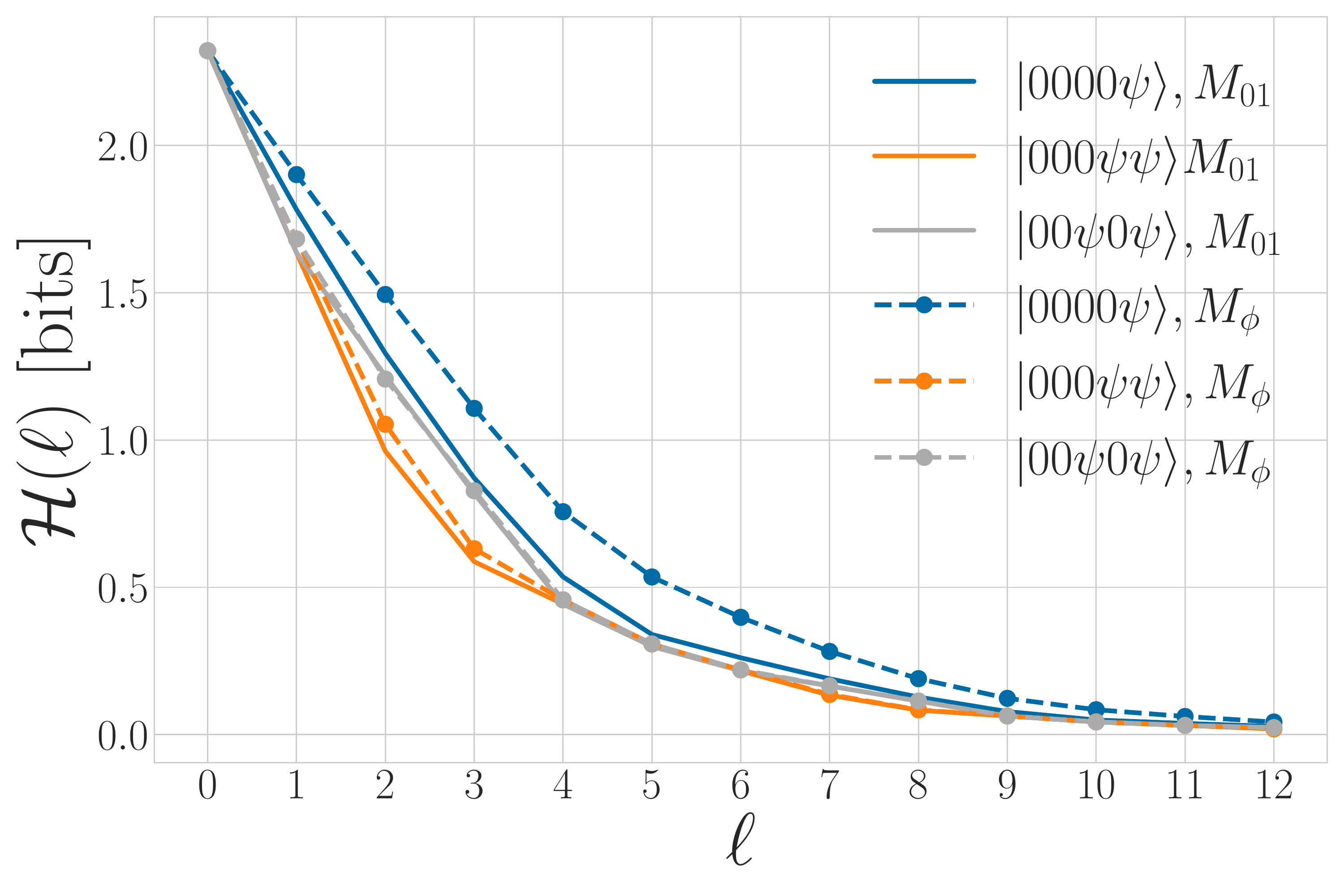}
\caption{Average state uncertainty $\mathcal{H}(\ell)$ for period-$5$ qudit
	processes: $\psi$ denotes state $\ket{\psi(\phi)}$. The associated PVM
	is $\Meas_\phi = \{\ket{\psi(\phi)},\ket{\psi_(\phi+\pi)}\}$. We set $\phi =
	3 \pi / 4$. The area under each curve is the synchronization information
	$\mathbf{S}$ for that process and measurement.
	}
\label{fig:periodic_H(l)}
\end{figure}

Figure \ref{fig:periodic_H(l)} shows the average state uncertainty for an
observer measuring three different period-$5$ qubit processes consisting of two
alphabet states: $\ket{0}$ and $\ket{\psi(\phi = 3 \pi / 4)}$. All other
period-$5$ sequences with this qubit alphabet are equivalent to these three
under shift and swap symmetries. As expected, $\mathcal{H}(\ell)$ decreases from
$\mathcal{H}(0) = \log_2 5$ to $C_\infty = 0$ for all combinations of sequence
and measurement basis. That noted, the rate at which synchronization
occurs---and the total amount of uncertainty seen by an observer, the
synchronization information---depends on both the particular sequence and the
particular repeated measurement applied to it.

For period-5 process with word `$0000\psi$', $\TI_q \approx 6.32$
bits$\times$time steps, using Eq. (\ref{eq:Tq_est}) with $\ell = 12$. When
measuring it in basis $\Meas_{01}$ ($\Meas_{\phi}$) we find that
$\mathbf{S}(\Meas_{01}) \approx 7.92$ bits ($\mathbf{S}(\Meas_{\phi}) \approx
9.30$ bits). These are also estimated by calculating up to $\ell = 12$.
Similarly, the bound between $\mathbf{S}(\qcChannel)$ and $\TI_q$ holds for the
other period-5 words. For word `$000\psi \psi$', $\TI_q \approx 4.86$
bits$\times$time steps, $\mathbf{S}(\Meas_{01}) \approx 6.84$ bits, and
$\mathbf{S}(\Meas_{\phi}) \approx 7.05$ bits. For word `$00\psi 0\psi$', $\TI_q
\approx 5.51$ bits$\times$time steps, $\mathbf{S}(\Meas_{01}) \approx 7.39$
bits, and $\mathbf{S}(\Meas_{\phi}) \approx 7.47$ bits.

At this point we wish to emphasize that $\qer$, $\EE_q$, and $\MOrder_q$ are
equal for generic period-5 processes with nonorthogonal alphabets $\AQ$. Despite
this, $\TI_q$ and $\mathbf{S}(\qcChannel)$ identify physically-relevant
differences between these processes for the task of synchronization. These
differences are intrinsic to the quantum process itself ($\TI_q$) and also
appear within the measurement outcomes an observer obtains
($\mathbf{S}(\qcChannel)$).

As a final comment on periodic processes, we note that while $\qer = 0$ for all
periodic quantum process, many measurement protocols (even those for which
$C_\infty = 0$) give a measured classical process with a nonzero entropy rate
$\hmu^Y$. In fact, no fixed-basis measurement of a periodic process results in
an observed process with $\hmu^Y = 0$ unless (i) all states in $\AQ$ are
orthogonal and (ii) the measurement is in an orthogonal basis which includes one
projector for each state in $\AQ$.

\begin{figure}
\centering
\includegraphics[width=\columnwidth]{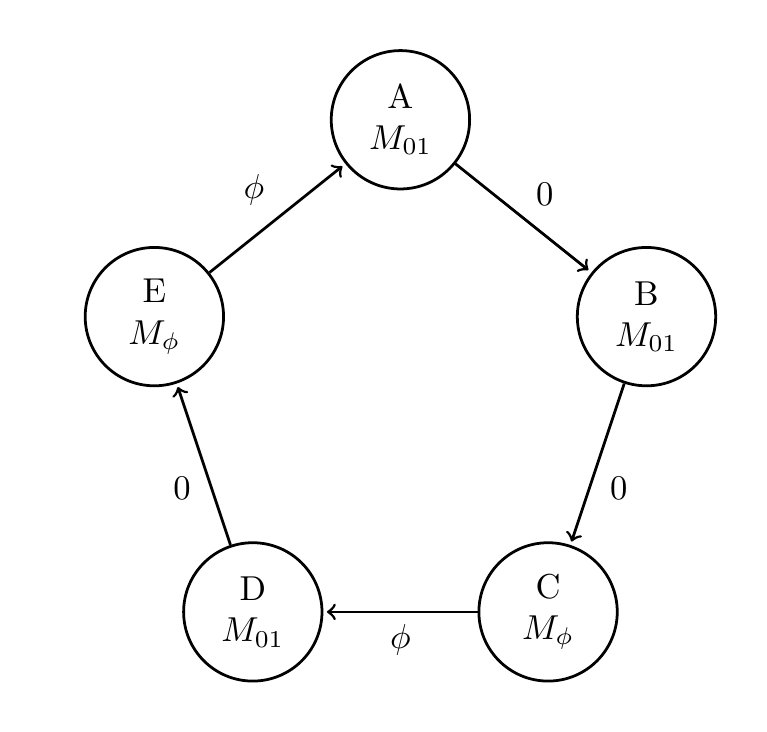}
\caption{Adaptive measurement protocol for period-5 sequence $\ket{00 \psi 0
	\psi}$. Each state is labeled with the next measurement to perform, either
	$\Meas_{01}$, with possible outcomes `$0$' and `$1$', or $\Meas_\phi$, the PVM
	with elements $\ket{\psi(\phi)}\bra{\psi(\phi)}$ and
	$\ket{\psi(\phi+\pi)}\bra{\psi(\phi+\pi)}$ and possible outcomes $\phi$ and
	$\left( \phi+\pi \right)$. The five states shown are the measurement protocol's
	recurrent states, that an observer only encounters when synchronized. An
	observer who is synchronized and using this protocol sees a measured period-$5$
	process with word `$00\phi 0 \phi$'.
}
\label{fig:period5_adaptive}
\end{figure}

In contrast, an observer using an adaptive measurement protocol can easily have
zero uncertainty in measurement outcomes once synchronized. For example, Fig.
\ref{fig:period5_adaptive} shows the recurrent states for a DQMP that swaps
between two different measurements---$\Meas_{01}$ and $\Meas_\phi$---depending
on the internal state of the source. The sequence of measurement outcomes
observed is deterministic, and the recurrent measured process is a period-$5$ process with
word `$00 \phi 0 \phi$' and $\hmu^{Y_{r}} = 0$.

\subsection{Synchronizing with PVMs}

The $\ket{0}$-$\ket{+}$ Quantum Golden Mean Process has multiple synchronizing
observations; see the generator in Fig. \ref{fig:0+_qgm}. Measuring with
$\Meas_{01}$ and seeing a `$1$' synchronizes the observer to internal state $B$;
measuring with $\Meas_{\pm}$ and seeing a `$-$' synchronizes the observer to
internal state $A$. Let's consider the mixed states produced by these two
repeated PVMs in turn, starting with $\Meas_{01}$.

\begin{figure*}
\centering
\includegraphics[width=\textwidth]{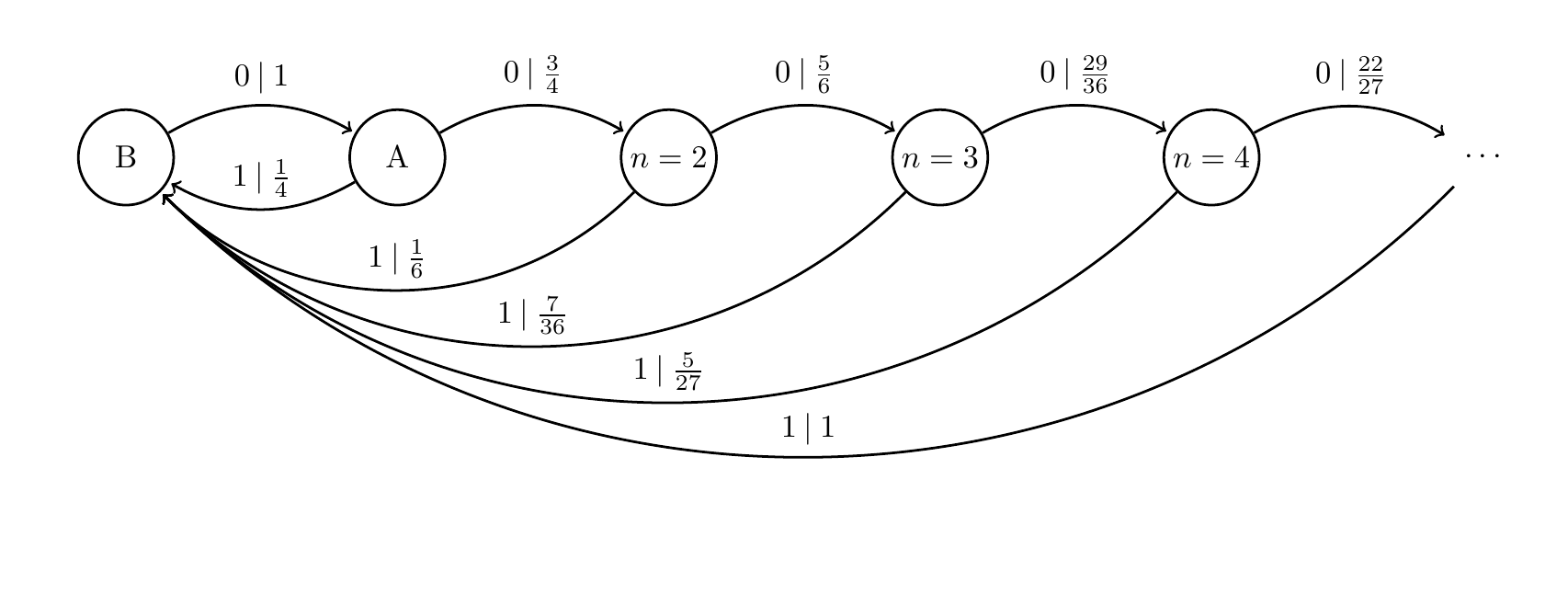}
\caption{Mixed state presentation for the $\ket{0}$-$\ket{+}$ Quantum Golden
	Mean Process measured with $\Meas_{01}$. $n$ refers to the number of
	consecutive `$0$'s since the most recent `$1$'.
	}
\label{fig:0+_qgm_01}
\end{figure*}

Figure \ref{fig:0+_qgm_01} shows the mixed states for the measured process
obtained by repeatedly applying $\Meas_{01}$. Observing synchronizing
measurement $y = \text{`1'}$ means that the source just transitioned to state
$B$ while emitting a $\ket{+}$ qubit; i.e., $\H{\eta(\text{`1'})} = 0$.
Additionally, the source in state $B$ can only transition to state $A$ while
emitting a $\ket{0}$ qubit, so the observer will see a `$0$', and
$\H{\eta(\text{`10'})} = 0$. However, once the source is in state $A$ an
observer easily desynchronizes from the source if they observe another `$0$', as
this is consistent with either of the two transitions to the source. As an
observer sees more `$0$'s they transition to mixed states further to the right
of Fig. \ref{fig:0+_qgm_01}. To summarize, measuring with $\Meas_{01}$ makes
use of synchronizing observations `$1$' and `$10$', but its MSP has a
countably-infinite number of recurrent states corresponding to sequences of $n$
`$0$'s.

This measured process is an infinite-state classical renewal process, whose
information properties can be estimated using methods from Ref. \cite{Marz15b}.
After seeing $n$ `$0$'s in a row, the next observation will be `$1$' with
probability:
\begin{align*}
\Pr(\text{`1'} \vert \text{`0'}^n) & = \tfrac{1}{4}* \Pr(A \vert \text{`0'}^n) \\
   & = \tfrac{3}{16} \left( 1 - \left( -\tfrac{1}{3} \right)^n \right)
  ~.
\end{align*}
We find $\hmu^Y \approx 0.60$ bits~per~symbol, $\EE^Y \approx 0.053$ bits, and a
single-symbol entropy of $\H{1} \approx 0.65$ bits. $\EE^Y$'s small value and
the fact that $\hmu^Y$ is not significantly lower than $\H{\ell = 1}$ indicate
that the infinite-state renewal process presentation provides only a small
predictive advantage over using a biased coin with $\Pr(\text{`0'}) = 5 / 6$.
This is further evidenced by how $\Pr(B \vert \text{`0'}^n)= \tfrac{1}{4} \left(
1 - \left( -1 / 3 \right)^{n-1} \right)$ converges exponentially quickly to the
its asymptotic value of $\Pr(B \vert \text{`0'}^n) = 1/4$.

\begin{figure*}
\centering
\includegraphics[width=\textwidth]{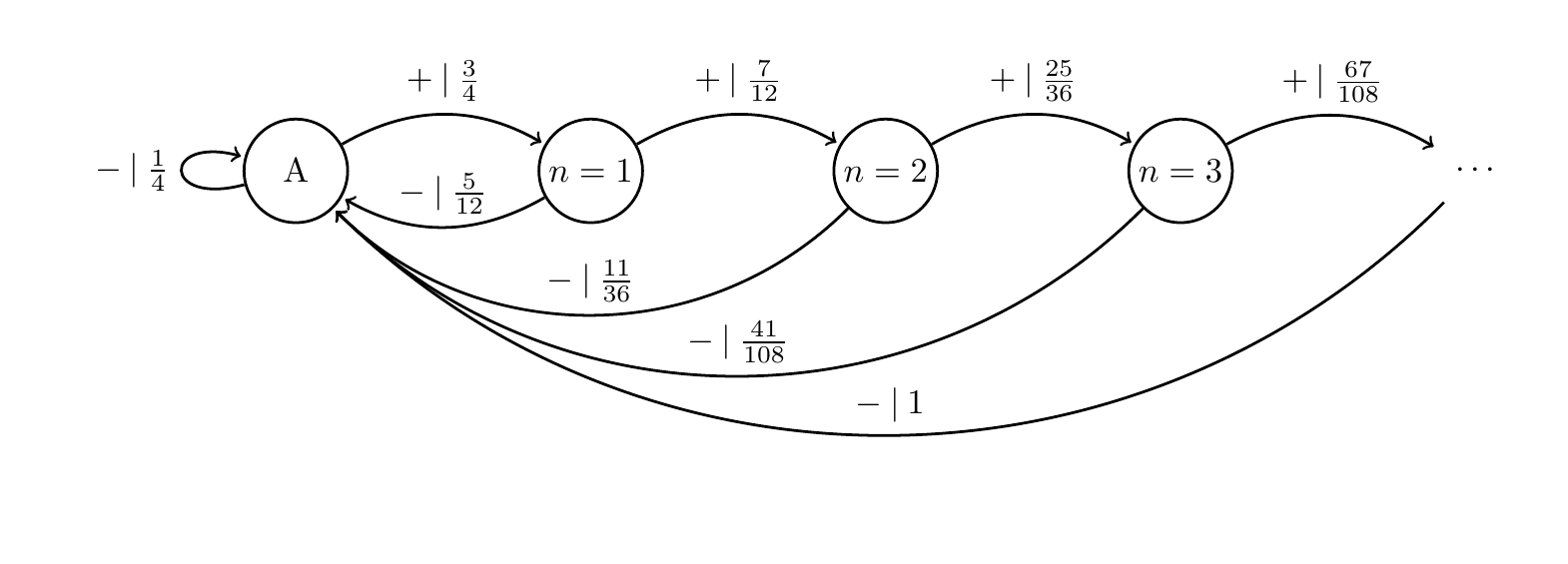}
\caption{Mixed state presentation for the $\ket{0}$-$\ket{+}$ Quantum Golden
	Mean Process measured with $\Meas_\pm$. $n$ refers to the number of
	consecutive `+'s since the most recent `$-$'.
	}
\label{fig:0+_qgm_pm}
\end{figure*}

Measuring with $\Meas_{\pm}$, symbol `$-$' is a synchronizing observation and
indicates the source is in state $A$. The recurrent mixed states for this
observed process are shown in Fig. \ref{fig:0+_qgm_pm}, and they also form a
classical renewal process. Observing $n$ `$+$'s since the last `$-$', the
probability that the generator is in state $A$ is $\Pr(A \vert \text{`+'}^n) =
\frac{3}{5} \left( 1 - \left(-2/3 \right) ^ {n+1} \right)$. The measured process
obtained with $\Meas_{\pm}$ has $\hmu^Y \approx 0.90$ bits~per~symbol, $\EE^Y
\approx 0.020$ bits, and $\H{\ell = 1} \approx 0.91$ bits. Again, the
infinite-state MSP provides only a small predictive advantage over a biased coin
with $\Pr(\text{`+'}) = 2 / 3$. Also, $\Pr(A \vert \text{`+'}^n)$ converges
exponentially quickly to $\Pr(A \vert \text{`+'}^n) = 3/5$.

\begin{figure}
\centering
\includegraphics[width=\columnwidth]{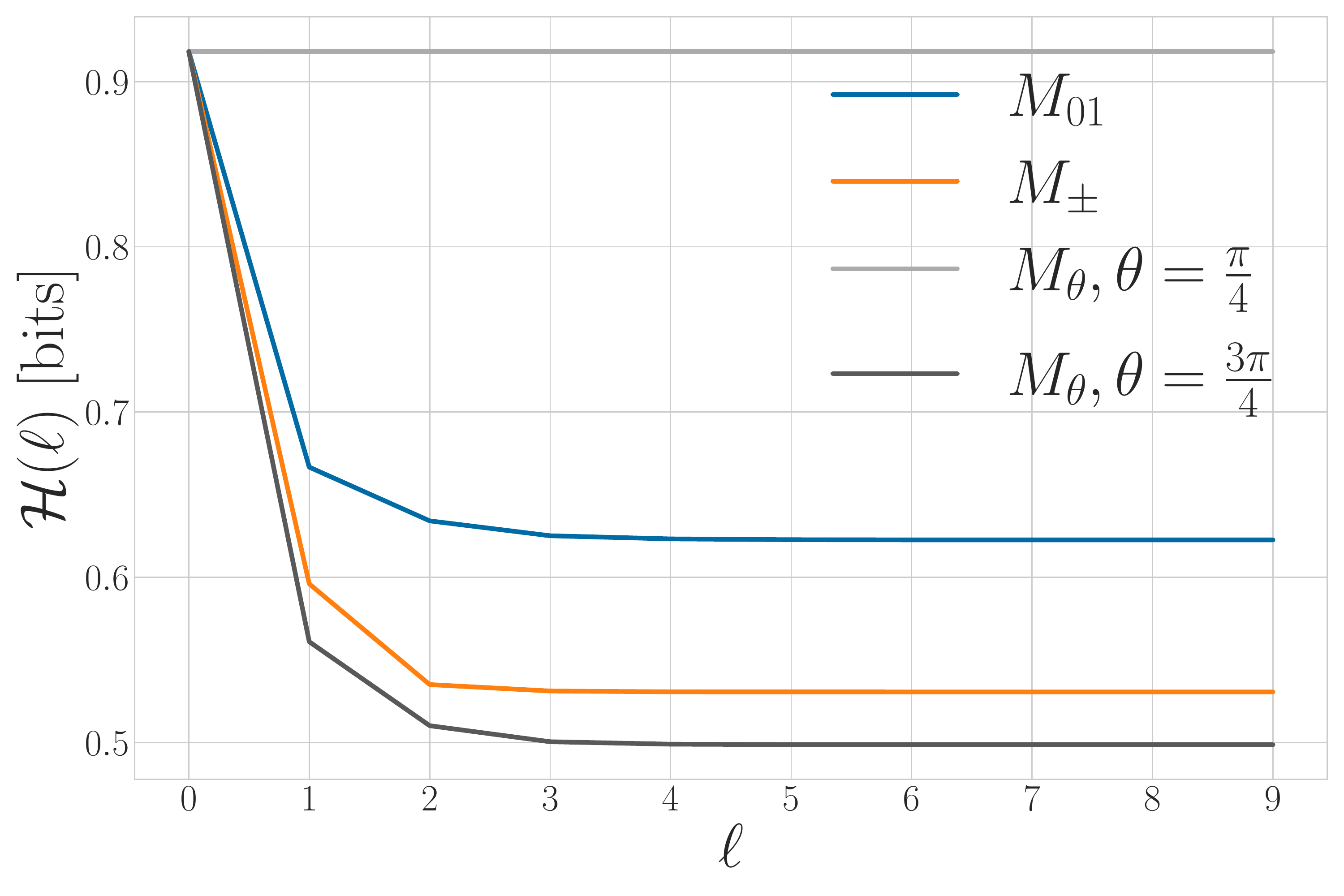}
\caption{Average state uncertainty $\mathcal{H}(\ell)$ for the
	$\ket{0}$-$\ket{+}$ Quantum Golden Mean generator after $\ell$ measurements.
	}
\label{fig:0+_qgm_H(l)}
\end{figure}

Applying either of these two fixed-basis measurements gives an average state
uncertainty that decreases monotonically with $\ell$, as shown in Fig.
\ref{fig:0+_qgm_H(l)}. Since this source is not quantum unifilar, an observer
repeatedly synchronizes and desynchronizes while measuring this process
regardless of basis, and $\mathcal{H}(\ell)$ does not approach $0$ as $\ell \to
\infty$. We find that $C_\infty(\Meas_{01}) \approx 0.62$ bits and
$C_\infty(\Meas_{\pm}) \approx 0.54$ bits. Measuring with $\Meas_{\pm}$ not only
results in less state uncertainty asymptotically but also results in less
average uncertainty for all values of $\ell$.

Figure \ref{fig:0+_qgm_H(l)} also displays the average state uncertainty for two
other relevant PVMs: $\Meas_\theta$ for $\theta = \pi / 4$ and $\theta = 3 \pi /
4$. Recall that $\Meas_\theta$ is the PVM consisting of projectors onto
orthogonal states $\ket{\psi(\theta)}$ and $\ket{\psi(\theta+\pi)}$. For
comparison, $\Meas_{01}$ corresponds to $\theta = 0$ and $\Meas_{\pm}$
corresponds to $\theta = \pi / 2$.

\begin{figure}
\centering
\includegraphics[width=\columnwidth]{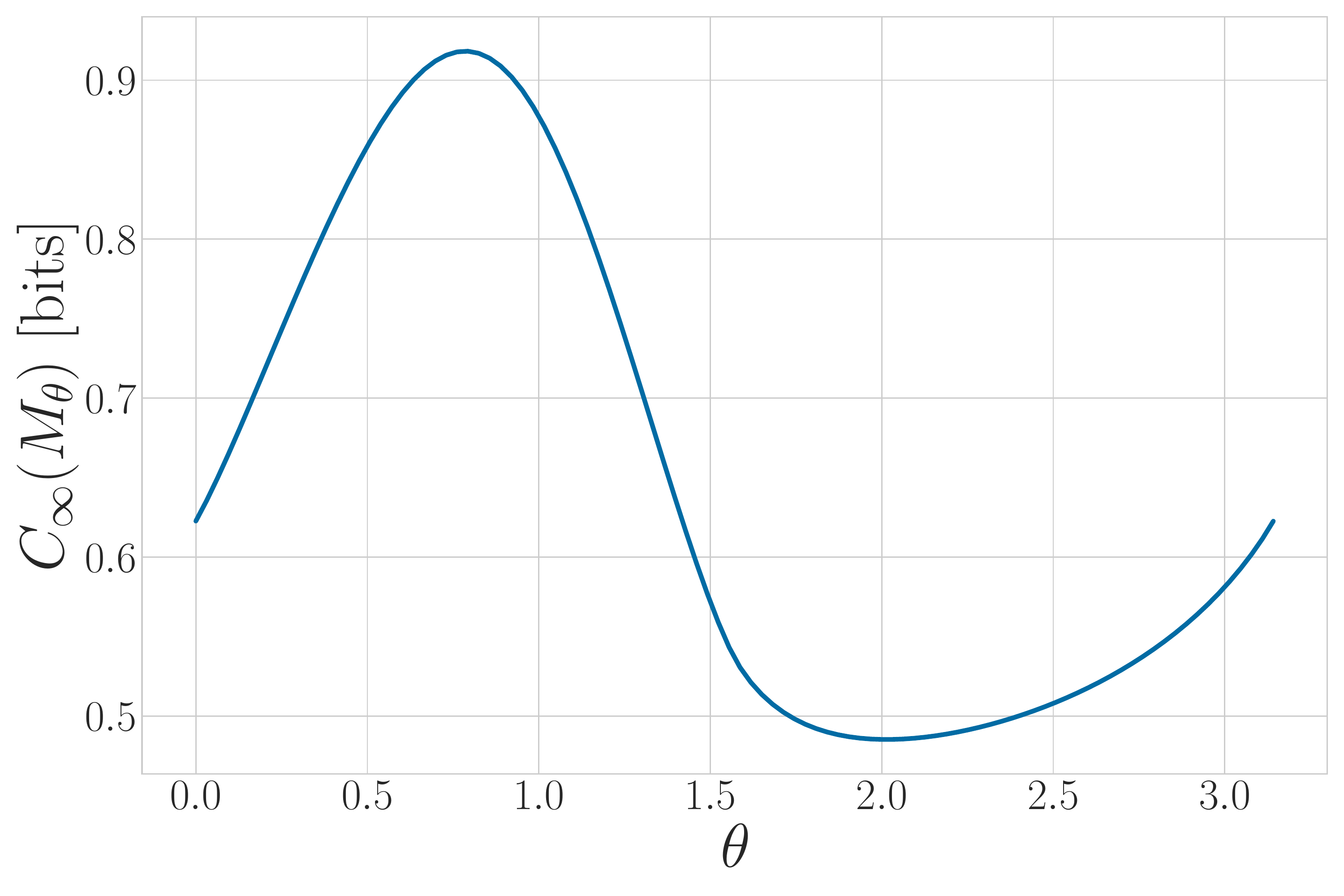}
\caption{Asymptotic state uncertainty when applying the PVM $\Meas_\theta$
	to the $\ket{0}$-$\ket{+}$ Quantum Golden Mean.
	}
\label{fig:0+_qgm_C_inf}
\end{figure}

For $\theta = \pi / 4$, the symbol states $\ket{0}$ and $\ket{+}$ give the
observer the exact same distribution of measurement outcomes, and the observer
cannot gain any information about the state of the source beyond the stationary
state distribution. This can also been seen as the maximum of the asymptotic
state uncertainty in Fig. \ref{fig:0+_qgm_C_inf}.

For $\theta = 3 \pi / 4$---and for the majority of values of
$\theta$---$\Meas_\theta$ has no synchronizing observations. Despite this, Fig.
\ref{fig:0+_qgm_H(l)} demonstrates that $\mathcal{H}(\ell|\Meas_{\theta = 3\pi /
4})$ is lower for all $\ell$ than both bases that can exactly synchronize.

A measured process for generic $\theta$ does not have the renewal process
structure of Figs. \ref{fig:0+_qgm_01} and \ref{fig:0+_qgm_pm}. Instead, in the
absence of synchronizing observations, the number of mixed states generically
grows exponentially with $\ell$. If we approximating the MSP using length-$\ell$
observations, then there will be $\vert \AY^\ell \vert$ mixed states---one for
each possible sequence of measurements---each corresponding to a different
distribution over the source's internal states. These infinite-state MSPs can be
characterized by their statistical complexity dimension \cite{Jurg20c}.

Despite the explosive complexity, many of these PVMs give lower average state
uncertainties than $\Meas_{01}$ and $\Meas_{\pm}$. $\theta = 3 \pi / 4$ is a
representative example. The asymptotic state uncertainty when applying
$\Meas_\theta$ is shown in Fig. \ref{fig:0+_qgm_C_inf}. Note that the maximum
value of $C_\infty(\qcChannel_\theta) = \H{\pi}$ occurs when $\theta = \pi / 4$,
as discussed. The minimum asymptotic state uncertainty for this set of PVMs is
$C_\infty \approx 0.49$ bits, which occurs for $\theta \approx 2.01$. An
observer may choose to use a basis with an exponential set of mixed states to
lower their uncertainty in the source's internal state, at the cost of having to
track the probability of a larger set of mixed states.

\subsection{Maintaining Synchrony with Adaptive Measurement}

Adaptive measurement protocols are also capable of maintaining synchronization
when no fixed-basis measurement can. To appreciate this, consider Fig.
\ref{fig:unifilar_qubit_source}'s unifilar qubit source.

For $0 < p < 1$ there are no synchronizing observations, however almost any
measurement reduces the average state uncertainty below $\H{\pi}$. Additionally,
if an observer comes to know the source's state by other means (perhaps the
source is initialized in state $A$ at time $t=0$), then they are able to
maintain synchronization with a simple adaptive measurement protocol
$\qcChannel_{adaptive}$.

This protocol, defined here only for recurrent states, is as follows. If the
source is in state $A$, apply measurement $\Meas_{01}$, and if the source is in
state $B$, apply measurement $\Meas_{\pm}$. It is only possible to define it
this simply and maintain synchronization since the source is quantum unifilar.

\begin{figure}
\centering
\includegraphics[width=\columnwidth]{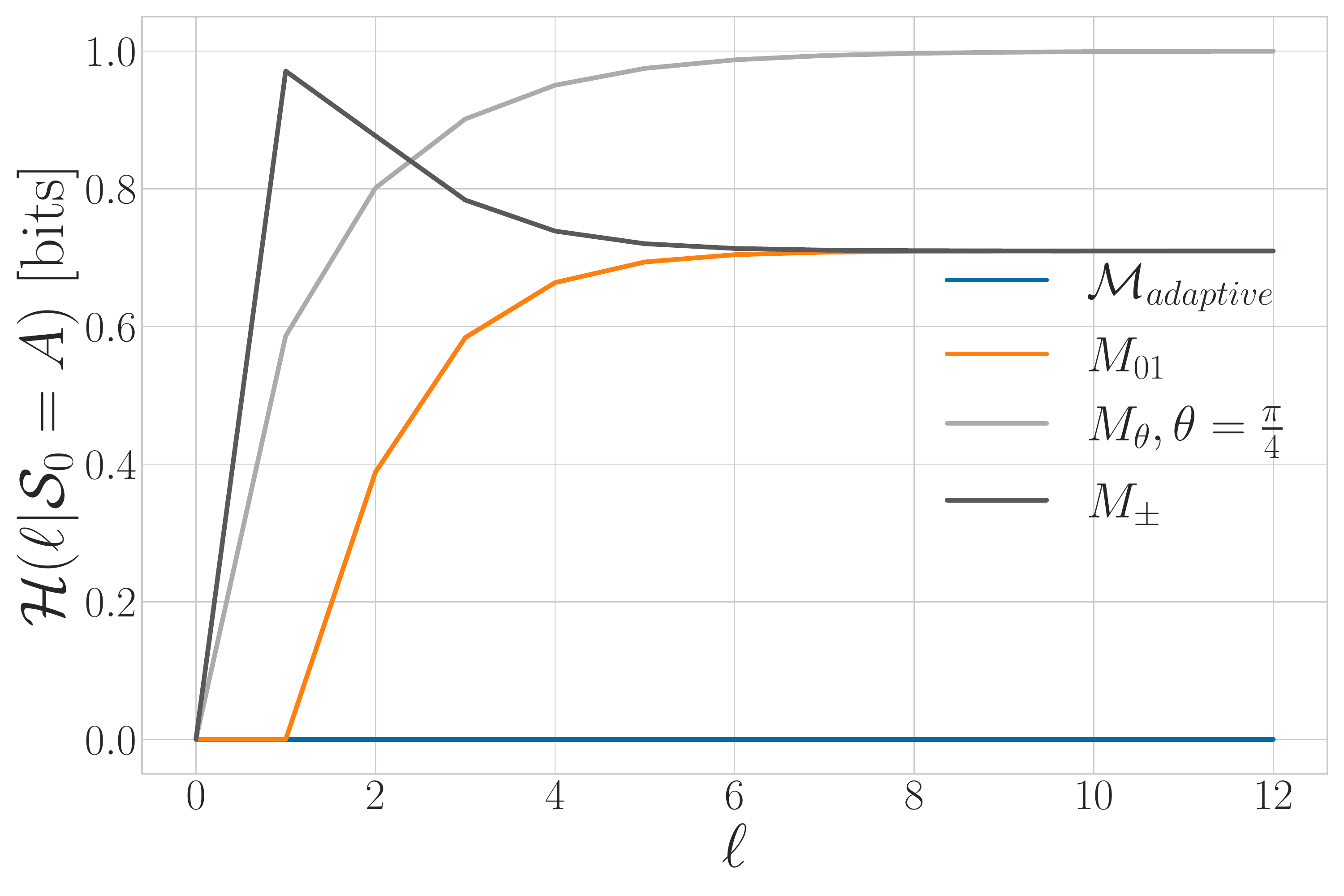}
\caption{Average state uncertainty $\mathcal{H}(\ell)$ for the unifilar qubit
	source ($p = 0.6$) initialized in state $A$. Only the adaptive measurement
	protocol (described in the text) is able to maintain synchronization.
	}
\label{fig:unifilar_qubit_source_state_uncertainty}
\end{figure}

Figure \ref{fig:unifilar_qubit_source_state_uncertainty} shows the behavior of
$\mathcal{H}(\ell)$ as the observer desynchronizes from the source initialized
in state $A$. For each repeated PVM measurement, it eventually reaches an
asymptotic value, though $\mathcal{H}(\ell)$ may be nonmonotonic.

\subsection{Synchronizing to a Qutrit Source}

A qubit source ($d=2$) presents limited opportunities for unambiguous state
discrimination and, hence, for synchronization. Qutrits allow for more general
behavior and have been suggested for applications in quantum communication
networks where one state is reserved specifically for synchronization
\cite{Fuji13}.

Let's apply this logic to the unifilar qutrit source in Fig.
\ref{fig:unifilar_qutrit_source}. Synchronizing observation sequences for this
process are `$2$', `$1+$', and `$1-$' that definitively place the source in
states $A$, $B$, and $C$, respectively. Once synchronized an observer may remain
synchronized by measuring with $\Meas_{012} =
\{\ket{0}\bra{0},\ket{1}\bra{1},\ket{2}\bra{2}\}$ when in state $A$, $\Meas_{\pm
2} = \{\ket{+}\bra{+},\ket{-}\bra{-},\ket{2}\bra{2}\}$ when in state $B$, and
either of the above when in state $C$.

\begin{figure}
\centering
\includegraphics[width=\columnwidth]{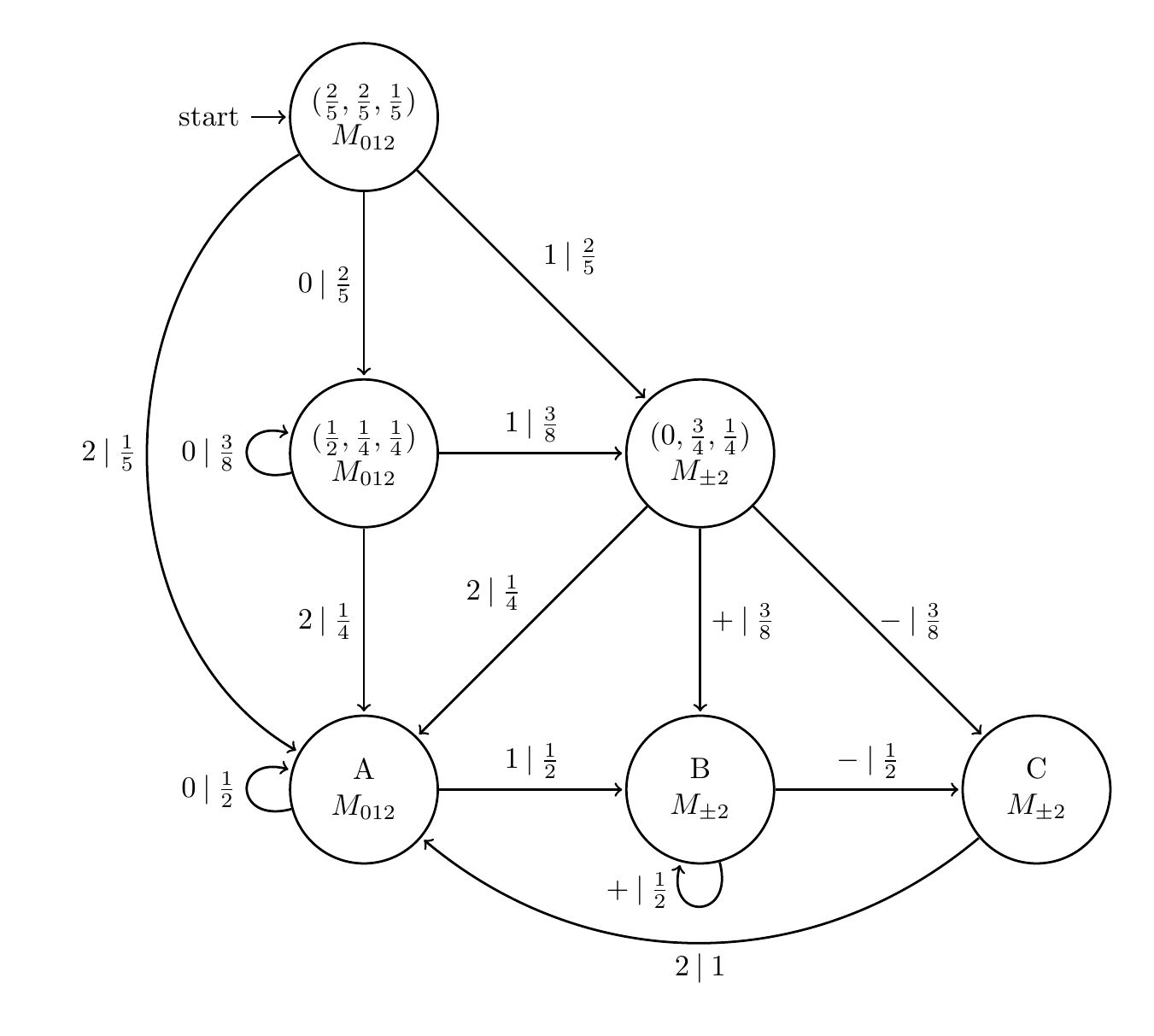}
\caption{Adaptive measurement protocol defined for the qutrit process generator
	in Fig. \ref{fig:unifilar_qutrit_source}. The three transient mixed states
	are labeled with the internal source states probabilities $\left(p_A, p_B,
	p_C\right)$ and the three recurrent states correspond exactly to those
	states. This adaptive protocol permanently synchronizes to the source since
	the source is quantum unifilar. Transitions with zero probability are
	omitted.
	}
\label{fig:qutrit_process_synchronizer}
\end{figure}

\begin{figure}
\centering
\includegraphics[width=\columnwidth]{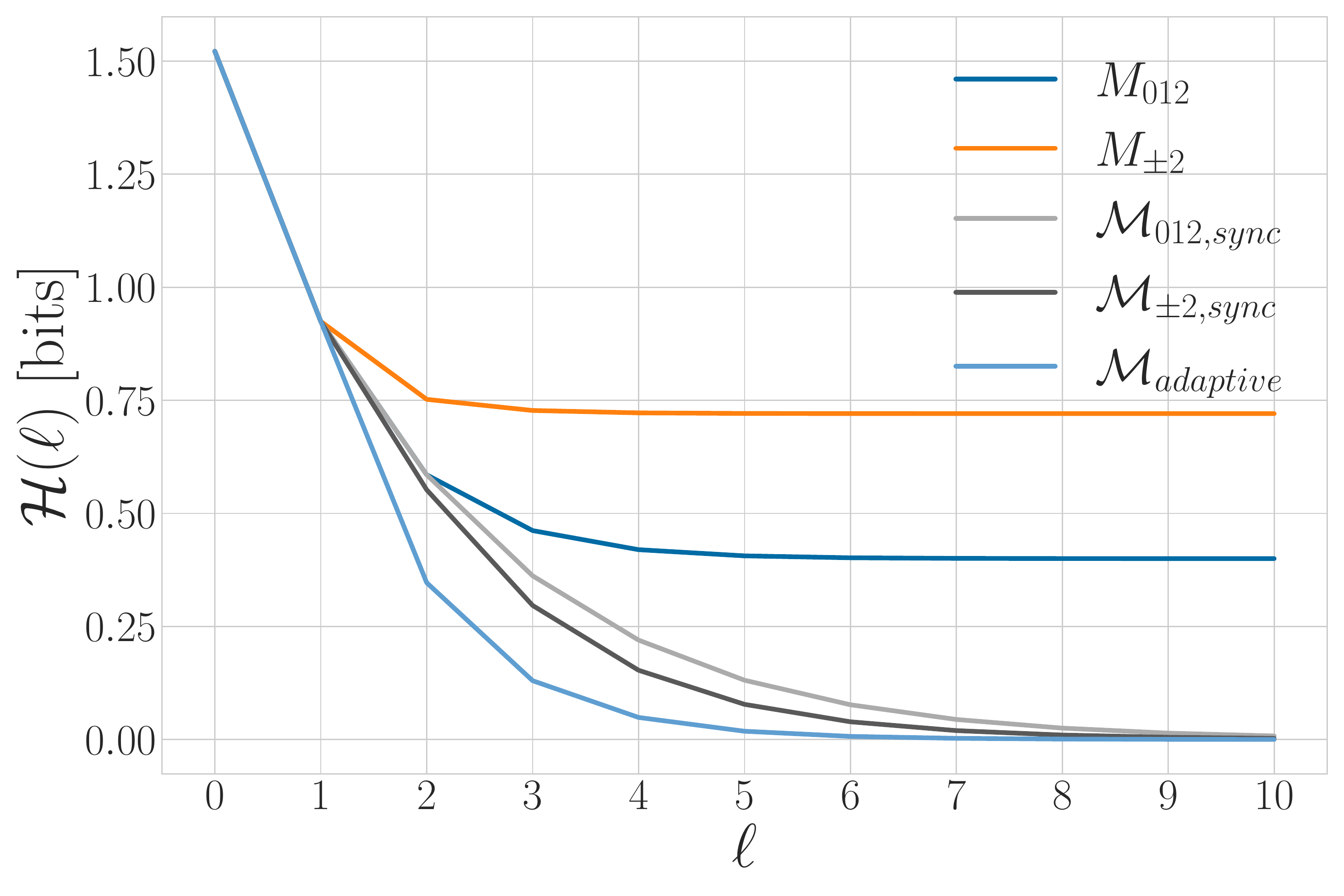}
\caption{Average state uncertainty $\mathcal{H}(\ell)$ while measuring the
	process generated by the unifilar qutrit source. $\Meas_{012}$ and
	$\Meas_{\pm 2}$ are fixed-basis measurements. $\qcChannel_{012,sync}$ and
	$\qcChannel_{\pm 2,sync}$ measure in a fixed basis until they observe a $2$
	and stay permanently synchronized afterwards. $\qcChannel_{adaptive}$
	refers to the protocol in Fig.  \ref{fig:qutrit_process_synchronizer}.
	}
\label{fig:qutrit_process_state_uncertainty}
\end{figure}

We explore synchronizing to this source with five different measurement
protocols and compare them in Fig. \ref{fig:qutrit_process_state_uncertainty}.
The first two are repeated PVMs in the $\Meas_{012}$ basis and the $\Meas_{\pm
2}$ basis. The other three are adaptive measurement protocols that share a
recurrent dynamic, but have different transient states. Consider measuring in
the $\Meas_{012}$ ($\Meas_{\pm 2}$) basis until observing a $2$ and therefore
synchronizing to source state $A$. Then, use $\Meas_{012}$ when the source is in
state $A$ and $\Meas_{\pm 2}$ when the source is in state $B$ or $C$. We refer
to this protocol as $\qcChannel_{012,sync}$ ($\qcChannel_{\pm 2,sync}$). The
fifth protocol $\qcChannel_{adaptive}$ is defined by the DQMP in Fig.
\ref{fig:qutrit_process_synchronizer} that uses an adaptive protocol over three
transient mixed states in addition to the protocol just defined for the
recurrent states.

The average state uncertainties for an observer implementing these five
measurement protocols are shown in Fig.
\ref{fig:qutrit_process_state_uncertainty}. The fixed basis measurements do not
lead to persistent synchronization and have a nonzero asymptotic state
uncertainty ($C_\infty(\Meas_{012}) \approx 0.40$ bits and $C_\infty(\Meas_{\pm
2}) \approx 0.72$ bits. If one measures in a fixed basis until synchronizing (by
observing a 2) and then takes advantage of the quantum unifilarity of the source
to stay synchronized, then the asymptotic state uncertainty vanishes. This is
the case for $\qcChannel_{012,sync}$ and $\qcChannel_{\pm 2,sync}$, which have
synchronization informations of $\mathbf{S}(\qcChannel_{012,sync}) \approx 3.91$
bits and $\mathbf{S}(\qcChannel_{012,sync}) \approx 3.60$ bits. The extra
complexity of the measurement protocol $\qcChannel_{adaptive}$ admits additional
synchronizing words (`$1+$' and `$1-$') and a lower synchronization information
($\mathbf{S}(\qcChannel_{adaptive}) \approx 3.00$ bits) than the simpler
strategy of waiting to see a `$2$'. Note that these three synchronization
informations are all greater than our estimate of the quantum transient
information for this process ($\TI_q \approx 2.16$ bits $\times$ symbols) These
values of $\mathbf{S}(\qcChannel)$ were estimated using $\ell = 10$.

\subsection{Discussion}

This section has detailed the process of synchronizing to a known qudit source.
In contrast to sources of classical random variables, an observer may attempt to
synchronize to a qudit source using a variety of measurement protocols. We can
compare different protocols with two informational quantities introduced above:
the asymptotic state uncertainty $C_\infty(\qcChannel)$ and the synchronization
information $\mathbf{S}(\qcChannel)$.

For sources that are not quantum unifilar, no protocol can remain synchronized
to the source. Nevertheless, for different measurement protocols $\qcChannel_0$
and $\qcChannel_1$ we can compare $C_\infty(\qcChannel_0)$ and
$C_\infty(\qcChannel_1)$. The protocol with a lower value is better at the task
of synchronization in the sense that an observer's uncertainty in the source's
internal state will be lower on average.

This leads to a natural question: What is the best measurement protocol for
synchronizing to a source? To answer it we introduce a new protocol-independent
property of a qudit process, the \emph{minimal asymptotic state uncertainty}:
\begin{align*}
  C_{min} = \min_{\qcChannel} C_\infty(\qcChannel)
  ~,
\end{align*}
where the minimum is taken over all possible measurement protocols defined via
DQMP. In practice, determining $C_{min}$ for a quantum process requires a proof
that no measurement protocol can achieve a lower value. Fully exploring the
space of measurement protocols is beyond the present scope. Nevertheless, we
introduced candidates for $C_{min}$ in the above examples and found
the minimal asymptotic state uncertainty for a restricted class of repeated PVMs
numerically; recall Fig. \ref{fig:0+_qgm_C_inf}.

For sources that are quantum unifilar, we explored several protocols that are
capable of persistent synchronization. For such processes, $C_{min} = 0$ bits.
We can compare different synchronizing measurement protocols $\qcChannel_0$ and
$\qcChannel_1$ through their synchronization informations
$\mathbf{S}(\qcChannel_0)$ and $\mathbf{S}(\qcChannel_1)$. A lower value means
that the observer experiences less state uncertainty while synchronizing. What
is the minimal amount of state uncertainty an observer can experience? We
define the \emph{minimal synchronization information} for a quantum process as:
\begin{align*}
\mathbf{S}_{min} = \min_{\qcChannel} \mathbf{S}(\qcChannel)
~.
\end{align*}
This minimum is also taken over all DQMPs. Establishing that a given protocol is
minimal is nontrivial.

One open question prompted by this work is `Is it always possible to synchronize
to a source that is quantum unifilar?' or equivalently `Does $C_{min} = 0$ for
all processes generated by quantum unifilar sources?'. Answering this question
will also require a greater understanding of the space of DQMPs. Progress may
involve proving the existence or nonexistence of a protocol that is able to
synchronize to the unifilar qubit source in Fig.
\ref{fig:unifilar_qubit_source}.

Finally, we recount several reasons why synchronization is an important task
not only for determining a source's internal state, but also for improving
predictions of future measurement outcomes. Generally, HMCQSs have inherent
stochasticity. Periodic sources are an exception. Through synchronization we
may substantially reduce the uncertainty in measurement outcomes. In the extreme
case where a source's internal state has only one possible transition, this
uncertainty vanishes, and we can measure the next qudit with a PVM which has a
deterministic outcome.

For example, if we know a $\ket{0}$-$\ket{+}$ Quantum Golden Mean generator is
in state $B$ the next qudit is $\ket{0}$, and we will always see `$0$' if we
apply the measurement $\Meas_{01}$. Thus, we consider synchronization as a form
of dynamical inference where an observer uses knowledge of both a source's
internal structure and a sequence of measurement outcomes to inform a more
accurate prediction of future measurement outcomes. Each mixed state corresponds
to a different prediction.

\section{Quantum Process System Identification}
\label{sec:tomography}

Synchronization is a task that performed when an observer has an accurate model
(here, a HMCQS) of the quantum information source. The next natural question
is: How does an observer create an accurate model of an \emph{unknown} quantum
information source? This section begins to answer this question. It starts by
briefly reviewing how one infers a classical information source from data. It
then discusses how to identify the state of a single qudit with quantum state
tomography. By combining these two ideas we arrive at an inference method for
stationary qudit sequences.

\subsection{Classical System Identification}

An observer of an unknown stationary classical process is limited to use only
the available data---distributions of words over symbol alphabet $\AX$---to
infer the source's structure. We denote the distribution of length-$\ell$ words
$\mathcal{P}(\ell) = \{w = \meassymbol_0 \cdots \meassymbol_{\ell-1} \vert w \in
\AX^\ell \}$.

For $\ell = 1$ we obtain a distribution over symbols in $\AX$. With
$\mathcal{P}(1)$ we can reconstruct a memoryless (i.i.d.) model of the source
with one internal state and each $\Pr(\meassymbol \in \AX)$ obtained directly
from $\mathcal{P}(1)$.

For $\ell = 2$ we use $\mathcal{P}(2)$ to create conditional probability
distributions for the next symbol conditioned on the previous one; i.e.,
$\Pr(\meassymbol_1 | \meassymbol_0)$, for all $\meassymbol_0$, $\meassymbol_1
\in \AX$. From these conditional distributions we may construct a Markov
approximation of the source with $\vert \AX \vert$ states, one corresponding to
each symbol. The conditional distributions set the transition matrices for
Markov approximation of the source: i.e., $\Pr(\meassymbol_1 | \meassymbol_0) =
T_{\meassymbol_0,\meassymbol_1}$.

Similarly, for $\ell > 2$ we obtain a length-$\ell$ HMM approximation of the
source by conditioning on length $\ell - 1$ words, each of which corresponds to
a different internal state. In this simplified picture, the number of internal
states of the model grows exponentially with $\ell$, as the number of possible
words of length $\ell$ is $\vert \AX \vert ^{\ell}$. This leads to numerical
problems when inferring processes with correlations over long periods of time.
That said, there are methods for both combining states that reflect identical
predictions of future symbols and for performing inference over machine
topologies with a certain number of states---known as Bayesian Structural
Inference \cite{Stre14}---to determine the most likely $\eM$ for the source.

\subsection{Tomography of a Qudit}

We cannot directly apply the procedure for inferring classical source dynamics
from word distributions to qudit processes since observations depend on the
measurement basis/protocol one uses. To fully characterize a stationary quantum
source we must instead take many measurements in different bases to reconstruct
the qudit density matrices. This task is known as \emph{quantum state
tomography}. We begin by inferring an individual qudit state $\rho_0$ before
introducing a general method for quantum system identification using separable
sequences of qudits.

Tomographic reconstruction of a single unknown qudit density matrix $\rho_0$
through measurement is challenging for two main reasons:
\begin{enumerate}
      \setlength{\topsep}{0pt}
      \setlength{\itemsep}{0pt}
      \setlength{\parsep}{0pt}
\item $\rho_0$'s complete description requires a number of parameters that
	scales exponentially with the state's Hilbert space dimension.
\item Quantum measurement is probabilistic, so one must prepare and measure
	many copies of $\rho_0$ to estimate a single parameter.
\end{enumerate}

Specific combinations of measurements are particularly useful for this task. For
example $\rho_0$ can be inferred by measuring with a set of mutually-unbiased
bases (MUB) \cite{Woot89} or a single informationally-complete POVM (IC-POVM)
\cite{Rene04}. For a qubit, one possible MUB consists of the x-, y-, and
z-bases. By measuring many copies of the qubit in each of these three bases one
obtains three probabilities---$\Pr(+x)$, $\Pr(+y)$, and $\Pr(+z)$---that
uniquely determine the density matrix $\rho_0$. We do not discuss the necessary
number of measurements to determine these parameters to within a desired
tolerance for general qudit tomography, a question which is well-studied
\cite{QCQI, Thew02}.

An example of an IC-POVM for a qubit consists of the projectors onto states:
\begin{align}
\ket{\phi_1} & = \ket{0} , \nonumber \\
\ket{\phi_2} & = \frac{1}{\sqrt{3}} \ket{0} + \sqrt{\frac{2}{3}} \ket{1} , \nonumber \\
\ket{\phi_3} & = \frac{1}{\sqrt{3}} \ket{0} + \sqrt{\frac{2}{3}}
e^{i 2\pi / 3}\ket{1} ,\text{~ and} \nonumber \\
\ket{\phi_4} & = \frac{1}{\sqrt{3}}\ket{0} + \sqrt{\frac{2}{3}} e^{i 4\pi
/ 3}\ket{1} 
~.
\label{eq:SIC-POVM}
\end{align}
This is also a symmetric IC-POVM, or SIC-POVM, because any combination of two
projectors has the same inner product.

By measuring many identical copies of $\rho_0$ with the same IC-POVM one obtains
a probability distribution over the $d^2$ possible measurement outcomes. This
provides $d^2-1$ parameters (due to normalization) that uniquely determine the
density matrix $\rho_0$.

The existence and properties of SIC-POVMs in higher dimensional Hilbert spaces
is an active area of research \cite{Beng17}.

\subsection{Tomography of a Qudit Process}

Now that we know how to estimate the density matrix for an individual qudit we
can begin analyzing length-$\ell$ density matrices $\rhoL$. We will measure each
qudit in turn rather than performing a joint measurement over the entire
length-$\ell$ block of qudits. This is a luxury afforded to us because we are
focusing on separable qudit sequences---the generic case of entangled qudit
processes requires measuring in nonlocal bases.

For $\ell = 1$ we reconstruct $\rho_0$ as described above and obtain a
memoryless (i.i.d.) estimate of the source that emits qudit $\rho_0$ at every
timestep, following Eq. (\ref{eq:iid_quantum_process}). Unless $\rho_0$ is a
pure state, there are many single-state HMCQS that generate this process since
many different pure-state ensembles correspond to the same density matrix. A
unique memoryless model of the source may be obtained by diagonalizing $\rho_0$
and having the source emit each pure eigenstate $\ketpsi[i]$ with probability
equal to the corresponding eigenvalue $\lambda_i$.

For $\ell = 2$ we must reconstruct the two-qudit density matrix $\rho_{0:2}$.
(Recall that our indexing is left-inclusive and right-exclusive, therefore
$\rho_{0:2}$ is the joint state of the qudits for $t=0,1$). Due to
stationarity, $\rho_{0:2}$ (the joint state of the two qudits) must be
consistent with the one-qudit marginals, i.e. $tr_{0}(\rho_{0:2}) = \rho_1 =
\rho_0 = tr_{1}(\rho_{0:2})$.
 
We will describe the iterative procedure for reconstructing $\rhoL$ for
\emph{qubits} in detail. When $d=2$, $\rho_{0:2}$ has $15$ real parameters that
must be determined via tomography. Tomography on the one-qubit marginals
determines 3 parameters, and the condition of stationarity fixes 3 more. The
state can be reconstructed fully by considering combinations of the set of
mutually-unbiased measurements. For two qubits, this means the 16 combinations
of Pauli matrices ($\sigma_{I_0} \otimes \sigma_{x_1}$, $\sigma_{x_0} \otimes
\sigma_{x_1}$, $\sigma_{x_0} \otimes \sigma_{y_1}$, and so on.) \cite{Lawr02}.
To fully characterize $\rho_{0:2}$ $9$ of these values must be
determined---those not involving the identity operators $\sigma_{I_0}$ and
$\sigma_{I_1}$ which are fixed by the one-qubit marginals. For $d > 2$, this
procedure can be modified by using a set of mutually-unbiased bases in that
higher-dimensional Hilbert space.

After determining $\rho_{0:2}$ one can continue on to determine $\rho_{0:3}$ (63
real parameters for qubits). Many of these parameters are fixed by the previous
tomography on the one-qubit marginals (3 parameters), the two-qubit marginals
(15 parameters), and their stationarity conditions (6 and 15 parameters
respectively). Combinations of three one-site Pauli matrices are sufficient for
full reconstruction. One may continue this procedure for larger $\ell$,
typically until the number of measurements becomes experimentally infeasible.

\subsection{Cost of I.I.D.}

Quantum information sources are often assumed to be i.i.d. \cite{Wild17, QCQI}.
If an observer performing quantum state tomography assumes that an unknown
quantum information source is i.i.d., then they will not go beyond determining
the one-qudit marginal $\rho_0$. If the qudits are instead correlated, this will
lead to an overestimate of the source's randomness. They will erroneously
conclude that each qudit is in state $\rho_0$ and that the entropy rate is
$S(\rho_0)$ bits~per~timestep. The latter overestimates the true entropy rate by
a factor of $S(1) - \qer$. An observer can obtain better estimates for the
entropy rate and other informational quantities by following the above procedure
and tomographically reconstructing blocks of qudits of length-$\ell$.

To demonstrate the degree to which this assumption may mislead an observer,
consider the process generated by nonunifilar qubit source in Fig.
\ref{fig:nonunifilar_qubit_source}. Assume the source begins in its stationary
state distribution: $(p_A = 1 / 2, p_B = 1 / 2)$. For any $p$ such that $0 \leq
p < 1$, $\rho_0$ is the maximally-mixed state and an observer assuming an
i.i.d. process estimates that $\qer = 1$ bit~per~timestep. This is only an
accurate description of the source for $p = 1 / 2$. Whereas, for many values of
$p$, this source has significant correlations between subsequent qubits.

Let's take a closer look at the extreme values of $p$. For $p=0$ the process is
period-$2$ with $\qer = 0$ bits~per~timestep and $\EE_q = 1$ bit. This pattern
can be easily detected by measuring in the $\{\ket{0},\ket{1}\}$ basis, where a
measurement of $0$ ($1$) immediately synchronizes an observer to state $B$
($A$). As $p \to 1$, the two source states become increasingly disconnected,
$\qer \to 0$, and $\EE_q \to 1$. An observer measuring in the
$\{\ket{+},\ket{-}\}$ basis observes a $+$ ($-$) is likely to measure another
$+$ ($-$). This source's rich and varied behavior at different values of $p$
will go entirely unappreciated when considering only the one-qubit density
matrix, $\rho_0$.

\subsection{Finite Length Estimation of Information Properties}

We just saw an example of how, if an observer assumes a source is i.i.d., they
will generally underestimate the structure and correlation of the quantum
process and will overestimate its entropy rate. The same is true if they only
perform tomography on blocks of qudits up to finite length $\ell$. Consider now
an experimenter who tomographically reconstructs the density matrix $\rhoL$ and
then assumes there are no additional (longer-range) correlations within the
qudit process.

Their estimates for the quantum entropy rate, quantum excess entropy and quantum
transient information are given by Eqs. \eqref{eq:qer_est}, \eqref{eq:Eq_est}, and
\eqref{eq:Tq_est} respectively.

\begin{figure}
\centering
\includegraphics[width=\columnwidth]{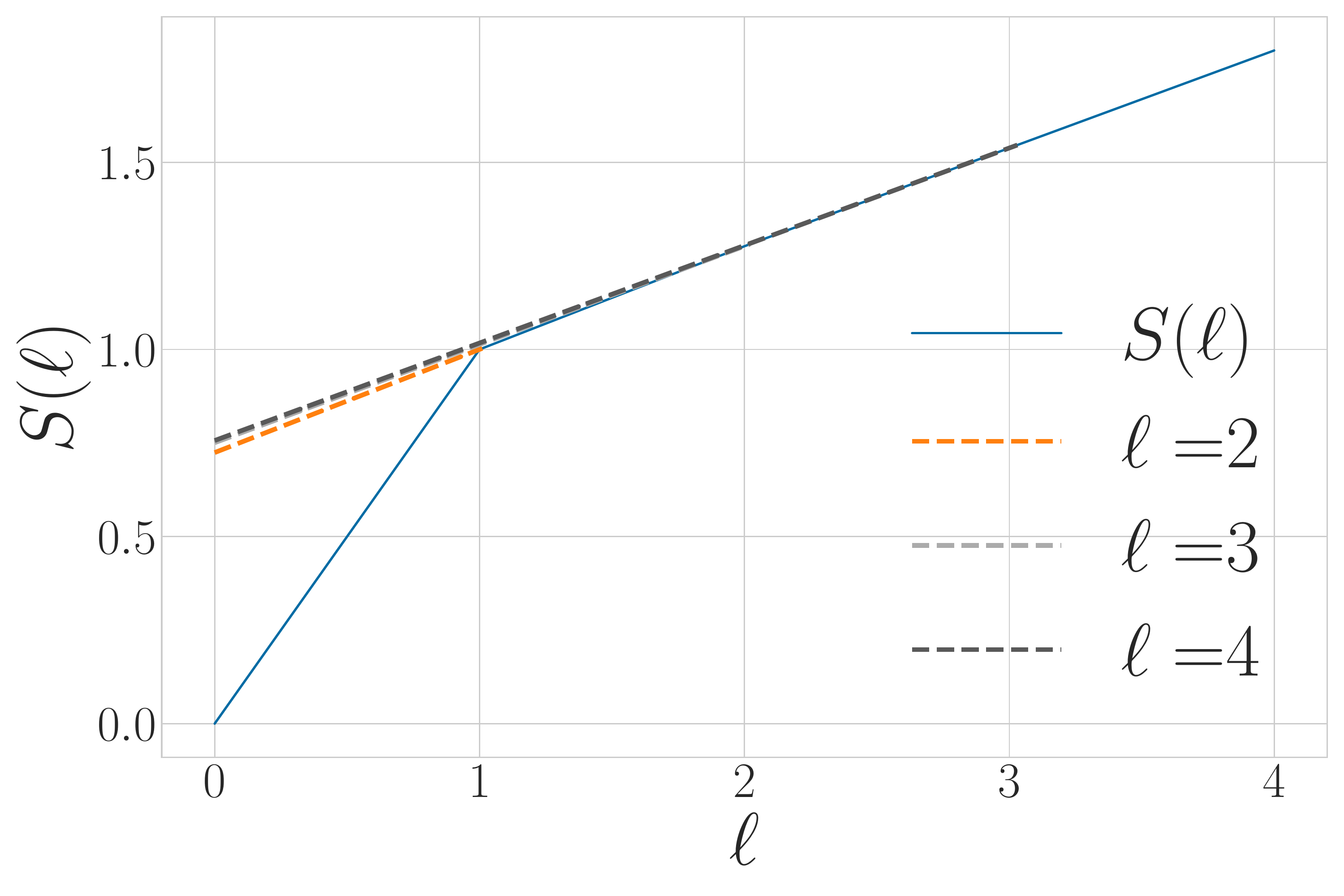}
\caption{Block entropy and length-$\ell$ estimates for information properties
	of the process generated by the nonunifilar qubit source in Fig.
	\ref{fig:nonunifilar_qubit_source} with $p = 0.05$. The slope and
	$y$-intercept of the linear estimates are $\qer$ and $\EE_q$, respectively.
	Note that the estimates do not improve significantly for $\ell > 2$,
	indicating that the two-qubit correlations are most significant for
	determining information properties of the process.
	}
\label{fig:nonunifilar_source_block_entropies}
\end{figure}

\begin{figure}
\centering
\includegraphics[width=\columnwidth]{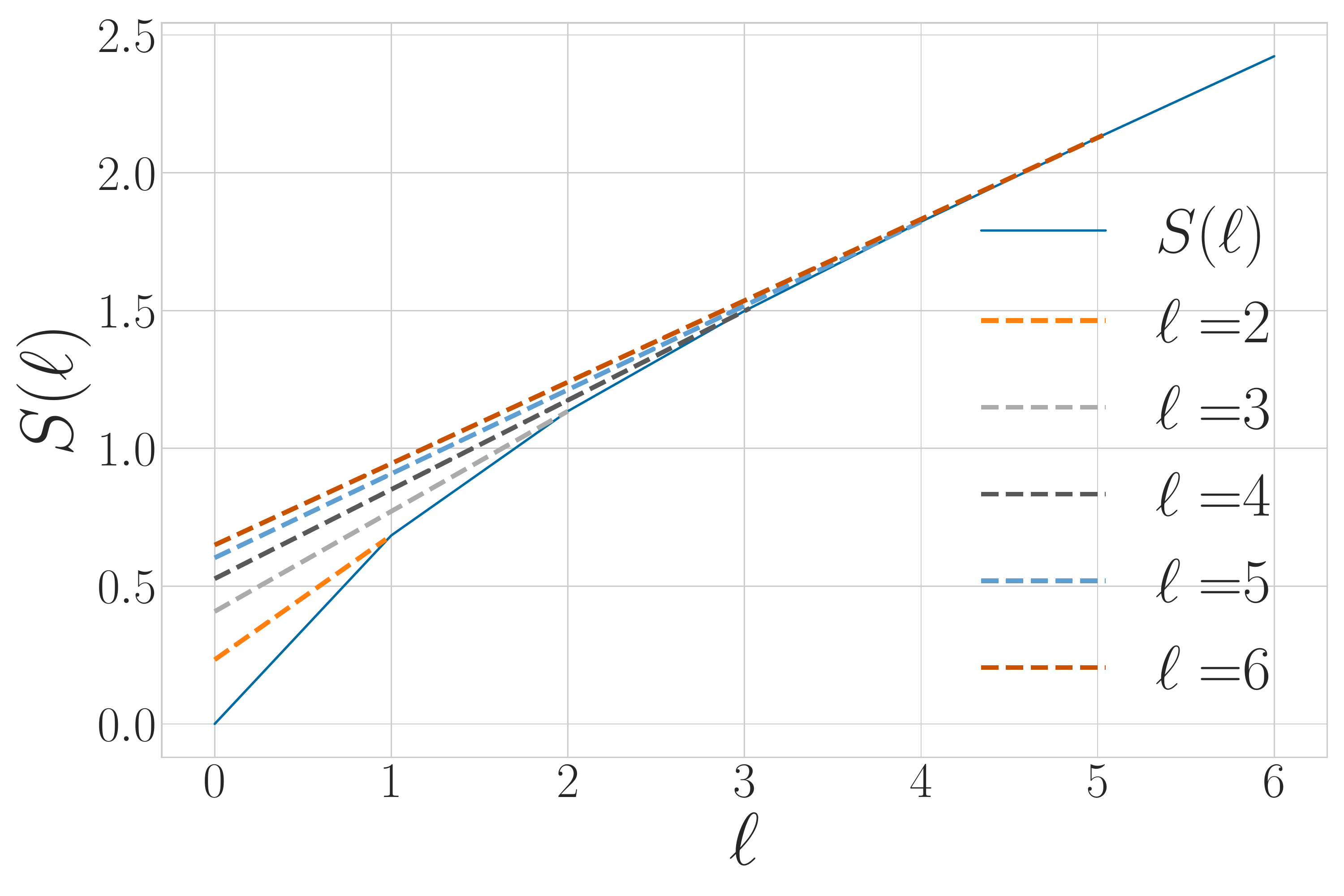}
\caption{Block entropy and length-$\ell$ estimates for information properties of
	the process generated by the unifilar qubit source in Fig.
	\ref{fig:unifilar_qubit_source} with $p = 0.05$. The slope and
	$y$-intercept of the linear estimates are $\qer$ and $\EE_q$, respectively.
	Note that the estimates improve steadily for larger $\ell$, indicating
	long-range correlations.
	}
\label{fig:unifilar_source_block_entropies}
\end{figure}

The difference between the length-$\ell$ estimate of an information property and
its true value depends on the process' internal structure and quantum alphabet.
The estimates for the nonunifilar and unifilar qubit sources in Figs.
\ref{fig:nonunifilar_source_block_entropies} and
\ref{fig:unifilar_source_block_entropies} represent two extremes in this
respect. As previously discussed, the single-qubit density matrix for the
nonunifilar qubit source is the maximally-mixed state. If an observer instead
reconstructs $\rho_{0:2}$, they significantly improve their estimate of
$\qer$, $\EE_q$, and $\TI_q$. However, for $\ell > 2$ these estimates do not
improve dramatically. There is always a trade-off between the number of
experiments necessary to reconstruct the process tomographically and the
accuracy of the estimates obtained from that reconstruction. For this source,
$\ell = 2$ strikes a balance between those two resources.

In contrast, for the unifilar qubit source the estimates of $\qer$, $\EE_q$, and
$\TI_q$ improve steadily as $\ell$ increases. Longer-range correlations are
captured by increasing the length of the reconstructed density matrices $\rhoL$.
When observing this source it is likely worth finding $\rhoL$ for the largest
$\ell$ that is experimentally feasible.

When faced with an unknown quantum source, how does one pick an appropriate
value of $\ell$? One strategy is to increase $\ell$ until the correction made to
the relevant information quantities by going to $\ell + 1$ is below some
threshold. Stationarity (and the resulting concavity of the quantum block
entropy) ensure that future corrections will also be below that threshold.

\subsection{Tomography with a Known Quantum Alphabet}

If an observer has additional knowledge of what possible pure states an HMCQS
may emit (i.e., the quantum alphabet $\AQ$), they can leverage this knowledge to
simplify the task of system identification by inferring the word probabilities
of the underlying classical process $\BiInfinity$ rather than performing full
tomographic reconstruction. For qubits we can represent this simplification
geometrically via the Bloch sphere; see Fig. \ref{fig:bloch_AQ}. Each point on
the surface of the Bloch sphere represents a pure qubit state in
$\HSpace^2$---such as, $\ket{0}$ and $\ket{1}$ on the poles of the z-axis---and
each interior point represents a possible qubit density matrix. The following
assumes each element of $\AQ$ is unique.

\subsubsection{$\ell = 1$}

The length-$1$ density matrix $\rho_0$ must satisfy the
equation
\begin{align*}
\rho_0 = \sum_{\ketpsi[\meassymbol] \in \AQ} \Pr(\ketpsi[\meassymbol]) \ketpsi[\meassymbol] \brapsi[\meassymbol]
~.
\end{align*}

After finding $\rho_0$ via tomography one may rearrange this equation to infer
the length-$1$ word distributions of $\BiInfinity$ given that $\Pr(\MeasSymbol =
\meassymbol) = \Pr(\ketpsi[\meassymbol])$. The feasibility of this task depends
on the relationship between $\vert \AQ \vert$ and $d$. For the qubit case
($d=2$) one can uniquely infer $\Pr(\MeasSymbol_0)$ if $\vert \AQ \vert \leq 3$;
assuming no degenerate states in $\AQ$.

When $\AQ$ is known it may not be necessary to fully reconstruct $\rho_0$. We
first present several simple examples before giving a general algorithm for finding
$\rhoL$ without performing full state tomography on $\rhoL$.

\begin{figure}
     \centering
     \subfloat[$\AQ = \{\ket{0},\ket{+} \}$ \label{fig:AQ2}]{\includegraphics[width=0.2\textwidth]{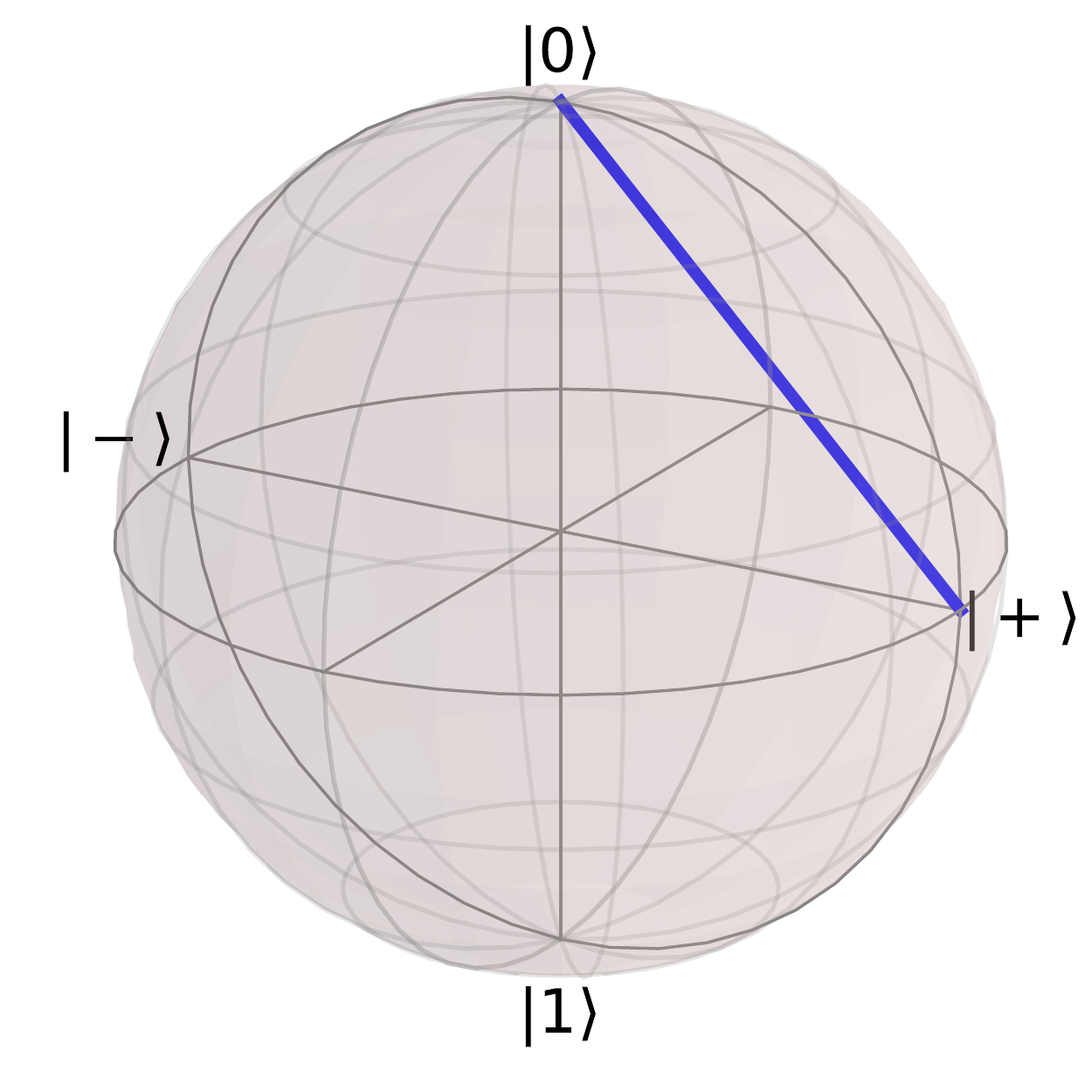}
     }
     \hfill
     \subfloat[$\AQ = \{\ket{0},\ket{1},\ket{+} \}$ \label{fig:AQ3}]{\includegraphics[width=0.2\textwidth]{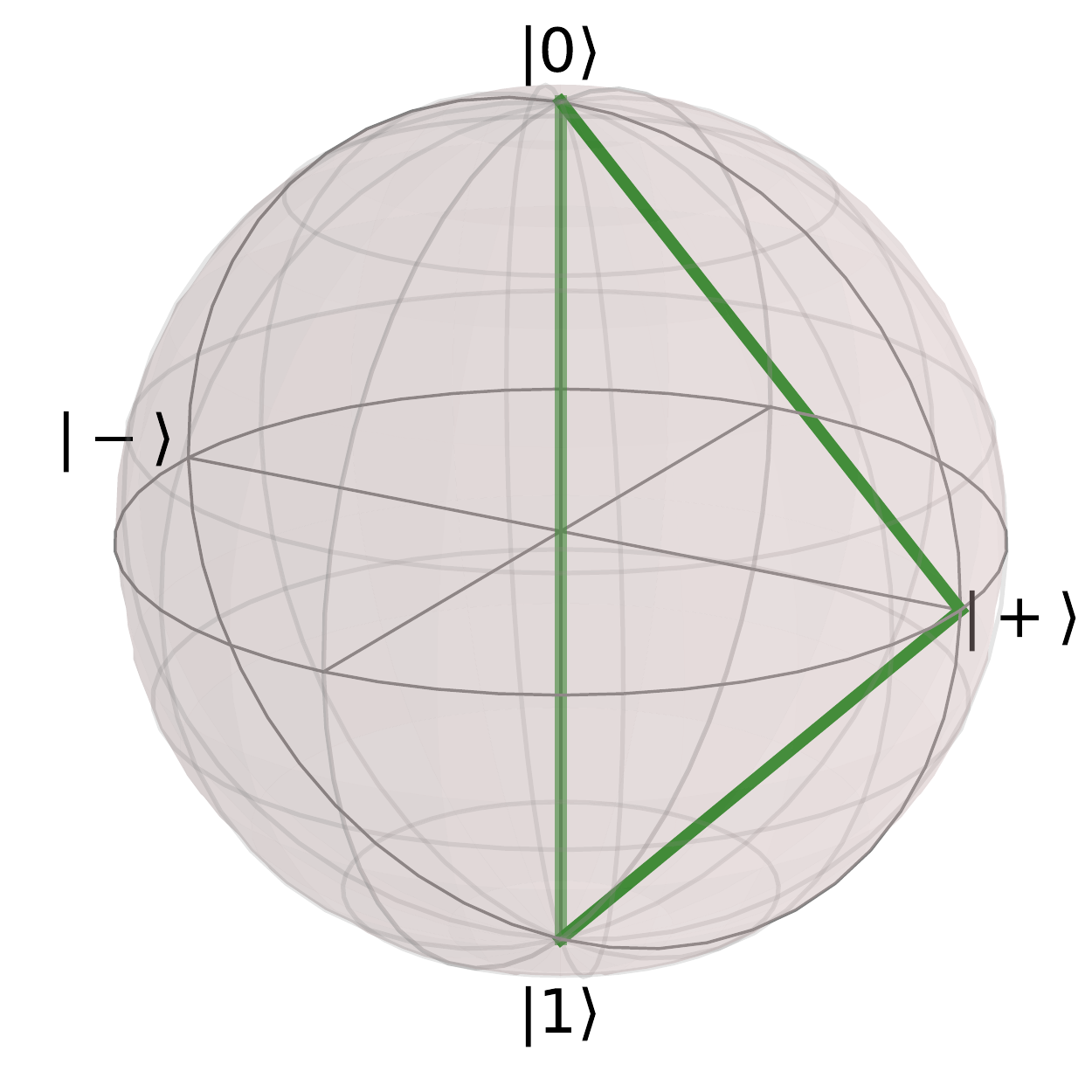}
     }
     \hfill
     \subfloat[$\AQ = \{\ket{0},\ket{1},\ket{+},\ket{y_+} \}$ \label{fig:AQ4}]{\includegraphics[width=0.2\textwidth]{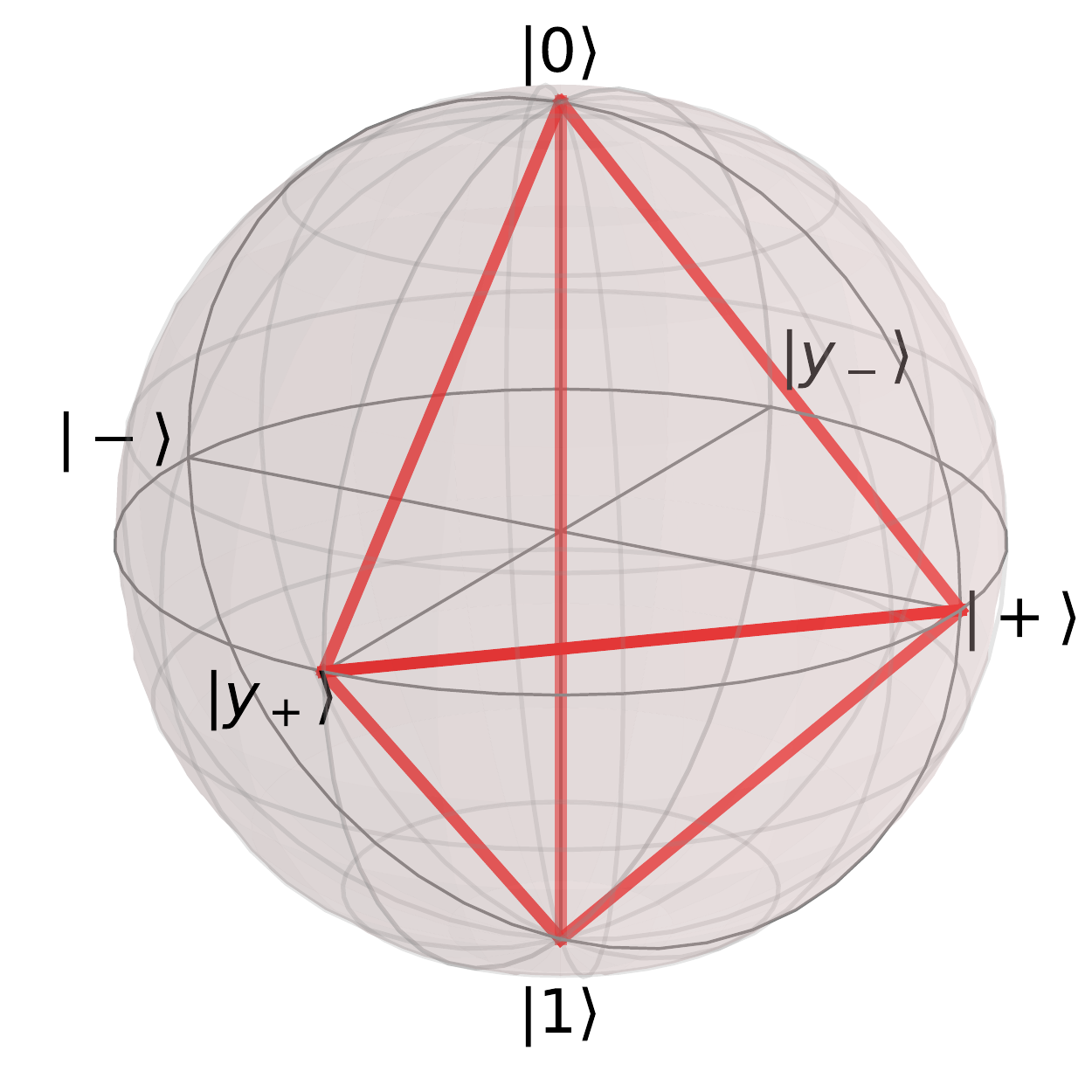}
     }
\caption{The Bloch sphere representation of the boundary of the set of possible
length-$1$ density matrices ($\rho_0$) for the given $\AQ$. (a) The set of valid
states is a line segment. An observer only needs to determine one parameter. (b)
The set of valid states is the interior of a triangle. An observer must
determine two parameters. (c) The set of valid states is the interior of a
tetrahedron. An observer must determine three parameters, and the decomposition
into an ensemble of basis states is not unique. Here $\ket{y_+} =
\frac{1}{\sqrt{2}} \left( \ket{0} - i \ket{1} \right)$.
} 
\label{fig:bloch_AQ}
\end{figure}

For $d = 2$ and $\vert \AQ \vert = 2$, the possible values of $\rho_0$ are
restricted to a chord within the Bloch sphere defined by $\rho_0 = p
\ketpsi[0]\brapsi[0]+ (1-p) \ketpsi[1]\brapsi[1]$ with $0 < p < 1$. One needs
only to determine the parameter $p$ rather than reconstruct $\rho_0$ in its
entirety. An observer can also pick a uniquely informative measurement to
determine $p$. The optimal PVM for doing so is one whose antipodal projectors
can be connected with the diameter of the Bloch sphere that runs parallel to
the line of possible values of $\rho_0$. For example, if $\AQ = \{ \ket{0},
\ket{+} \}$ then the set of possible density matrices lies on the line segment
in Fig. \ref{fig:AQ2}. The best PVM is then $\Meas_{\theta}$ with $\theta =
\frac{3\pi}{4}$ and measurement outcomes $y_0$ (corresponding to a projection on
pure state $\ket{\psi(\theta = \frac{3\pi}{4})}$) and $y_1$ (corresponding to
the orthogonal projector). In this case it can easily be shown that $p =
\frac{\sqrt{2}+1}{2} - \sqrt{2}\Pr(y_0 \vert \rho_0)$ and that
$\cos^2(\frac{3\pi}{8}) \leq \Pr(y_0 \vert \rho_0) \leq
\sin^2(\frac{3\pi}{8})$. $\Meas_{01}$ and $\Meas_{\pm}$ would also be able to
determine $p$, but require more samples to determine $p$ to within some
desired tolerance.

For $d = 2$ and $\vert \AQ \vert = 3$, the possible values of
$\rho_0$ are confined to a simplex in the Bloch sphere defined by $\rho_0 = p_0
\ketpsi[0]\brapsi[0] + p_1 \ketpsi[1]\brapsi[1] + (1-p_0-p_1)
\ketpsi[2]\brapsi[2]$ with $0 < p_0, p_1 < 1$ and $p_0 + p_1 < 1$. One may
determine the parameters $p_0$ and $p_1$ rather than the 3 parameters usually
required to characterize a qubit mixed state. There are two simple choices of
measurements to do so: using a IC-POVM or using two different PVMs.

For the first case, consider measuring $\rho_0$ with a SIC-POVM with elements
$E_y = \frac{1}{2} \ket{\phi_y} \bra{\phi_y}$, with each $\ket{\phi_y}$
described by Eq. \ref{eq:SIC-POVM}. The probability of observing measurement
$y$ can be written as:
\begin{align*}
\Pr(y \vert \rho_0) & = \sum_{x} \Pr(\ketpsi[x]) \Pr(y \vert \ketpsi[x]) \\
&= \sum_{x} \frac{p_{x}}{2} \lvert \braket{\psi_x|\phi_{y}} \rvert ^2
~.
\end{align*}
One can rearrange the system of $4$ equations (one for each POVM element) to
obtain a unique set of $p_x$'s.

Alternatively, one uses two PVMs whose projectors can be connected by
(ideally orthogonal) diameters of the Bloch sphere that are parallel to the
simplex of possible $\rho_0$ values. This will yield 2 parameters that uniquely
determine a point on the $\rho_0$ simplex. An example with $\AQ =
\{\ket{0},\ket{1},\ket{+}\}$ is shown in Fig. \ref{fig:AQ3}, for which the
possible values of $\rho_0$ are confined to the interior of a triangle in the
Bloch sphere. One can determine $p_0$ and $p_1$ by measuring with orthogonal
PVMs $\Meas_{01}$ and $\Meas_{\pm}$ (among many other combinations), in which
case $(1-p_0-p_1) = 2 \Pr($`$+$'$ \vert \rho_0, \Meas_{\pm}) - 1$ and $(p_0 - p_1) =
2 \Pr($`$0$'$ \vert \rho_0, \Meas_{01}) - 1$.

For $d=2$ and $\vert \AQ \vert = 4$ the possible values of $\rho_0$ are confined
to a tetrahedron in the Bloch sphere whose vertices are the elements of $\AQ$,
and one cannot uniquely infer the classical symbol distribution from a
fully-reconstructed $\rho_0$. For example, if $\AQ = \{\ket{0}, \ket{1},
\ket{+}, \ket{-}\}$ and $\rho_0$ is the maximally-mixed state, this could
correspond to any mixture of the form $\rho_0 = p \ket{0}\bra{0} + p
\ket{1}\bra{1} + (1-p) \ket{+}\bra{+} + (1-p)\ket{-}\bra{-}$ with $0 < p < 1$.
Thus, for $d=2$ and $\vert \AQ \vert \geq 3$, the tomographic advantage to
knowing $\AQ$ is reduced but not eliminated as an observer can immediately
exclude any value of $\rho_0$ that lies outside the convex polyhedron defined
by the elements of $\AQ$. This is shown in Fig. \ref{fig:AQ4}, where the region
of possible $\rho_0$ values is confined to less than $\frac{1}{4}$ of the volume
of the Bloch sphere.

Similar simplifications apply for $d=3$ (qutrits) when $\AQ$ is known. Full
tomography of an arbitrary mixed qutrit state requires the determination of 8
parameters, whereas determining the classical distribution given $\AQ$ requires
$\vert \AQ \vert - 1$ parameters. This presents an advantage in general when
$\vert \AQ \vert \leq 8$. We do not explicitly construct measurements that can
realize this advantage, as a geometric understanding of mixed states over the
8-dimensional space $\HSpace^3$ is significantly more involved than the
3-dimensional Bloch sphere describing mixed states over $\HSpace^2$.

For a generic qudit the number of parameters required for full tomography is
$d^2 - 1$. And so, we expect that knowledge of $\AQ$ gives a clear tomographic
advantage (fewer parameters must be determined) when $\vert \AQ \vert < d^2$.

We are now prepared to give a general protocol for $\vert \AQ \vert = n$ and
arbitrary $d$. We wish to recover the underlying distribution of alphabet states
($p_x = \Pr(\ketpsi[x])$, $0 \leq x < n$) from measurement statistics alone.
First, we construct a POVM with $n+1$ elements: $E_y = c_y \ketpsi[x]
\brapsi[x]$ for each $\ketpsi[x]$ in $\AQ$, and $E_{n} = \mathbb{I} -
\sum_{y=0}^{n-1} E_y$. Each $c_y$ is a parameter which can be varied to ensure
that $E_{n}$ is positive semi-definite. By applying this POVM to $\rho_0$ we
obtain a distribution ($\Pr(y \vert \rho_0)$) over the $n+1$ possible
measurements. The first $n$ are related to our desired distribution by:
\begin{align*}
\Pr(y \vert \rho_0) & = \sum_x p_x \Pr(y \vert \ketpsi[x]) \\
&= \sum_x p_x c_y \lvert \braket{\psi_x|\psi_y} \rvert ^2
~,
\end{align*}
with one equation for each $y < n$. If a set of $p_x$'s is a solution to this
system of linear equations, it is consistent with the observed measurements. The
solution will be unique for $\vert \AQ \vert < d^2$.

\subsubsection{$\ell = 2$}

Knowing $\AQ$ provides further advantage when considering tomography of multiple
qudits. The distribution over classical words of length $\ell$ has $\vert \AQ
\vert ^ \ell - 1$ parameters, whereas full tomography of $\ell$ qudits requires
the determination of $d^{2\ell} - 1$ parameters.

For $\ell = 2$:
\begin{align}
\begin{split}
\rho_{0:2} & = \sum_{\ketpsi[\meassymbol_0], \ketpsi[\meassymbol_1] \in \AQ} \Bigg[ \Pr(\ketpsi[\meassymbol_0] \otimes \ketpsi[\meassymbol_1])  \\
& \qquad\qquad \bigl( \ketpsi[\meassymbol_0]  \otimes \ketpsi[\meassymbol_1] \bigr) \bigl( \brapsi[\meassymbol_0] \otimes \brapsi[\meassymbol_1] \bigr) \Bigg]
~.
\end{split}
\label{eq:rho2_separable}
\end{align}

Once again we consider the case of $d = 2$ and $\vert \AQ \vert = 2$ explicitly.
There are 4 length-2 classical word probabilities, but there are 3 constraints
imposed by (i) normalization, (ii) stationarity, and (iii) consistency with the
one-qubit marginal. Thus, one only needs to determine a single parameter to
reconstruct $\rho_{0:2}$.

Consider the task of reconstructing the length-2 density matrix produced by
the $\ket{0}$-$\ket{+}$ Quantum Golden Mean generator in Fig. \ref{fig:0+_qgm}
with the knowledge that $\AQ = \{ \ket{0}, \ket{+} \}$. One would first analyze
the one-qubit density matrix to find that $\Pr(\ket{0}) = \frac{2}{3}$ and
$\rho_0 = \frac{2}{3} \ket{0}\bra{0}+ \frac{1}{3} \ket{+}\bra{+}$. 

The word probability $\Pr(\ket{00})$ is the only necessary additional
information to find $\rho_{0:2}$. The following simple measurement protocol can
determine $\Pr(\ket{00})$: Measure two consecutive qubits with $\Meas_{\pm}$.
If the qubit is in state $\ket{+}$, one will never see outcome `$-$'. If the
qubit is in state $\ket{0}$, one will see outcome `$-$' with probability
$\frac{1}{2}$. And so, $\Pr(\ket{00}) = 4 \Pr($`$--$'$ \vert \rho_{0:2},
\Meas_{\pm} \otimes \Meas_{\pm})$.

This procedure can be generalized to arbitrary 2-element alphabets $\AQ = \{
\ketpsi[0], \ketpsi[1] \}$. First, measure two consecutive qudits with a PVM
$\Meas_{\tilde{0}1}$, where one element is a projector onto $\ketpsi[1]$ with
outcome `$1$' and the orthogonal projector corresponds to measurement outcome
`$\tilde{0}$'. Second:
\begin{align*}
\Pr(&\ketpsi[0] \otimes \ketpsi[0]) \\
&= (\Pr(\text{`$\tilde{0}$'} | \ketpsi[0],
\Meas_{\tilde{0}1}))^{-2} \Pr(\text{`$\tilde{0}\tilde{0}$'}| \rho_{0:2}, \Meas_{\tilde{0}1} \otimes \Meas_{\tilde{0}1}) \\
&= (1 - \lvert \braket{\psi_0|\psi_1} \rvert )^{-2} \Pr(\text{`$\tilde{0}\tilde{0}$'}| \rho_{0:2}, \Meas_{\tilde{0}1} \otimes \Meas_{\tilde{0}1})
~.
\end{align*}
This provides a clear advantage over the usual $9$ parameters necessary to
reconstruct $\rho_{0:2}$ as it takes into account that the one-qubit marginals
and stationarity each impose 3 constraints. 

For $d = 2$ and $\vert \AQ \vert \geq 3$ we must determine additional parameters
of the underlying classical distribution over $\AQ$. We do so by repeatedly
applying the SIC-POVM with elements $E_y = \frac{1}{2} \ket{\phi_y}
\bra{\phi_y}$, with each $\ket{\phi_y}$ described by Eq. \ref{eq:SIC-POVM}.

The probability of observing the length-2 measurement sequence $y_0 y_1$ can be
written as:
\begin{align*}
\Pr(y_0 y_1 \vert \rho_{0:2}) & = \sum_{x_0,x_1} p_{x_0,x_1} \Pr(y_0 y_1 \vert \ketpsi[x_0] \otimes \ketpsi[x_1]) \\
&= \sum_{x_0,x_1} \frac{p_{x_0,x_1}}{4} \lvert \braket{\psi_{x_0}|\phi_{y_0}} \rvert ^2 \lvert \braket{\psi_{x_1}|\phi_{y_1}} \rvert ^2
~.
\end{align*}
There are $\vert \AQ \vert^2$ $p_{x_0,x_1}$'s. If a set of $p_{x_0,x_1}$'s is a
solution to this system of equations it is consistent with the measurement
statistics, and the solution will be unique for $\vert \AQ \vert = 3$.

For $\vert \AQ \vert = n$ and arbitrary $d$ we may construct a POVM from $\AQ$
as we did for $\ell = 1$. It is then possible infer a set of $p_{x_0,x_1}$ that
solves the resulting system of equations from measurement statistics. For
length-2 words the equations are:
\begin{align*}
\Pr(y_0 y_1 \vert \rho_{0:2}) & = \sum_i p_{x_0,x_1} \Pr(y_0 y_1 \vert \ketpsi[x_0] \otimes \ketpsi[x_1]) \\
&= \sum_{x_0,x_1} p_{x_0,x_1} c_{y_0} c_{y_1} \lvert \braket{\psi_{x_0}|\psi_{y_0}} \rvert ^2 \lvert \braket{\psi_{x_1}|\psi_{y_1}} \rvert ^2
~.
\end{align*}
There are $n^2$ equations, one for each length-2 measurement sequence
corresponding to the POVM elements $E_{y_0} \otimes E_{y_1}$. Calculating the
other possible outcomes (corresponding to $E_n$) is redundant due to
normalization.

\subsubsection{$\ell \geq 3$}

Extending this analysis to a length-$\ell$ density matrix $\rhoL$ takes the
form:
\begin{align}
\nonumber
\rhoL = \sum_{\ketpsi[w] \in \AQ^\ell} \Bigg[ \Pr(\ketpsi[w]) \ketpsi[w] \brapsi[w] \Bigg]
~,
\end{align}
where each $\ketpsi[w]$ has the form of Eq. (\ref{eq:separable_word}). One can
determine the length-$\ell$ word distributions uniquely for the general case
where $\vert \AQ \vert ^\ell < d^{2\ell}$.

The various measurement strategies explored above for $\ell = 2$ can be extended
to arbitrary values of $\ell$. We will explicitly describe two: using a SIC-POVM
for $d=2$ and using a POVM constructed from $\AQ$ for arbitrary $d$.

Consider repeatedly applying the SIC-POVM with elements $E_y = \frac{1}{2}
\ket{\phi_y} \bra{\phi_y}$, with each $\ket{\phi_y}$ described by Eq.
(\ref{eq:SIC-POVM}). Taking $\ell$ measurements, one observes a length-$\ell$
word $y_{0:\ell}$. $\AY$ is the $4$-element alphabet of measured symbols.

The probability of observing the length-$\ell$ measurement sequence $y_{0:\ell}$
can be written as:
\begin{align*}
\Pr(y_{0:\ell} \vert \rhoL) & = \sum_{\ketpsi[w]} \Pr(\ketpsi[w]) \Pr(y_{0:\ell} \vert \ketpsi[w]) \\
&= \sum_{\ketpsi[w]} \frac{\Pr(\ketpsi[w])}{2^\ell} \lvert \braket{\psi_w| \phi_{y_{0:\ell}}} \rvert ^2
~,
\end{align*}
where $\ket{\phi_{y_{0:\ell}}} = \bigotimes_{t = 0}^\ell \ket{\phi_{y_t}}$ and
the factor of $2^\ell$ comes from the POVM elements. Each of the $4^\ell$
sequences has a probability that can be estimated from measurement. This system
of equations can be solved to find underlying length-$\ell$ word probabilities
that are consistent with measurements. If $\vert \AQ \vert \leq 3$ this solution
is unique.

We can also measure $\ell$ times with a POVM constructed directly from $\AQ$. In
this case the resulting equations are:
\begin{align*}
\Pr(y_{0:\ell} \vert \rhoL) & = \sum_{\ketpsi[w]} \Pr(\ketpsi[w]) \Pr(y_{0:\ell}) \vert \ketpsi[w]) \\
& = \sum_{\ketpsi[w]} p_w \left(\prod_{t=0}^{\ell-1} c_{y_t} \right) \lvert \braket{\psi_{w}|\psi_{y_{0:\ell}}} \rvert ^2
~.
\end{align*}
As before, one infer the $\vert \AQ \vert ^ \ell$ underlying word
probabilities ($p_w$'s) uniquely in the case where $\vert \AQ \vert < d^2$.

\subsection{Source Reconstruction}

After observing a separable qudit process and finding the length-$\ell$ density
matrix $\rhoL$ can one infer the HMCQS that generated it? The following
reconstructs the source that generates $\rhoL$ of an unknown process for
different values of $\ell$. Note that a source which generates $\rhoL$ may fail
to generate $\rho_{0:\ell+1}$.

\subsubsection{$\ell = 1$}

After determining $\rho_0$ an observer may construct an i.i.d. approximation of
the quantum information source. Given any decomposition of $\rho_0$ into pure
states $\ketpsi[\meassymbol]$---as in Eq. (\ref{eq:iid_density_matrix})---the
corresponding HMCQS consists of one internal state $\AQ =
\{\ketpsi[\meassymbol]\}$ and the single transition probability for each
$\ketpsi[\meassymbol]$ is its corresponding probability
$\Pr(\ketpsi[\meassymbol])$ in the decomposition of $\rho_0$. A unique model may
be obtained by taking $\rho_0$'s eigendecomposition. In this case the
model emits each eigenstate $\ketpsi[i]$ with probability equal to the
corresponding eigenvalue: $\Pr(\ketpsi[i]) = \lambda_i$.

\subsubsection{$\ell = 2$}

With $\rho_{0:2}$ an observer begins to model a source that generates
correlations between qudits. Doing so requires finding a separable decomposition
of $\rho_{0:2}$ of Eq. (\ref{eq:rho2_separable})'s form. If $\AQ$ is known,
multiple procedures for finding such a separable decomposition have been
introduced above. For a generic two-qudit density matrix, determining whether it
is separable or entangled is generally NP-hard \cite{Ghar08}. There are also
many necessary and sufficient conditions for separability; for example, the
Positive-Partial Transpose (PPT) criterion \cite{Horo97}.

The following assumes that the observer has a separable decomposition of
$\rho_{0:2}$ into alphabet states $\AQ '$ that may or may not be the source's
alphabet $\AQ$. In general, the sets of basis states in the decomposition of a
two-qudit density matrix may differ, but we require a symmetric decomposition
such that the basis states of both qudits are $\AQ'$. From this decomposition
they can construct an HMCQS with an underlying Markov dynamic described by Eq.
(\ref{eq:quantum_markov}). This HMCQS has $\vert \AQ ' \vert$ internal states
and transition probabilities $T_{\sigma_x, \sigma_{x'}} = \Pr(\ketpsi[x'] \vert
\ketpsi[x])$ that can be calculated from $\rho_{0:2}$. Different separable
decompositions of $\rho_{0:2}$ yield different HMCQS, whose statistics over
longer-length sequences may differ. Determining which is a more accurate model
of the source requires performing tomography on $\rho_{0:3}$ to refine the
model.

\subsubsection{$\ell \geq 3$}

Finding a separable decomposition becomes more computationally expensive as the
number of qudits increases \cite{Horo01}. Nevertheless, if one obtains a
separable decomposition of $\rho_{0:3}$ where the basis states for all $3$
qudits are $\AQ '$, then one can create a length-$2$ Markov approximation of
the source. Each length-$2$ sequence of qudits in $\AQ'$ corresponds to a
different internal HMCQS state. Thus, it has $\vert \AQ ' \vert ^2$ internal
states, unless some length-$2$ sequences are forbidden. The transition
probabilities take the form $T_{\sigma_{x_0,x_1},\sigma_{x_1,x_2}} =
\Pr(\ketpsi[x_2] | \ketpsi[x_0] \ketpsi[x_1])$. Only transition probabilities
that obey concatenation are nonzero.

One can continue in this manner, approximating the source given a separable
decomposition of $\rhoL$ to obtain an HMCQS with $\vert \AQ ' \vert ^ {\ell-1}$
states or fewer, if some sequences are forbidden. Each state corresponds to a
word of length $\ell-1$ where the pure states composing the word are drawn from
$\AQ '$. The number of internal states in the model grows exponentially with
$\ell$, but not all of these states may lead to unique future predictions. If
so, they can be combined without a loss of predictivity, as is done in
classical computational mechanics. A general algorithm for doing so is beyond
the present scope.

\subsection{Discussion}

The preceding detailed many aspects of identifying an unknown quantum process by
tomographically reconstructing the length-$\ell$ density matrices $\rhoL$. When
correlations exist between qudits, this provides a predictive advantage over the
common assumption that sources are i.i.d.. Starting with the method for
reconstructing classical processes, we developed a variety of measurement
protocols for different values of $d$ and $\vert \AQ \vert$ and $\ell$ to find a
process' statistics when $\AQ$ is known. We then introduced effective models for
quantum sources derived entirely from separable decompositions of the density
matrices $\rhoL$ reconstructed via tomography.

Since our aim is to harness correlations to improve predictions of future
measurement outcomes, it is worth asking, How accurately can future measurement
outcomes be predicted? To begin to answer this the following now defines
maximally-predictive measurements for the special case of $\ell = 2$ and for
arbitrary $\ell$.
 
For a correlated qudit process, an observer with knowledge of the length-2
density matrix $\rho_{0:2}$ may perform a measurement on the first qudit and
condition upon the outcome to reduce their uncertainty when measuring the second
qudit. They apply $\Meas_0$ (with possible outcomes $y_i$) on the first qubit
$\rho_{0}$. This leaves the joint system in the classical-quantum state:
\begin{align*}
\rho^{\Meas_0}_{0:2} = \sum_{y_i} \Pr(y_i) \ket{y_i} \bra{y_i} \otimes \rho^{y_i}_1
~,
\end{align*}
where the $\rho^{y_i}_1$ are the qudit density matrices conditioned on outcome
$y_i$.

The conditional von Neumann entropy of the second qubit is then:
\begin{align*}
S(\rho_1| \Meas_0(\rho_0)) = \sum_{y_i} \Pr(y_i) S(\rho^{y_i}_1)
~.
\end{align*}
The rank-one measurement with the minimal uncertainty in measurement outcomes is
in the eigenbasis of $\rho^{y_i}_1$. Different $y_i$ values generally
correspond to different minimal-entropy measurements on the second qubit.

For two qubits, we can find the PVM for which the conditional von
Neumann entropy of the second qubit is minimized. This leads to a
basis-independent property of the process:
\begin{align*}
S_{min}(\rho_1|\Meas_{min}(\rho_0)) = \min_{\Meas_0} S(\rho_1| \Meas_0(\rho_0))
~,
\end{align*}
where the minimum is taken over all PVMs on $\rho_0$.

If an experimenter reconstructs $\rhoL$, a measurement on all but the last qudit
in the block with a measurement protocol $\qcChannel$ with possible outcomes
$y_{0:\ell-1}$ leaves the block in the classical-quantum state:
\begin{align*}
\rho^{\qcChannel}_{0:\ell} = \sum_{y_{0:\ell-1}} \Pr(y_{0:\ell-1}) \ket{y_{0:\ell-1}} \bra{y_{0:\ell-1}} \otimes \rho^{y_{0:\ell-1}}_{\ell}
~.
\end{align*}

To minimize the conditional von Neumann entropy of the $\ell$-th qubit, we
calculate:
\begin{align*}
S_{min}(\rho_\ell|\qcChannel_{min}(\rho_{0:\ell-1})) = \min_{\qcChannel} S(\rho_\ell| \qcChannel(\rho_0))
~,
\end{align*}
where the minimum is taken over all local measurement protocols on $\rho_{0:\ell-1}$.

Further development necessitates exploring the space of measurement protocols to
find those with the greatest predictive advantage over i.i.d. models for
arbitrary separable qudit processes.

Finally, we note that our procedure for source reconstruction required a number
of model states that grows exponentially with $\ell$. Future work on finding
\emph{minimal} models from density matrices will also require combining states
with identical predictions and performing inference over possible model
topologies, as with classical Bayesian Structural Inference.

\section{Conclusion}
\label{sec:conclusion}

Inspired by prior information-theoretic studies of classical stochastic
processes, we introduced methods to quantify structure and information
production for stationary quantum information sources. We identified properties
related to the quantum block entropy $S(\ell)$ that allow one to determine the
amount of randomness and structure within a given qudit process. We gave bounds
on informational properties of the resulting measured classical processes. In
particular, we showed that they cannot have a lower entropy rate or block
entropy (at any $\ell$) than the original quantum process.

We analyzed a number of hidden Markov chain quantum sources (HMCQSs),
explaining how an observer synchronizes to a source's internal states via
measuring the emitted qudits. If the source allows synchronizing observations,
then we showed that adaptive measurement protocols are capable of synchronizing
and maintaining synchronization when fixed-basis measurements cannot.

Sequels will extend these methods and results in a number of ways. Despite
focusing here on separable quantum sequences for simplicity, entangled qudit
sequences can similarly be studied by combining a HMCQS and a $D$-dimensional
quantum system capable of sequentially generating matrix product states
\cite{Scho05a}. Doing so will open up the study of entropy convergence of matrix
product operators \cite{Pirv10}.

Many results exist for classical stochastic processes that may be extended to
quantum processes. For example, there exist closed-form expressions for
informational measures for nondiagonalizable classical dynamics
\cite{Crut13a,Riec13a,Riec16a,Riec17a}. Extending these to quantum dynamics
would allow for more accurate determination of the quantum information
properties introduced here. Similarly, the preceding lays the groundwork for
fluctuation theorems and large deviation theory of separable quantum processes.
Finally, it will be worthwhile to develop a causal equivalence relation for
quantum stochastic processes and develop quantum \eMs by extending classical
results \cite{Shal01}.

Separable quantum sequences also serve as a resource for information processing
by finite-state quantum information transducers that transform one quantum
process to another. Beyond interest in their own right, such operations have
thermodynamic consequences, either requiring work to operate (as overcoming
dissipation induced by Landauer erasure \cite{Land61a}) or acting as a quantum
version of information-powered engines capable of leveraging environmental
correlations to perform useful work \cite{Boyd16c, Boyd16d, Jurg20a}. This
behavior has already been demonstrated for certain quantum processes
\cite{Huan22}.

Finally, though spatially-extended, ground and thermal states of spin chains
under various Hamiltonians are quantum processes. And so, the quantum
information measures introduced here can serve to classify these states
according to the source complexity required to generate them.

\section*{Acknowledgments}
\label{sec:acknowledgments}

We thank Fabio Anza, Sam Loomis, Alex Jurgens, and Ariadna Venegas-Li and
participants of the Telluride Science Research Center Information Engines
Workshops for helpful discussions. JPC acknowledges the kind hospitality of the
Telluride Science Research Center, Santa Fe Institute, and California Institute
of Technology for their hospitality during visits. This material is based upon
work supported by, or in part by, Grant Nos. FQXi-RFP-IPW-1902 and
FQXi-RFP-1809 from the Foundational Questions Institute and Fetzer Franklin
Fund (a donor-advised fund of Silicon Valley Community Foundation) and grants
W911NF-18-1-0028 and W911NF-21-1-0048 from the U.S. Army Research Laboratory
and the U.S. Army Research Office.

\appendix

\section{Information in Classical Processes}
\label{app:classical_properties}

This section briefly recounts properties that quantify the randomness and
correlation of classical stochastic processes as developed in Ref.
\cite{Crut01a}. Details and complete proofs can be found there. The main text
here introduces versions appropriate to quantum processes.

\subsection{Shannon Entropy}

We quantify the amount of uncertainty in a discrete random variable $X$
with possible outcomes $\{x_1,x_2,...x_n\}$ by its \emph{Shannon entropy}:
\cite{Shan48a}:
\begin{align*}
\H{X} \equiv -\sum_{i=1}^{n} \Pr(x_i) \log_2 \Pr(x_i)
~.
\end{align*}
We use $\log_2$, in which case the units for Shannon entropy are \emph{bits}.

To study correlations between multiple random variables we use several
additional information quantities related to the Shannon entropy. First, the
\emph{joint entropy} of two discrete random variables---$X$ and $Y$, with
possible outcomes $\{x_1,x_2,...x_n\}$ and $\{y_1,y_2,...y_m\}$,
respectively---is defined as:
\begin{align*}
\H{X,Y} \equiv -\sum_{i=1}^{n}\sum_{j=1}^{m}
	\Pr(x_i,y_j) \log_2 \Pr(x_i,y_j)
~.
\end{align*}
$X$ and $Y$ are statistically independent if and only if the joint entropy
decomposes as $\H{X,Y} = \H{X} + \H{Y}$.

Second, the \emph{conditional entropy} of $X$ conditioned on $Y$ is:
\begin{align*}
\H{X \vert Y} \equiv -\sum_{i=1}^{n} \sum_{j=1}^{m}
	\Pr(x_i,y_j) \log_2 \Pr(x_i \vert y_j)
~.
\end{align*}
$\H{X \vert Y}$ is the uncertainty in the value of $X$ after already knowing
the value of $Y$. Note that $\H{X \vert Y} \geq 0$ and is not symmetric: $\H{X
\vert Y} \neq \H{Y \vert X}$. If $X$ and $Y$ are uncorrelated, then $\H{X \vert
Y} = \H{X}$.

From the above definitions one can derive the following identity, linking
conditional and joint entropies:
\begin{align*}
\H{X \vert Y} = \H{X,Y} - \H{Y}
~.
\end{align*}

Third and finally, the \emph{mutual information} between two random variables
is:
\begin{align*}
    \I{X:Y} \equiv \H{X} - \H{X \vert Y}
    ~.
\end{align*}
This is the amount of information one can gain about $X$ by having complete
knowledge of $Y$. The mutual information is symmetric, and $\I{X:Y} = 0$ if and
only if $X$ and $Y$ are statistically-independent.

\subsection{Block Entropy}

For a classical stochastic process we quantify the amount of uncertainty in a
block of $\ell$ consecutive random variables by taking the joint entropy
$\H{\LPresent{\ell}}$. This is the \emph{block entropy}:
\begin{align*}
\H{\ell} & \equiv \H{\LPresent{\ell}} \nonumber \\
	& = - \sum_{\lpresent{\ell}} \Pr(\lpresent{\ell}) \log_2 \Pr(\lpresent{\ell})
  ~,
\end{align*}
where the sum is taken over all words of length $\ell$ and $\H{0} \equiv 0$.

$\H{\ell}$ is monotonically increasing and concave down, and its behavior as
$\ell \to \infty$ is indicative of a process' correlations and randomness
\cite{Crut01a}. The generic behavior of $\H{\ell}$ and its relation to other
information properties that we will define can be seen in Fig.
\ref{fig:process_properties}.

\subsection{Shannon Entropy Rate}

The following briefly summarizes Ref. \cite{Crut01a}'s results. Refer there
for a more thorough exploration of information-theoretic quantities
related to multivariate systems and the block entropy.

First among these is the \emph{Shannon entropy rate} $\hmu$:
\begin{align}
\hmu = \lim_{\ell \to \infty} \frac{\H{\ell}}{\ell}
  ~.
\label{eq:hmu}
\end{align}
The limit in Eq. (\ref{eq:hmu}) is guaranteed to exist for all stationary
processes \cite{Cove06a}. $\hmu$ is irreducible randomness produced by an
information source. Its units are \emph{bits~per~symbol}.

The Shannon entropy rate can equivalently be written using the \emph{conditional
entropy}:
\begin{align}
\hmu = \lim_{\ell \to \infty} \H{\Present \vert \MeasSymbol_{-\ell:0}}
~,
\label{eq:hmu2}
\end{align}
Therefore, $\hmu$ can equivalently be thought of as the average uncertainty in
the next symbol if all preceding symbols are known.

To better appreciate $\hmu$ we consider a few simple cases.

For an i.i.d. process the block entropy trivially is $\H{\ell} = \ell
\H{\Present}$ and, therefore, $\hmu = \H{\Present}$. Since there are no
correlations between variables, knowledge of past symbols cannot reduce the
uncertainty of the next.

If a process is periodic (for example consisting of alternating $0$'s and
$1$'s), then a keen observer will note this pattern and be able to predict with
certainty all future symbols of the process. In this case $\hmu = 0$.

For stationary Markov and hidden Markov processes, an observer can leverage past
observations to reduce their uncertainty about succeeding symbols. $\hmu$ will
be their average uncertainty about the next symbol once they have accounted for
all of the correlations with past symbols.

Graphically, $\hmu$ corresponds to the slope of the block entropy curve as
$\ell \to \infty$ as shown in Fig. \ref{fig:process_properties}. 

\subsection{Redundancy}

For a stochastic process with alphabet $\AX$ the maximum entropy rate is
$\log_2 \vert \AX \vert$, corresponding to i.i.d. random variables
$\MeasSymbol_t$ with uniform distributions over all measurement outcomes
$\meassymbol_t$. Any other stationary process can be \emph{compressed} down to
its entropy rate $\hmu < \log_2 \vert \AX \vert$.

The amount that a particular source can be compressed is known as its
\emph{redundancy}, defined as:
\begin{align*}
\mathbf{R} \equiv \log_2 \vert \AX \vert - \hmu
~.
\end{align*}
$\mathbf{R}$ includes two very different effects: bias within individual random
variables and correlations between different random variables. To determine the
relative importance of those two factors requires closer examination of
$\H{\ell}$.

\subsection{Block Entropy Derivatives and Integrals}

Since the limit in Eq. (\ref{eq:hmu}) exists, $\H{\ell}$ scales (at most)
linearly. We are interested in how $\H{\ell}$ converges to its linear
asymptote, and we will see that taking discrete derivatives of $\H{\ell}$ (and
integrals of those derivatives) provides us with useful quantities for
classifying processes.

Consider applying a discrete derivative operator $\Delta$ to a function $F:
\mathbb{Z} \to \mathbb{R}$:
\begin{align*}
    \Delta F (\ell) = F(\ell) - F(\ell - 1)
    ~.
\end{align*}

We can apply $\Delta$ to $F$ multiple times to obtain higher-order derivatives:
\begin{align*}
    \Delta^n F (\ell) = (\Delta \circ \Delta^{n-1}) F(\ell)
    ~.
\end{align*}

Taking discrete derivatives of $\H{\ell}$ yields a set of functions
$\Delta^n \H{\ell}$. We then study how these discrete derivatives themselves
converge to their asymptotic values. To do so, we take ``integrals'' of a
discrete function $\Delta F(\ell)$ in the following manner:
\begin{align}
\sum_{\ell=A}^B \Delta F (\ell) = F(B) - F(A - 1)
~.
\label{eq:discrete_integral}
\end{align}

To study the convergence properties of each $\Delta^n \H{\ell}$, we
compare it at each $\ell$ to its asymptotic value $\lim_{\ell \to \infty}
\Delta^n \H{\ell}$. We do so with the following general integral form:
\begin{align}
    \mathcal{I}_n \equiv \sum_{\ell=\ell_0}^\infty [\Delta^n H (\ell) - \lim_{\ell \to \infty} \Delta^n \H{\ell}]
    ~,
    \label{eq:discrete_integral_block_entropy}
\end{align}
where $\ell_0$ is the first value of $\ell$ for which $\Delta^n \H{\ell}$ is defined.

\subsection{Entropy Gain}

The first derivative of $\H{\ell}$ is known as the \emph{entropy gain}.
It is defined as:
\begin{align*}
\Delta \H{\ell} \equiv \H{\ell} - \H{\ell - 1}
~,
\end{align*}
for $\ell > 0$. We set $\Delta \H{0} \equiv \log_2(\AX)$. The entropy
gain is the amount of additional uncertainty introduced by increasing the block
size by one random variable, and its units are \emph{bits~per~symbol}. Note that,
because $\H{\ell}$ is monotone increasing and concave, $\Delta \H{\ell} \geq
\Delta \H{\ell+1} \geq 0$, for all $\ell$. This behavior is shown in Fig.
\ref{fig:entropy_convergence_properties}.

The entropy gain can be rewritten as a conditional entropy:
\begin{align}
\Delta \H{\ell} = \H{\Present \vert \MeasSymbol_{-\ell:0}}
~,
\label{eq:entropy_gain_conditional}
\end{align}
in which case its relation to the entropy rate becomes clear. Using Eq.
(\ref{eq:hmu2}) we see that:
\begin{align*}
\hmu = \lim_{\ell \to \infty} \Delta \H{\ell}
~.
\end{align*}

For an observer with no prior knowledge of an information source, it will often
be necessary to estimate the entropy rate of a source using finite sequences of
data. In this case the entropy gain can serve as a finite-$\ell$ approximation
of the true entropy rate:
\begin{align}
\hmu(\ell) & \equiv \Delta \H{\ell} \nonumber \\
            & \equiv \H{\ell} - \H{\ell - 1}
  ~.
\label{eq:hmu_est}
\end{align}

\subsection{Predictability Gain}

The second derivative of $\H{\ell}$ is the \emph{predictability gain}, defined
as:
\begin{align*}
\Delta^2 \H{\ell} & \equiv \Delta \hmu(\ell) \\
    & = \hmu(\ell) - \hmu(\ell-1) 
~,
\end{align*}
where $\ell > 0$. Note that $\Delta^2 \H{\ell} \leq 0$.

$ \vert \Delta^2 \H{\ell} \vert $ is the average amount of additional predictive information
an observer obtains when expanding their observations from blocks of length
$\ell-1$ to blocks of length $\ell$, and its units are \emph{bits~per~symbol$^2$}.
Large values of $ \vert \Delta^2 \H{\ell} \vert $ imply that the $\ell$-th measurement is
particularly informative to an observer and therefore greatly improves their
estimate of the entropy rate as given by Eq. (\ref{eq:hmu_est}).

One can also calculate higher-order discrete derivatives of $\H{\ell}$. For our
purposes this is not necessary except to note that, for stationary processes:
\begin{align*}
\lim_{\ell \to \infty} \Delta^n \H{\ell} = 0, n \geq 2
~.
\end{align*}
For $n=2$ this follows from convergence of $\hmu(\ell)$, and the argument for
all $n > 2$ is similar.

\subsection{Total Predictability}
We can now integrate the functions $\Delta^n \H{\ell}$. As a general heuristic,
the larger the magnitude of these integrals, the more correlation or
statistical bias exists within the process.

We begin by studying how the predictability gain $\Delta^n \H{\ell}$ converges
to its asymptotic value $\lim_{\ell \to \infty} \Delta^2 \H{\ell} = 0$. Since
$\Delta^2 \H{0}$ is undefined, we will integrate using Eq.
(\ref{eq:discrete_integral_block_entropy}) with $\ell_0 = 1$ to obtain the total
predictability $\mathbf{G}$:
\begin{align}
    \mathbf{G} \equiv \mathcal{I}_2 = \sum_{\ell=1}^\infty \Delta^2 H (\ell)
    ~.
\label{eq:total_predictability}
\end{align}
Since $\Delta^2 \H{\ell} < 0$ for all $\ell$, $\mathbf{G} < 0$ as well. The
units of $\mathbf{G}$ are \emph{bits~per~symbol}. Graphically, it is the the
area between the predictability gain curve and its linear asymptote of $0$, as
seen in Fig. \ref{fig:predictability_convergence_properties}.

To interpret $\mathbf{G}$'s value, we apply Eq. (\ref{eq:discrete_integral}) to
get:
\begin{align*}
\mathbf{G} & = -\Delta \H{0} + \lim_{\ell \to \infty} \Delta \H{\ell} \\
            & = - \log_2( \vert \AX \vert ) + \hmu \\
            & = - \mathbf{R}
~.
\end{align*}
A process' total predictability is then equal in magnitude to its redundancy,
and we can interpret $ \vert \mathbf{G} \vert $ as the amount of predictable
information per symbol for a process. Here, we emphasize once more that for a
given process with alphabet $\AX$, any random variable has a maximum entropy of
$\log_2( \vert \AX \vert )$, that consists of two kinds of information: $\hmu$,
the irreducible randomness, and $ \vert \mathbf{G} \vert $, the amount of
information that an observer can possibly predict about it.

The total predictability is thus a function of the entropy rate for a given
process and shares $\hmu$'s weakness: it cannot identify statistical
correlations between random variables. A large value of $ \vert \mathbf{G}
\vert $ could be the result of either an i.i.d. process with heavily-biased
random variables or strong correlations between subsequent variables.
Fortunately, our next quantity does distinguish between these cases.

\subsection{Excess Entropy}

Investigating the convergence of the entropy gain to the asymptotic entropy rate
$\hmu$ leads to a well-studied process property, the excess entropy:
\begin{align*}
\EE \equiv \mathcal{I}_1 & = \sum_{\ell=1}^\infty \left[ \Delta \H{\ell} - \lim_{\ell \to \infty} \Delta \H{\ell} \right] \\
        & = \sum_{\ell=1}^\infty \left[ \Delta \H{\ell} - \hmu \right]
~.
\end{align*}
Since $\Delta \H{\ell} \geq \hmu$, $\EE \geq 0$ and its units are \emph{bits}.

Our interpretation of $\EE$ becomes clearer after applying Eq.
(\ref{eq:discrete_integral}) to obtain:
\begin{align}
\EE = \lim_{\ell \to \infty} \left[ \H{\ell} - \hmu \ell \right]
~.
\label{eq:EE_block}
\end{align}
While $\hmu$ determines the linear asymptotic behavior of $\H{\ell}$, $\EE$
encapsulates all sublinear effects. We largely discuss processes for which
$\EE$ is finite, known as \emph{finitary} processes. A process for which $\EE$
is infinite (for example if $\H{\ell}$ scales logarithmically with $\ell$), is
known as an \emph{infinitary} process. Graphically, $\EE$ corresponds to the
$\ell = 0$ intercept of the linear asymptote to the block entropy curve, as
shown in Fig. \ref{fig:process_properties}, as well as the area between the
area between the entropy gain curve and its asymptote $\hmu$ as seen in Fig.
\ref{fig:entropy_convergence_properties}.

The excess entropy can also be written as a mutual information between two
halves of a process' chain of random variables:
\begin{align}
\EE = \lim_{\ell \to \infty} \I{\MeasSymbol_{-\ell:0}:\MeasSymbol_{0:\ell}}
  ~.
\label{eq:EE_mutual_info}
\end{align}
This suggests it is the total amount of information in a process' past useful
for predicting the future. It is therefore considered an indicator of the
amount of process \emph{memory}. A more structural approach reveals that $\EE$
is a lower bound on the actual amount of memory required to predict a
stochastic process \cite{Shal01}.

Importantly, $\EE$ easily distinguishes between i.i.d. processes (for which
$\EE = 0$) and processes with correlations between random variables ($\EE >
0$). For all processes that are periodic with period $p$, $\hmu = 0$ and
$\H{\ell}$ reaches a maximum value of $\log_2 p$ for $\ell = p$. Therefore, all
period-$p$ processes have $\EE = \log_2 p$.

As with $\hmu$, it is often useful for an observer to make an approximation of
the true excess entropy using only finite length-$\ell$ symbol sequences. Using
Eq. (\ref{eq:EE_block}) we can estimate $\EE$ as:
\begin{align}
\EE(\ell) & \equiv \H{\ell} - \ell\hmu(\ell)
~.
\label{eq:EE_est}
\end{align}

\subsection{Transient Information}

Based upon our analysis of the excess entropy, we can now say that as $\ell \to
\infty$ the block entropy curve has a linear asymptote:
\begin{align}
\H{\ell} \sim \EE + \hmu \ell
~.
\label{eq:block_asymptote}
\end{align}

We capture the way that $\H{\ell}$ converges to this asymptote by taking
another discrete integral to obtain the transient information:
\begin{align*}
\TI \equiv -\mathcal{I}_0 = \sum_{\ell=0}^\infty \left[ \EE + \hmu \ell - \H{\ell} \right]
~.
\end{align*}
The units of $\TI$ are \emph{bits$\times$symbol}.  $\TI$ is represented graphically in Fig.
\ref{fig:process_properties} as the area between the $\H{\ell}$ curve and its
linear asymptote for $\ell \to \infty$.

The transient information can be rewritten as:
\begin{align*}
\TI = \sum_{\ell = 1}^{\infty}\ell \left[ \hmu(\ell) - \hmu \right]
~,
\end{align*}
indicating that $\TI$ is a measure of how difficult it is to
\emph{synchronize} to a process. An observer is considered ``synchronized'' when
they are able to infer exactly which internal state the source currently
occupies. If they remain synchronized, then they can optimally predict future
measurement outcomes. That is, their estimate of the source's entropy rate
$\hmu(\ell)$ is equal to its actual entropy rate $\hmu$.

$\TI$ is a notable quantity since, unlike $\EE$, it is capable of distinguishing
between different period-$p$ processes.

\subsection{Markov Order}

The final property we introduce for classical stochastic processes is the
Markov order. A process has Markov order $\MOrder$ if $\MOrder$ is the minimum
value for which the following condition holds:
\begin{align*}
\Pr(\Present  \vert  \MeasSymbol_{\MOrder:0}) = \Pr(\Present  \vert  \MeasSymbol_{-\infty:0})
~.
\end{align*}

From Eq. $\eqref{eq:markov}$ we conclude that a Markov process is one for
which $\MOrder \leq 1$. More generally, for a process with Markov order
$\MOrder$ the distribution for $\MeasSymbol_t$ depends only on the previous
$\MOrder$ symbols. Equivalently, $\MOrder$ is the value of $\ell$ for which the
block entropy reaches its linear asymptote:
\begin{align*}
\H{\MOrder} = \EE + \hmu \MOrder
~.
\end{align*}
This fact is represented graphically in Fig.
\ref{fig:process_properties}.

The overwhelming majority of finite-state hidden Markov processes have infinite
Markov order, meaning that $\H{\ell} < \EE + \hmu \ell$ for all $\ell$
\cite{Jame10a}. Nevertheless, an observer can still make use of finite-$\ell$
estimates of the information properties defined here. In fact, these
estimates typically converge exponentially fast with $\ell$ \cite{Trav14a}.

\section{Quantum Channels for Preparation and Measurement}
\label{appendix:cq_channels}

The main text claims that a separable qudit process is the result of passing
realizations of a classical process $\BiInfinity$ with symbol alphabet $\AX$
through a classical-quantum channel that takes $x \in \AX \to
\ketpsi[\meassymbol] \in \HSpace$. It also claims that the act of measurement
can be described as a quantum-classical channel taking $\rhoL \to y_{0:\ell}$.
The following elaborates on both claims, adopting the formalism of Ref.
\cite{Wild17}.

We first describe passing realizations of a classical process $\BiInfinity$
through a $\emph{conditional quantum encoder}$. Consider a classical-quantum
state composed of a classical register of dimension $\vert \AX \vert$ and a
qudit in $\HSpace$ initialized in the state $\ket{0}$. Furthermore, let the
classical register store the outcome of some classical random variable $X_t$
with outcomes $x \in \AX$ such that:
\begin{align}
\rho_{XA} = \sum_x \Pr(x) \ket{x}\bra{x}_{X} \otimes \ket{0}\bra{0}_{A}
~.
\end{align}

This state will serve as input to a conditional quantum encoder $\cqChannel_{XA
\to B}$ that consists of a set $\{ \cqChannel^x_{A\to B}\}$ of $\vert \AX
\vert$ completely-positive trace-preserving (CPTP) maps. We construct
each map such that it transforms the initial quantum state $\ket{0}$ to the
desired pure qudit state $\ketpsi[\meassymbol]$; i.e.:
\begin{align}
\cqChannel^x_{A \to B}(\ket{0}\bra{0}_A) = \ketpsi[\meassymbol]\brapsi[\meassymbol]_B
~,
\end{align}
with $\cqChannel^x_{A \to B}$ being a unitary operation. (For $d=2$ we can
consider rotations on the Bloch sphere.)

The encoding is:
\begin{align*}
\rho_B &= 
\cqChannel_{XA \to B}(\rho_{XA}) \\
& = tr_X\left(\sum_x \Pr(x) \ket{x}\bra{x} \otimes \cqChannel^x_{A \to B}\left(\ket{0}\bra{0}\right) \right) \\
& = \sum_x \Pr(x) \ketpsi[\meassymbol]\brapsi[\meassymbol]_B
~.
\end{align*}

Now, taking $\ell$ classical registers that consist of length-$\ell$ words
$\meassymbol_{0:\ell}$ of a classical process $\BiInfinity$, we can use
$\cqChannel_{XA \to B}$ to encode $\MeasSymbol_{0:\ell}$ and obtain:
\begin{align*}
\rhoL &= tr_{X_{0:\ell}} \left( \sum_{x_{0:\ell}} \Pr(x_{0:\ell}) \prod_{t=0}^{\ell-1} \ket{x_t}\bra{x_t} \otimes \cqChannel^{x_t}_{A \to B} \left(\ket{0}\bra{0}\right) \right) \\ 
& = \sum_{w} \Pr(w) \ketpsi[w]\brapsi[w]
~,
\end{align*}
where $\ket{\psi_w}$ take the separable form of Eq.
(\ref{eq:separable_word}) and $\rhoL$ therefore matches Eq.
(\ref{eq:finite_rho}).

Measurement of qudits is described similarly. Let $\Meas$ be a POVM with
elements $\{E_y\}$ that acts on a qudit state $\rho$ in such a way that it
records each measurement outcome in a classical register $Y$. The distribution
over values in $Y$ is determined by:
\begin{align*}
Y & = \Meas(\rho) \\
    & = \sum_y tr(E_y \rho) \ket{y} \bra{y}
~.
\end{align*}

Likewise consider $\qcChannel_{0:\ell}$ to be a length-$\ell$ sequence of
measurements with possible measurement outcome sequences $y_{0:\ell}$ determined
according to some protocol $\qcChannel$ . When applying $\qcChannel_{0:\ell}$ to
$\ell$ consecutive qudits in the joint state $\rhoL$ we assume
$\qcChannel_{0:\ell}$ consists of local measurements in the form of Eq.
(\ref{eq:local_measurement}). In this case we factor the POVM elements
corresponding to particular sequences of measurement outcomes: $E_{y_{0:\ell}} =
\bigotimes_{t=0}^{\ell-1} E_{y_t}$. A length-$\ell$ measurement outcome can be
stored in $Y_{0:\ell}$, a set of $\ell$ classical registers, with its associated
probability $tr(E_{y_{0:\ell}}\rhoL)$ so that:
\begin{align*}
Y_{0:\ell} &= \qcChannel_{0:\ell}(\rhoL) \\
    & = \sum_{y_{0:\ell}} tr(E_{y_{0:\ell}}\rhoL) \ket{y_{0:\ell}}\bra{y_{0:\ell}} \\
    & = \sum_{y_{0:\ell}} tr(\bigotimes_{t=0}^{\ell-1} E_{y_t} \rhoL) \ket{y_{0:\ell}} \bra{y_{0:\ell}}
~,
\end{align*}
where the last line assumes local measurements. Each $\ket{y_{0:\ell}}$ is then
a separable state and all $\ket{y_{0:\ell}}$ are orthogonal.

\bibliography{chaos}

\end{document}